\documentclass[reprint,nofootinbib,aps,prd,superscriptaddress]{revtex4-2}

\pdfoutput=1

\usepackage[T1]{fontenc}

\usepackage{graphicx}
\usepackage[dvipsnames]{xcolor}

\usepackage[hidelinks]{hyperref}
\hypersetup{colorlinks=true,linkcolor=Blue,citecolor=Blue,urlcolor=Blue}

\usepackage{amsmath}
\usepackage{amssymb}
\usepackage[normalem]{ulem}
\DeclareMathAlphabet{\mathbfrak}{U}{euf}{b}{n}

\usepackage{amsthm}

\newtheorem*{corollary*}{Corollary}

\usepackage{calrsfs}

\usepackage[cal=cm,scr=rsfs]{mathalfa}

\newcommand{\dif}{\mathrm{d}}
\newcommand{\Dif}{\mathrm{D}}
\newcommand{\const}{\mathrm{const.}}

\newcommand{\tr}{\mathrm{tr}}
\newcommand{\Left}{\mathopen{}\mathclose\bgroup\left}
\newcommand{\Right}{\aftergroup\egroup\right}
\newcommand{\sform}{\boldsymbol{\omega}}

\newcommand{\lin}{\mathrm{lin}}
\newcommand{\hor}{\mathrm{H}}
\newcommand{\ver}{\mathrm{V}}
\newcommand{\thor}{{\tilde{\mathrm{H}}}}
\newcommand{\tver}{{\tilde{\mathrm{V}}}}

\newcommand{\hthor}{\hat{{\tilde{\mathrm{H}}}}}
\newcommand{\htver}{\hat{{\tilde{\mathrm{V}}}}}
\newcommand{\sasaki}{\boldsymbol{\mathfrak{g}}}
\newcommand{\tsasaki}{\tilde{\sasaki}}
\newcommand{\htsasaki}{\hat{\tilde{\mathbfrak{g}}}}
\newcommand{\induced}{\boldsymbol{\mathfrak{h}}}
\newcommand{\tinduced}{\tilde{\induced}}
\newcommand{\htinduced}{\hat{\tilde{\mathbfrak{h}}}}
\newcommand{\xp}{(x,\boldsymbol{p})}
\newcommand{\stress}{\mathcal{T}}
\newcommand{\normal}{\boldsymbol{n}}
\newcommand{\hnormal}{\tilde{\boldsymbol{n}}}
\newcommand{\nnormal}{\boldsymbol{N}}
\newcommand{\hnnormal}{\hat{\boldsymbol{N}}}
\newcommand{\tieta}{\tilde{\boldsymbol{\eta}}}
\newcommand{\htieta}{\hat{\tieta}}
\newcommand{\diver}{\mathrm{div}}
\newcommand{\hdiver}{\widehat{\mathrm{div}}}
\newcommand{\dvolp}{\boldsymbol{\mathrm{d}}\widetilde{\boldsymbol{\mathrm{vol}}}_x(p)}
\newcommand{\hdvolp}{\boldsymbol{\mathrm{d}}\widehat{\widetilde{\boldsymbol{\mathrm{vol}}}}_x(p)}
\newcommand{\massshell}{\mathscr{C}}
\newcommand{\masshyp}{\mathscr{P}}

\newcommand{\wrongeq}{\overset{!}{=}}

\begin{document}

\title{Relativistic Kinetic Theory of Collisionless Particles with Variable Mass and Frame Transformations}

\author{Adri\'{a}n Casado-Turri\'{o}n}
\email{adricasa@ucm.es}
\affiliation{%
    Fundamentini\k{u} Tyrim\k{u} Skyrius,
    Fizini\k{u} ir Technologijos Moksl\k{u} Centras (FTMC), \\
    Saul\.{e}tekio al.~3, 10257 Vilnius, Lithuania.
}%
\author{Paulo B.~Ferraz}
\email{paulo.ferraz@pi.infn.it}
\affiliation{INFN, Sezione di Pisa, Largo B. Pontecorvo 3, 56127 Pisa, Italy}%
\affiliation{%
    Faculdade de Ci\^{e}ncias e Tecnologia and CFisUC, Departamento de F\'{i}sica, \\ 
    Universidade de Coimbra, Rua Larga, P-3004-516 Coimbra, Portugal.
}%
\author{Mindaugas Kar\v{c}iauskas}
\email{mindaugas.karciauskas@ftmc.lt}
\affiliation{%
    Fundamentini\k{u} Tyrim\k{u} Skyrius,
    Fizini\k{u} ir Technologijos Moksl\k{u} Centras (FTMC), \\
    Saul\.{e}tekio al.~3, 10257 Vilnius, Lithuania.
}%
\author{Jos\'{e} Jaime Terente D\'{i}az}
\email{jterente@uc.pt}
\affiliation{%
    Faculdade de Ci\^{e}ncias e Tecnologia and CFisUC, Departamento de F\'{i}sica, \\ 
    Universidade de Coimbra, Rua Larga, P-3004-516 Coimbra, Portugal.
}%

\date{\today}

\begin{abstract}

In this work, we present the relativistic kinetic theory of a gas of collisionless point particles with non-constant mass, which can also be understood as being subject to a scalar force. Our results hold in an arbitrary conformal frame, and we explicitly derive how they transform between frames. The mathematical construction relies on a tilted Sasaki metric, whose associated phase-space geometry is adapted to particles of variable mass, as well as on the one-particle distribution function, from which macroscopic observables are derived. Finally, we show that the one-particle distribution function can always be taken to be gauge- and frame-invariant, and the corresponding generalisation of the Liouville (or collisionless Boltzmann) equation is satisfied in each frame.
\end{abstract}

\maketitle

\section{Introduction}

By combining the principles of statistical mechanics with a geometric description of spacetime and gravitation, relativistic kinetic theory provides the framework to describe the macroscopic behaviour of relativistic many-particle systems from their underlying microscopic dynamics \cite{Synge:1934zzb,synge1957relativisticgas,Ehlers1971}. Since its inception, which saw the generalisation of the Boltzmann equation to relativistic settings \cite{Israel1972,Cercignani2002}, the mathematical foundations of the theory, formulated in terms of phase-space geometry, have been thoroughly investigated for gases of free massive point particles \cite{Sarbach:2013fya,Sarbach:2013uba,Acuna-Cardenas:2021nkj}, as well as charged particles of constant mass \cite{Sarbach:2013hna}. 

Moreover, relativistic kinetic theory has become an indispensable tool for modelling a variety of physical phenomena across different scales. For instance, in cosmology, it underpins our understanding of the thermal history of the Universe \cite{Kolb:1990vq}, including primordial nucleosynthesis \cite{Iocco:2008va}, relic-neutrino freeze-out \cite{Birrell:2014uka,Bond:2024ivb}, anisotropies in the cosmic microwave background \cite{Ma:1995ey} and structure formation \cite{Mo2010, Nascimento:2024hle}. Meanwhile, in the context of strong-field gravity, relativistic kinetic theory is crucial for the precise modelling of compact-object dynamics \cite{Chavanis:2019ouz,Chavanis:2019mkk} such as neutron stars and black-hole accretion discs \cite{Galishnikova:2022mjg,ThesisRioseco}. In recent times, studies have surpassed the standard Einsteinian paradigm, venturing into the realm of modified theories of gravity \cite{Wojnar:2022dvo}, where compact objects constitute a natural arena where deviations from General Relativity can be probed \cite{AparicioResco:2016xcm,Wojnar:2018hal,Lope-Oter:2023urz}. In particular, research on accretion discs has has drawn increasing attention with the advent of instruments such as the Event Horizon Telescope, enabling direct imaging of the surroundings of ultracompact objects \cite{EventHorizonTelescope:2019dse}, which may then be compared with simulations of their optical appearance in any given theory of gravity \cite{Guerrero:2021ues}.

Practical applications in a number of theories beyond Einstein's gravity, such as scalar-tensor theories, often rely on conformal rescalings of the metric tensor \cite{Faraoni:1998qx,Sotiriou:2008rp}, specifically those leading to the so-called Einstein frame, where the action is rendered as that of General Relativity plus additional terms \cite{Maeda:1988ab}. Conformal rescaling have been recognised as mathematically equivalent to spacetime-dependent transformations of units of measure \cite{Dicke:1961gz,Brans:1962wsm,Brans:1996tp,Catena:2006gk}. While, historically, there has been much debate in the literature concerning the equivalence of the different conformal frames \cite{Quiros:2012rnn}, such equivalence is now well established at the classical level \cite{Postma:2014vaa,Jarv:2015kga,Domenech:2016yxd,Magnano:1993bd,Faraoni:2006fx,Deruelle:2010ht,Postma:2014vaa}, although the situation is not as clear once quantum aspects are taken into account \cite{Shapiro:1995kt,Steinwachs:2011zs,Kamenshchik:2014waa,Pandey:2016unk}. In this work, we are only concerned with entirely classical scenarios where frame equivalence holds. However, even in such cases, the conformal transformation manifests itself in the form of interpretational differences between frames \cite{Deruelle:2010ht}, such as coupling constants \cite{Fujii:2003pa} and temperature \cite{Faraoni:2023gqg} becoming spacetime-dependent after the transformation has been carried out. Properly accounting for these differences is crucial when translating results between conformally related frames \cite{Karciauskas:2022jzd,Diaz:2023tma,Casado-Turrion:2026qvs}.

In relativistic kinetic theory, the most relevant interpretational subtlety associated with conformal transformations is the fact that, if non-vanishing, a particle's mass generally differs between frames. Indeed, if a particle has a non-vanishing constant mass as measured in one frame, it will then generally acquire a spacetime-dependent one in another \cite{Catena:2006bd}. In other words, the notion of \emph{free} massive particles becomes frame-dependent. This is problematic because the phase-space formulation of relativistic kinetic theory relies on such a notion. In particular, the statistical interpretation derived from a gas of massive particles which are free in some frame is inequivalent to that stemming from a gas of particles which have constant mass in another. So far, the issue has apparently been overlooked in the literature. We point out that this subtlety does not arise in the case of photons or other massless particles, which remain massless after conformal transformation. In fact, a number of previous investigations have considered ultrarelativistic systems as seen from different frames \cite{Baier:2007ix,Anguige:1999gh,Denicol:2014tha,Joudioux:2019ecs,Lee:2023cbn}.

In light of the aforementioned gap in the literature, in this work we provide a consistent phase-space formulation of the relativistic kinetic theory of a gas of collisionless point particles of non-constant mass. We carry out the construction in an arbitrary conformal frame and provide the transformation rules relating it to other frames. It is shown that such a gas can always be described statistically using a one-particle distribution function which is both frame- and gauge-invariant, and obeys a modified version of the collisionless Boltzmann (or Liouville) equation in each frame. It is possible to define macroscopic observables from the one-particle distribution function, namely the particle current density and the stress-energy tensor, satisfying the expected conformal-transformation rules. The aforementioned results are achieved by providing a suitable generalisation of the mathematical structures necessary to describe particles which are free in a given frame, such as the metric and volume elements in phase space, in order to render them conformally covariant. An additional advantage of our formalism is that it is independent of the parametrisation employed to describe the particle's world-line, in contrast with most of the preceding literature, where the gauge is fixed to be affinely-related to proper time.

The article is organised as follows. First, in Section \ref{sec:massive particles}, we summarise the action principle and Hamiltonian formulation for a particle of non-constant mass, as seen from two distinct (yet arbitrary) conformal frames. We pay special attention to the primary constraint the particle is subject to, which turns out to correspond to the same hypersurface (the mass shell) in all conformal frames. Next, in Section \ref{sec:horizontal and vertical}, we generalise the notion of horizontal subspace of phase-space vectors (already present in the description of free particles) by tilting the so-called connection map in each frame, thus obtaining a distinct tilted horizontal subspace associated to every conformal rescaling. The preceding construction will allows us to define a metric in phase space per conformal frame, the tilted Sasaki metrics, in Section \ref{sec:sasaki metric}. Subsequently, in Section \ref{sec:volume elements}, we define the volume elements in phase space derived from the tilted Sasaki metrics of each frame. Once these volume forms are established, it is possible to define the integrals appearing in the statistical interpretation of a gas of varying-mass particles. This is done in Section \ref{sec:statistical interpretation}, where we introduce the one-particle distribution function from where the macroscopic properties of the gas (particle current density and stress-energy tensor) emerge. The generalised Liouville (or collisionless Boltzmann) equation in each frame is obtained by requiring the conservation of the particle current density. Finally, Appendix \ref{appendix:Lie derivative} contains a general result (valid for every Hamiltonian system) concerning the conservation of the volume form of the phase-space hypersurface tangent to a given Hamiltonian vector field; this finding is employed in the derivation of the generalised collisionless Boltzmann equation. A summary of the notational and conventional choices employed hereafter may be found in the ensuing Subsection \ref{sec:notation}.

\subsection{Notation and Conventions} \label{sec:notation}

\subsubsection{Conformal Frames and Hat Notation}

Let $\mathscr{M}$ be an oriented four-dimensional differentiable manifold equipped with a Lorentzian metric tensor field; such a pair constitutes a \emph{spacetime}. Consider now a family of spacetimes sharing the same base manifold $\mathscr{M}$, but whose respective metrics are different, yet related through \emph{conformal transformations}, i.e.~rescalings of the metric tensor by strictly positive smooth scalar functions on $\mathscr{M}$. Each of those conformally-related spacetimes is said to be a \emph{(conformal) frame}. In other words, if $(\mathscr{M},\boldsymbol{g})$ and $(\mathscr{M},\hat{\boldsymbol{g}})$ are two distinct conformal frames, the relation between their respective metric tensors $\boldsymbol{g}$ and $\hat{\boldsymbol{g}}$ is then given by
\begin{equation} \label{eq:def conformal}
    \hat{\boldsymbol{g}}:=\Omega^2\,\boldsymbol{g},
\end{equation}
where the non-vanishing scalar field $\Omega=\Omega(x)$ is known as the \emph{conformal factor} relating $(\mathscr{M},\boldsymbol{g})$ and $(\mathscr{M},\hat{\boldsymbol{g}})$, and represents a point-wise rescaling of the metric tensor \cite{Dabrowski:2008kx}. In what follows, the arbitrary frame $(\mathscr{M},\boldsymbol{g})$ shall be dubbed the \emph{`unhatted' frame}, whereas $(\mathscr{M},\hat{\boldsymbol{g}})$ will be known as the \emph{`hatted' frame}.

Given any (tensorial or non-tensorial) quantity $Q$ in the unhatted frame, we denote by $\hat{Q}$ the corresponding quantity in the hatted frame playing \emph{the same physical role} in that frame. To further clarify what we mean by this, let us consider, as a first example, the metrics $\boldsymbol{g}$ and $\hat{\boldsymbol{g}}$ themselves. These do indeed play the same physical role in their respective frames, allowing for lengths and angles to be defined therein. Another instance would be the respective Levi-Civita connection coefficients $\Gamma^\rho{}_{\mu\nu}$ and $\hat{\Gamma}^\rho{}_{\mu\nu}$ associated to $\boldsymbol{g}$ and $\hat{\boldsymbol{g}}$, which endow each frame with a natural affine structure obtained from the corresponding metric. The explicit relation between coefficients $\Gamma^\rho{}_{\mu\nu}$ and $\hat{\Gamma}^\rho{}_{\mu\nu}$ is
\begin{equation} \label{eq:connection transformation}
    \hat{\Gamma}^\rho{}_{\mu\nu}=\Gamma^\rho{}_{\mu\nu}+C^\rho{}_{\mu\nu},
\end{equation}
where $\boldsymbol{C}$ is known as the \emph{disformation tensor} \cite{BeltranJimenez:2015pnp,BeltranJimenez:2017tkd}, whose components in a coordinate basis read
\begin{equation} \label{eq:disformation}
    C^\rho{}_{\mu\nu}\equiv\Omega^{-1}\Left(\delta^\rho{}_{\mu}\dfrac{\partial\Omega}{\partial x^{\nu}}+\delta^\rho{}_{\nu}\dfrac{\partial\Omega}{\partial x^{\mu}}-g_{\mu\nu}g^{\rho\sigma}\dfrac{\partial\Omega}{\partial x^{\sigma}}\Right),
\end{equation}
and $\delta^\mu{}_{\nu}=g^{\mu\rho}g_{\rho\nu}=\hat{g}^{\mu\rho}\hat{g}_{\rho\nu}$.

\subsubsection{Other Notational and Conventional Choices}

The sign convention for the metric and associated geometrical objects (for instance, curvature tensors) is the same as in Refs.~\cite{Carroll:2004st,Misner:1973prb}, i.e.~the `mostly-pluses' convention. Undotted Greek indices run from $0$ to $3$, and represent components in a coordinate basis. Meanwhile, dotted indices refer instead to a non-coordinate basis, and run from either $\dot{0}$ to $\dot{3}$, if Greek, or from $\dot{1}$ to $\dot{3}$, if Latin. When using index-free notation, abstract tensors other than functions, exterior derivatives of scalar functions and some basis vectors such as $\partial/\partial x^\mu$ are written in boldface, whereas non-tensorial objects, scalar functions and their gradients, as well as the aforementioned basis elements are written in non-bold characters. Tensor components are likewise represented using non-bold characters. In general, given a tensor $\boldsymbol{T}$, its components in a certain basis are denoted as $T^{\cdots}_{\cdots}$ throughout. Finally, $\nabla$ and $\hat{\nabla}$ refer, respectively, to the covariant derivatives associated to the Levi-Civita connections of the spacetime metrics $\boldsymbol{g}$ and $\boldsymbol{\hat{g}}$.

\section{Massive Point Particles} \label{sec:massive particles}

\subsection{Action, Symmetries and Equations of Motion}

Imagine a massive test point particle propagating in a spacetime with base manifold $\mathscr{M}$. Its mass is allowed to vary at each point $x$ in $\mathscr{M}$, but required to be everywhere non-vanishing. The particle is furthermore assumed to be electrically neutral. Its world-line is independent of the conformal frame chosen to fix the physical units of measurement \cite{Dicke:1961gz}, and can be expressed in terms of an arbitrary parameter $\lambda$ as $x^\mu(\lambda)$. The vector with components $\dif x^\mu/\dif\lambda$ in a coordinate basis is thus tangent to the particle's trajectory.

In the unhatted frame $(\mathscr{M},\boldsymbol{g})$, an action for such a particle is
\begin{equation} \label{eq:action m(x) unhatted}
    S=-\int\dif\lambda\,m(x(\lambda))\,\sqrt{-g_{\mu\nu}(x(\lambda))\dfrac{\dif x^\mu}{\dif\lambda}\dfrac{\dif x^\nu}{\dif\lambda}},
\end{equation}
where the function $m=m(x)$ represents the particle's mass as measured in this particular frame at any point $x$. This spacetime-dependent mass can be understood as a scalar field on $\mathscr{M}$ to which the particle couples. Thus, for a particle, a non-constant mass signifies a scalar charge. The action in Eq.~\eqref{eq:action m(x) unhatted} is invariant under arbitrary reparametrisations $\lambda\mapsto\tilde{\lambda}(\lambda)$. A noteworthy parameter choice is given by the particle's proper time in the unhatted frame, denoted by $\tau$ and characterised by
\begin{equation}
    g_{\mu\nu}\dfrac{\dif x^\mu}{\dif\tau}\dfrac{\dif x^\nu}{\dif\tau}:=-1.
\end{equation}
Using $\tau$, the unhatted-frame action in Eq.~\eqref{eq:action m(x) unhatted} simplifies to 
\begin{equation} \label{eq:action m(x) unhatted tau}
    S=-\int\dif\tau\,m(x(\tau)).
\end{equation}

Under a conformal transformation from the unhatted frame to an arbitrary hatted frame, as given by Eq.~\eqref{eq:def conformal}, the action in Eq.~\eqref{eq:action m(x) unhatted} becomes
\begin{equation}
    S=-\int\dif\lambda\,\dfrac{m(x(\lambda))}{\Omega(x(\lambda))}\,\sqrt{-\hat{g}_{\mu\nu}(x(\lambda))\dfrac{\dif x^\mu}{\dif\lambda}\dfrac{\dif x^\nu}{\dif\lambda}}.
\end{equation}
As such, we see that the action is conformally invariant, i.e.
\begin{equation} \label{eq:action m(x) hatted}
    S=-\int\dif\lambda\,\hat{m}(x(\lambda))\,\sqrt{-\hat{g}_{\mu\nu}(x(\lambda))\dfrac{\dif x^\mu}{\dif\lambda}\dfrac{\dif x^\nu}{\dif\lambda}},
\end{equation}
provided that the particle's mass is rescaled as
\begin{equation} \label{eq:m(x) transformation}
    \hat{m}(x):=\Omega^{-1}(x)\,m(x),
\end{equation}
which is the well-established transformation rule for masses due to a change of conformal frame \cite{Dicke:1961gz,Faraoni:2006fx}. The particle's action remains invariant under arbitrary reparametrisations even when expressed in terms of the hatted variables. However, proper time $\hat{\tau}$ in the hatted frame does not coincide with $\tau$: from
\begin{equation}
    \hat{g}_{\mu\nu}\dfrac{\dif x^\mu}{\dif\hat{\tau}}\dfrac{\dif x^\nu}{\dif\hat{\tau}}:=-1
\end{equation}
one has
\begin{equation} \label{eq:tau transformation}
    \dif\hat{\tau}=\Omega\,\dif\tau,
\end{equation}
rendering \eqref{eq:action m(x) hatted} as
\begin{equation} \label{eq:action m(x) hatted tau hat}
    S=-\int\dif\hat{\tau}\,\hat{m}(x(\hat{\tau})).
\end{equation}

Requiring the action in Eq.~\eqref{eq:action m(x) unhatted} to be stationary under arbitrary variations of the particle's world-line $x^\mu(\lambda)$ with fixed endpoints, one obtains the equations of motion in the unhatted frame:
\begin{eqnarray}
    0 &=& \dfrac{\dif^2 x^\mu}{\dif\lambda^2}+\Gamma^\mu{}_{\rho\sigma}\dfrac{\dif x^\rho}{\dif\lambda}\dfrac{\dif x^\sigma}{\dif\lambda}-\dfrac{\dif\ln e(\lambda)}{\dif\lambda}\dfrac{\dif x^\mu}{\dif\lambda} \nonumber \\
    &-& g_{\rho\sigma}\dfrac{\dif x^\rho}{\dif\lambda}\dfrac{\dif x^\sigma}{\dif\lambda}g^{\mu\nu}\dfrac{\partial\ln{m(x)}}{\partial x^\nu}, \label{eq:unhatted generalised geodesic}
\end{eqnarray}
where we have introduced the so-called `einbein' \cite{Green:1987sp,Ortin:2015hya},
\begin{equation} \label{eq:unhatted einbein}
    e(\lambda)\equiv\dfrac{1}{m(x(\lambda))}\sqrt{-g_{\mu\nu}(x(\lambda))\dfrac{\dif x^\mu}{\dif\lambda}\dfrac{\dif x^\nu}{\dif\lambda}}.
\end{equation}
Eqns.~\eqref{eq:unhatted generalised geodesic} generalise the geodesic equations of $\boldsymbol{g}$ (expressed in terms of arbitrary parameter $\lambda$) by adding a force term proportional to the gradient of the mass scalar field $m(x)$. Consequently, the force term vanishes if $m(x)=\const$, in which case the standard geodesic equations in the unhatted frame are recovered. Likewise, one may impose the stationarity of the action expressed using hatted variables in Eq.~\eqref{eq:action m(x) hatted}, which yields the equations of motion in the hatted frame:
\begin{eqnarray}
    0 &=& \dfrac{\dif^2 x^\mu}{\dif\lambda^2}+\hat{\Gamma}^\mu{}_{\rho\sigma}\dfrac{\dif x^\rho}{\dif\lambda}\dfrac{\dif x^\sigma}{\dif\lambda}-\dfrac{\dif\ln\hat{e}(\lambda)}{\dif\lambda}\dfrac{\dif x^\mu}{\dif\lambda} \nonumber \\
    &-& \hat{g}_{\rho\sigma}\dfrac{\dif x^\rho}{\dif\lambda}\dfrac{\dif x^\sigma}{\dif\lambda}\hat{g}^{\mu\nu}\dfrac{\partial\ln{\hat{m}(x)}}{\partial x^\nu}, \label{eq:hatted generalised geodesic}
\end{eqnarray}
with
\begin{equation} \label{eq:hatted einbein}
    \hat{e}(\lambda)\equiv\dfrac{1}{\hat{m}(x(\lambda))}\sqrt{-\hat{g}_{\mu\nu}(x(\lambda))\dfrac{\dif x^\mu}{\dif\lambda}\dfrac{\dif x^\nu}{\dif\lambda}}.
\end{equation}
Substituting \eqref{eq:def conformal} and \eqref{eq:m(x) transformation} in this expression, we find that the transformation rule for the einbein is
\begin{equation} \label{eq:einbein transformation rule}
    \hat{e}(\lambda)=\Omega^2\,e(\lambda).
\end{equation}
As in the unhatted case, Eq.~\eqref{eq:hatted generalised geodesic} generalises the geodesic equation of $\hat{\boldsymbol{g}}$ by supplementing it with a force term proportional to the spacetime derivative of $\hat{m}(x)$. It may be readily verified that this equation can be recast as Eq.~\eqref{eq:unhatted generalised geodesic} by rewriting all hatted quantities in terms of their unhatted counterparts. Stated differently, the particle's equations of motion have the same form and admit the same solutions $x^{\mu}(\lambda)$ in both frames, in accordance with the conformal invariance of the action.

At this point, it is important to note that Eq.~\eqref{eq:hatted generalised geodesic} reduces to the standard geodesic equation in the hatted frame if $\hat{m}(x)=\const$, a condition which is incompatible with $m(x)=\const$ unless the conformal factor $\Omega$ is itself also constant. We therefore arrive at the following crucial conclusion: the concept of \emph{free massive particle} (i.e.~one having a non-zero constant mass, and therefore following time-like geodesics) is, in general, frame-dependent. Under a non-constant conformal rescaling, massive particles which are free in a given frame transform are represented by particles of variable mass (which, as per the discussion above, can be understood as carrying a scalar charge). Similarly, given any particle of non-constant mass in a certain frame, there exists a class of conformally related frames where the particle is free. For instance, suppose (without loss of generality) that the particle satisfies $m(x)\neq\const$ in the unhatted frame. Then, by virtue of Eq.~\eqref{eq:m(x) transformation}, the one-parameter family of conformal transformations
\begin{equation}
    \Omega(x)=\dfrac{m(x)}{\hat{m}_{\hat{0}}},~~~~\hat{m}_{\hat{0}}=\const\neq 0,
\end{equation}
produces a class of hatted frames where the particle is free, having $\hat{m}(x)=\hat{m}_{\hat{0}}$.

\subsection{Hamiltonian Formulation}

\subsubsection{Relativistic Phase Space and Hamiltonian Systems}

As widely known, given any differentiable manifold $\mathscr{M}$ (including those lacking a metric tensor field), its associated cotangent bundle is defined as
\begin{equation}
    \mathscr{T}^*\mathscr{M}:=\Left\{\xp\,|\,x\in\mathscr{M},\,\boldsymbol{p}\in \mathscr{T}^*_x\mathscr{M}\Right\},
\end{equation}
where $\mathscr{T}^*_x\mathscr{M}$ is the cotangent space at each point $x$ in $\mathscr{M}$, and $\boldsymbol{p}$ is accordingly a one-form at $x$. Spacetimes sharing the same base manifold $\mathscr{M}$, such as conformally related frames, hence possess the same cotangent bundle $\mathscr{T}^*\mathscr{M}$ and cotangent spaces $\mathscr{T}^*_x\mathscr{M}$. The physical importance of cotangent bundles stems from the fact that they are naturally endowed with a metric-independent symplectic structure $\sform$ \cite{Sarbach:2013fya,Sarbach:2013uba,Acuna-Cardenas:2021nkj},\footnote{
    We remind the reader that a symplectic structure is a two-form field $\sform:\mathscr{X}(\mathscr{T}^*\mathscr{M})\times\mathscr{X}(\mathscr{T}^*\mathscr{M})\rightarrow\mathscr{F}(\mathscr{T}^*\mathscr{M})$, which is both non-degenerate (i.e.~$\sform(\boldsymbol{X},\,\cdot\,)=0$ if and only if $\boldsymbol{X}=0$) and closed ($\dif\sform=0$). Here and hereafter, $\mathscr{X}(\mathscr{N})$ and $\mathscr{F}(\mathscr{N})$ denote the sets of smooth vector fields and smooth functions on a given differentiable manifold $\mathscr{N}$, respectively.
} thus permitting a Hamiltonian formulation of classical relativistic systems in arbitrary spacetimes. For this reason, $\mathscr{T}^*\mathscr{M}$ is identified with relativistic \emph{phase space}, while one-forms $\boldsymbol{p}$ (that is to say, elements of $\mathscr{T}^*_x\mathscr{M}$) correspond to \emph{physical momenta}, and are conformally invariant:
\begin{equation}
    \hat{\boldsymbol{p}}=\boldsymbol{p}.
\end{equation}
The metric independence of the symplectic structure $\sform$ can be inferred from its definition as the exterior derivative of the so-called canonical phase-space one-form $\boldsymbol{\Theta}$,\footnote{
    We remark that, in this work, no additional typographical distinction is made between tensors in phase space or in spacetime. This is the case of $\sform$ and $\boldsymbol{\Theta}$ above, for example, which are written using the standard boldface notation for abstract tensors.
}
\begin{equation}
    \sform := \dif\boldsymbol{\Theta}.
\end{equation}
where the action of $\boldsymbol{\Theta}$ on any phase-space vector $\boldsymbol{X}$ at phase-space point $\xp$ is given by\footnote{
    For further details on the geometry of relativistic phase space, we refer the reader to \cite{Sarbach:2013fya,Sarbach:2013uba,Acuna-Cardenas:2021nkj} and references therein.
}
\begin{equation}
    \Left.\boldsymbol{\Theta}(\boldsymbol{X})\Right|_{\xp}:=\boldsymbol{p}\Left(\dif\pi\Left.(\boldsymbol{X})\Right|_{\xp}\Right).
\end{equation}
Here, $\pi:\mathscr{T}^*\mathscr{M}\rightarrow\mathscr{M}$ is the projection map assigning to each point $\xp$ in $\mathscr{T}^*\mathscr{M}$ the corresponding point $x$ in $\mathscr{M}$. Because neither $\boldsymbol{p}$, nor the exterior derivative, nor $\pi$ require a metric to be defined, one has that both $\boldsymbol{\Theta}$ and $\sform$ are metric-independent, and consequently frame-invariant:
\begin{equation}
    \hat{\boldsymbol{\Theta}}=\boldsymbol{\Theta},~~~~\hat{\sform}=\sform.
\end{equation}
This may be explicitly verified by choosing a system of `local adapted coordinates' $(x^\mu,p_\mu)$ on $\mathscr{T}^*\mathscr{M}$, where $p_\mu$ are the components of $\boldsymbol{p}$ with respect to the (coordinate) basis of co-vectors $\{\dif x^\mu|_x\}$ of $\mathscr{T}^*_x\mathscr{M}$, i.e.~$\boldsymbol{p}=p_\mu \dif x^\mu|_x$. Then,
\begin{equation} \label{eq:adapted one-form basis}
    \Left\{\dif x^\mu|_{\xp},\,\dif p_\mu|_{\xp}\Right\}
\end{equation}
constitutes a (conformally invariant) coordinate basis for phase-space covectors at every point $\xp$, in terms of which
\begin{eqnarray}
    \boldsymbol{\Theta} &=& p_\mu\dif x^\mu, \\
    \sform &=& \dif p_\mu\wedge\dif x^\mu,
\end{eqnarray}
where `$\wedge$' indicates the standard exterior (or wedge product). It is clear that these expressions are frame-independent¡. For future reference, we include here the coordinate basis of phase-space vectors at $\xp$ dual to \eqref{eq:adapted one-form basis}, which is
\begin{equation} \label{eq:adapted vector basis}
    \Left\{\Left.\dfrac{\partial}{\partial x^\mu}\Right|_{\xp},\,\Left.\dfrac{\partial}{\partial p_\mu}\Right|_{\xp}\Right\}
\end{equation}

A \emph{Hamiltonian} $\mathcal{H}$ is any smooth scalar function on the cotangent bundle. Its associated \emph{Hamiltonian vector field} $\boldsymbol{X}_\mathcal{H}$ is defined implicitly by setting
\begin{equation} \label{eq:implicit hamiltonian vector def}
    \dif\mathcal{H}=:-\iota_{\boldsymbol{X}_\mathcal{H}}\sform,
\end{equation}
with $\iota$ denoting the contraction operation, in this case $\iota_{\boldsymbol{X}_\mathcal{H}}\sform=\sform(\boldsymbol{X}_\mathcal{H},\,\cdot\,)$. Expanding $\boldsymbol{X}_\mathcal{H}$ in the basis \eqref{eq:adapted vector basis}, one finds
\begin{equation} \label{eq:explicit hamiltonian vector def}
    \boldsymbol{X}_\mathcal{H}=\dfrac{\partial\mathcal{H}}{\partial p_\mu}\dfrac{\partial}{\partial x^\mu}-\dfrac{\partial\mathcal{H}}{\partial x^\mu}\dfrac{\partial}{\partial p_\mu}.
\end{equation}
The integral curves of $\boldsymbol{X}_\mathcal{H}$ correspond to Hamilton's equations for a relativistic physical system described by the Hamiltonian $\mathcal{H}$:
\begin{eqnarray}
    \dfrac{\dif x^\mu}{\dif\lambda} &=& \dfrac{\partial\mathcal{H}}{\partial p_\mu}, \label{eq:Hamilton x} \\
    \dfrac{\dif p_\mu}{\dif\lambda} &=& -\dfrac{\partial\mathcal{H}}{\partial x^\mu}. \label{eq:Hamilton p}
\end{eqnarray}
Moreover, $\boldsymbol{X}_\mathcal{H}$ generates canonical transformations, i.e.~coordinate changes on $\mathscr{T}^*\mathscr{M}$ leaving the symplectic form $\sform$ invariant. This is because
\begin{equation} \label{eq:Lie derivative of omega}
    \pounds_{\boldsymbol{X}_\mathcal{H}}\sform=\dif(\iota_{\boldsymbol{X}_\mathcal{H}}\sform)+\iota_{\boldsymbol{X}_\mathcal{H}}(\dif\sform)=-\dif^2\mathcal{H}=0,
\end{equation}
where $\pounds$ denotes the Lie derivative. This result is achieved using Cartan's formula in combination with Eq.~\eqref{eq:implicit hamiltonian vector def} and the fact that $\sform$ is closed. Finally, we note that, for any Hamiltonian function $\mathcal{H}$,
\begin{equation} \label{eq:generic Hamiltonian vector tangency}
    \boldsymbol{X}_\mathcal{H}(\mathcal{H})=0=\dif\mathcal{H}(\boldsymbol{X}_\mathcal{H}),
\end{equation}
and, as a result, $\boldsymbol{X}_\mathcal{H}$ is necessarily tangent to the phase-space hypersurface described by $\mathcal{H}=0$ (whose normal one-form is simply $\dif\mathcal{H}$).

\subsubsection{Canonical Momenta and Constraint of a Massive Point Particle}

The canonical momenta for a particle with variable mass may be readily computed from either Eq.~\eqref{eq:action m(x) unhatted} or Eq.~\eqref{eq:action m(x) hatted}: 
\begin{equation} \label{eq:canonical momenta}
    p_\mu = \dfrac{m(x)\,g_{\mu\nu}\dfrac{\dif x^\nu}{\dif\lambda}}{\sqrt{-g_{\rho\sigma}\dfrac{\dif x^\rho}{\dif\lambda}\dfrac{\dif x^\sigma}{\dif\lambda}}} =  \dfrac{\hat{m}(x)\,\hat{g}_{\mu\nu}\dfrac{\dif x^\nu}{\dif\lambda}}{\sqrt{-\hat{g}_{\rho\sigma}\dfrac{\dif x^\rho}{\dif\lambda}\dfrac{\dif x^\sigma}{\dif\lambda}}}.
\end{equation} 
As expected, these conjugate momenta are frame-invariant. However, not all of them are independent. More specifically, they satisfy
\begin{equation}
    \boldsymbol{g}^{-1}(\boldsymbol{p},\boldsymbol{p})=g^{\mu\nu}p_\mu p_\nu=-m^2(x),
\end{equation}
or, equivalently,
\begin{equation}
    \hat{\boldsymbol{g}}^{-1}(\boldsymbol{p},\boldsymbol{p})=\hat{g}^{\mu\nu}p_\mu p_\nu=-\hat{m}^2(x),
\end{equation}
since the conformal factor $\Omega$ is everywhere non-vanishing. The existence of this (primary) constraint implies that not all phase space $\mathscr{T}^*\mathscr{M}$ is accessible to the particle under consideration. Instead, its dynamics is confined to the seven-dimensional hypersurface in phase space known as the \emph{mass shell}, which is given by
\begin{eqnarray}
    \massshell &:=& \{\xp\in\mathscr{T}^*\mathscr{M}\,|\,\boldsymbol{g}^{-1}(\boldsymbol{p},\boldsymbol{p})=-m^2(x)\} \nonumber \\
    &=& \{\xp\in\mathscr{T}^*\mathscr{M}\,|\,\hat{\boldsymbol{g}}^{-1}(\boldsymbol{p},\boldsymbol{p})=-\hat{m}^2(x)\}.
\end{eqnarray}
The mass shell can be decomposed into two disconnected pieces $\massshell^\pm$ (mathematically, $\massshell\equiv\massshell^+\sqcup\massshell^-$), respectively known as the \emph{future} and \emph{past mass shells}, consisting of the restriction of $\massshell$ to future- and past-directed canonical momenta, in that order:
\begin{eqnarray}
    \massshell^+ &:=& \Left\{\xp\in\massshell\,|\,\boldsymbol{p} \text{ future-directed}\Right\}, \label{eq:future mass shell} \\
    \massshell^- &:=& \Left\{\xp\in\massshell\,|\,\boldsymbol{p} \text{ past-directed}\Right\}. \label{eq:past mass shell}
\end{eqnarray}
The corresponding future and past \emph{mass hyperboloids} $\masshyp^\pm_x$ are then defined as the fibres of the fibre bundles $\massshell^\pm$, respectively:
\begin{eqnarray}
    \masshyp^+_x &:=& \Left\{\boldsymbol{p}\in\mathscr{T}^*_x\mathscr{M}\,|\,\boldsymbol{p} \in\massshell^+\Right\}, \label{eq:future mass hyperboloid} \\
    \masshyp^-_x &:=& \Left\{\boldsymbol{p}\in\mathscr{T}^*_x\mathscr{M}\,|\,\boldsymbol{p} \in\massshell^-\Right\}. \label{eq:past mass hyperboloid}
\end{eqnarray}
Of these sets, the ones that are relevant to the statistical interpretation developed later are the future mass shell \eqref{eq:future mass shell} and future mass hyperboloid \eqref{eq:future mass hyperboloid}, as they describe the region of phase space available to particles of non-constant mass propagating towards the future.

We stress at this point that the mass shells and hyperboloids are guaranteed to be frame-independent by the strict positivity of $\Omega^2$, and for this reason it can be equivalently expressed in terms of either hatted or unhatted quantities. Equalities that are valid only on the mass shell are henceforth referred to as \emph{weak equalities}, and denoted by `$\approx$.' For instance,
\begin{equation} \label{eq:constraint m(x) unhatted}
    \mathcal{C}:=\boldsymbol{g}^{-1}(\boldsymbol{p},\boldsymbol{p})+m^2(x)\approx 0
\end{equation}
represents the mass-shell constraint equation in the unhatted frame. Similarly, in the hatted frame, the mass shell can be expressed implicitly via
\begin{equation} \label{eq:constraint m(x) hatted}
    \hat{\mathcal{C}}:=\hat{\boldsymbol{g}}^{-1}(\boldsymbol{p},\boldsymbol{p})+\hat{m}^2(x)\approx 0,
\end{equation}
noting that
\begin{equation} \label{eq:constraint transformation rule}
    \hat{\mathcal{C}}=\Omega^{-2}\mathcal{C}.
\end{equation}
Weak equalities contrast with the standard \emph{strong equalities} denoted by `$=$,' which hold everywhere, including on $\massshell$.

\subsubsection{Hamiltonian for a Massive Point Particle}

As expected for any physical system exhibiting reparametrisation invariance, the particle's canonical Hamiltonian is strongly vanishing:
\begin{equation} \label{eq:canonical hamiltonian unhatted}
    \mathcal{H}_\mathrm{c}:=p_\mu\dfrac{\dif x^\mu}{\dif\lambda}-m(x)\,\sqrt{-g_{\mu\nu}\dfrac{\dif x^\mu}{\dif\lambda}\dfrac{\dif x^\nu}{\dif\lambda}}=0,
\end{equation}
where, in the last step, we have used the definition of canonical momenta in Eq.~\eqref{eq:canonical momenta}. It can be easily shown that the canonical Hamiltonian is conformally invariant,
\begin{equation}
    \hat{\mathcal{H}}_\mathrm{c}:=p_\mu\dfrac{\dif x^\mu}{\dif\lambda}-\hat{m}(x)\,\sqrt{-\hat{g}_{\mu\nu}\dfrac{\dif x^\mu}{\dif\lambda}\dfrac{\dif x^\nu}{\dif\lambda}}=\mathcal{H}_\mathrm{c}=0,
\end{equation}
in agreement with the conformal invariance of the action in Eq.~\eqref{eq:action m(x) unhatted} or Eq.~\eqref{eq:action m(x) hatted}. Therefore, according to the Dirac-Bergmann formalism \cite{Dirac:1950pj,Anderson:1951ta} for constrained systems, the total one-particle Hamiltonian is proportional to the constraint, and hence only weakly vanishing. In consequence, the unhatted-frame one-particle Hamiltonian has to be of the form
\begin{equation} \label{eq:unhatted H_1}
    \mathcal{H}_1(\lambda):=\mathcal{H}_\mathrm{c}+\mathcal{N}(\lambda)\,\mathcal{C}=\mathcal{N}(\lambda)\,\mathcal{C},
\end{equation}
with $\mathcal{N}(\lambda)$ a Lagrange multiplier and $\mathcal{C}$ given by Eq.~\eqref{eq:constraint m(x) unhatted}. Analogously, the hatted one-particle Hamiltonian is
\begin{equation} \label{eq:hatted H_1}
    \hat{\mathcal{H}}_1(\lambda):=\hat{\mathcal{H}}_\mathrm{c}+\hat{\mathcal{N}}(\lambda)\,\hat{\mathcal{C}}=\hat{\mathcal{N}}(\lambda)\,\hat{\mathcal{C}},
\end{equation}
where $\hat{\mathcal{N}}(\lambda)$ is another Lagrange multiplier and $\hat{\mathcal{C}}$ was defined back in \eqref{eq:constraint m(x) hatted}. Observe that, by virtue of \eqref{eq:constraint transformation rule}, Eq.~\eqref{eq:hatted H_1} becomes
\begin{equation} \label{eq:hatted H_1 partial conformal transformation}
    \hat{\mathcal{H}}_1(\lambda)=\Omega^{-2}\,\hat{\mathcal{N}}(\lambda)\,\mathcal{C}.
\end{equation}
In order to find the relation between $\mathcal{N}(\lambda)$ and $\hat{\mathcal{N}}(\lambda)$, we resort to the first of Hamilton's equations, namely \eqref{eq:Hamilton x}. Particularising this expression for $\mathcal{H}_1$ as given in \eqref{eq:unhatted H_1}, one obtains 
\begin{equation} \label{eq:Hamilton x H_1 unhatted}
    \dfrac{\dif x^\mu}{\dif\lambda}=\dfrac{\partial\mathcal{H}_1}{\partial p_\mu}=2\mathcal{N}(\lambda)\,g^{\mu\nu}p_\nu.
\end{equation}
Combining the preceding equality with \eqref{eq:canonical momenta}, the ensuing unhatted-frame relation follows:
\begin{equation} \label{eq:N on-shell}
    \mathcal{N}(\lambda)=\dfrac{1}{2m(x(\lambda))}\sqrt{-g_{\mu\nu}(x(\lambda))\dfrac{\dif x^\mu}{\dif\lambda}\dfrac{\dif x^\nu}{\dif\lambda}}=\dfrac{e(\lambda)}{2},
\end{equation}
where $e(\lambda)$ was introduced in equation Eq.~\eqref{eq:unhatted einbein}. The corresponding hatted-frame on-shell equation is found by substituting $\hat{\mathcal{H}}_1$, as provided in \eqref{eq:hatted H_1}, into \eqref{eq:Hamilton x}, yielding
\begin{equation} \label{eq:Hamilton x H_1 hatted}
    \dfrac{\dif x^\mu}{\dif\lambda}=\dfrac{\partial\hat{\mathcal{H}}_1}{\partial p_\mu}=2\hat{\mathcal{N}}(\lambda)\,\hat{g}^{\mu\nu}p_\nu,
\end{equation}
whence
\begin{equation} \label{eq:N hat on-shell}
    \hat{\mathcal{N}}(\lambda)=\dfrac{1}{2\hat{m}(x(\lambda))}\sqrt{-\hat{g}_{\mu\nu}(x(\lambda))\dfrac{\dif x^\mu}{\dif\lambda}\dfrac{\dif x^\nu}{\dif\lambda}}=\dfrac{\hat{e}(\lambda)}{2},
\end{equation}
where, once more, we have made use of \eqref{eq:canonical momenta}. Recalling the previously derived transformation rule \eqref{eq:einbein transformation rule} for the einbein, or by direct comparison of Eqns.~\eqref{eq:Hamilton x H_1 unhatted} and \eqref{eq:Hamilton x H_1 hatted}, we have
\begin{equation} \label{eq:N transformation}
    \hat{\mathcal{N}}(\lambda)=\Omega^2\,\mathcal{N}(\lambda),
\end{equation}
from where we finally deduce that the one-particle Hamiltonian is conformally invariant:
\begin{equation} \label{eq:H_1 conformal invariance}
    \hat{\mathcal{H}}_1(\lambda)=\mathcal{H}_1(\lambda),
\end{equation}
see Eq.~\eqref{eq:hatted H_1 partial conformal transformation}. Again, this result could have been foreseen, owing to the manifest conformal invariance of the particle's action. Replacing either \eqref{eq:N on-shell} or \eqref{eq:N hat on-shell} in the second Hamilton equation \eqref{eq:Hamilton p}, one respectively gets the unhatted or hatted equations of motion \eqref{eq:unhatted generalised geodesic} or \eqref{eq:hatted generalised geodesic}.

Three fundamental observations are in order at this stage. First, starting from Eqns.~\eqref{eq:N on-shell} and \eqref{eq:N hat on-shell}, it is clear that choosing a specific functional form for either of the Lagrange multipliers is tantamount to selecting a specific parameter $\lambda$, that is to say, to fixing the gauge. For example, setting $\mathcal{N}=1/(2m(x))$ gives $\lambda=\tau+\const$, whereas taking $\hat{\mathcal{N}}=1/(2\hat{m}(x))$ yields $\lambda=\hat{\tau}+\const$.

Second, even though we have just shown that the Hamiltonian for a single point particle with varying mass is conformally invariant, we note that it is \emph{not} gauge-invariant.\footnote{
    In fact, using either \eqref{eq:N on-shell} or \eqref{eq:N hat on-shell}, one finds that the one-particle Hamiltonian transforms under reparametrisations as
    \begin{equation*}
        \mathcal{H}_1(\tilde{\lambda}) = \dfrac{\dif\lambda}{\dif\tilde{\lambda}}\mathcal{H}_1(\lambda(\tilde{\lambda})).
    \end{equation*}
} For this reason, the conformal invariance rule \eqref{eq:H_1 conformal invariance} only holds provided that the same parameter $\lambda$ is taken in both frames.

Last but not least, we would like to point out that, in deriving the conformal invariance of $\mathcal{H}_1$, it was essential to use an arbitrary parametrisation, as spurious results may be (incorrectly) obtained by trying to transform a gauge-fixed one-particle Hamiltonian. For instance, suppose that we choose $\tau$ as our parameter in both the unhatted and hatted frames.\footnote{
    Let us stress that curves can be parametrised by $\tau$ in \emph{any} frame, but only in the unhatted frame this parametrisation will coincide with proper time therein.
} As mentioned above, this `unhatted proper-time gauge' corresponds to $\mathcal{N}=1/(2m(x(\tau)))$, which implies
\begin{equation}
    \mathcal{H}_1(\tau)=\dfrac{\mathcal{C}}{2m(x(\tau))}.
\end{equation}
Taking the last expression at face value, without knowing there is a hidden square root containing the metric that has simplified to $1$ in this particular gauge which we only know is there because we have derived the generic, gauge-free expression \eqref{eq:N on-shell}), one would have obtained the \emph{incorrect} transformation rule
\begin{equation}
    \hat{\mathcal{H}}_1(\tau)\wrongeq\Omega^{-1}\,\mathcal{H}_1(\tau)~~~~\text{(incorrect!)}.
\end{equation}
It is essential to bear this warning in mind, as most of the existing literature on the kinetic theory of massive particles which are free in one specific frame (e.g.~Refs.~\cite{Sarbach:2013fya,Sarbach:2013uba,Acuna-Cardenas:2021nkj}) considers precisely this proper-time gauge, and not the more general formulation in terms of an arbitrary $\lambda$.

\subsubsection{The Liouville Vector}

We define the \emph{Liouville vector} $\boldsymbol{L}$ as the Hamiltonian vector field (as defined in equation Eq.~\eqref{eq:implicit hamiltonian vector def} above) associated with the one-particle Hamiltonian function $\mathcal{H}_1$:
\begin{equation}
    \boldsymbol{L}(\lambda):=\boldsymbol{X}_{\mathcal{H}_1(\lambda)}=\mathcal{N}(\lambda)\,\boldsymbol{X}_\mathcal{C}+\mathcal{C}\,\boldsymbol{X}_{\mathcal{N}(\lambda)},
\end{equation}
where the last equality can be easily checked using \eqref{eq:explicit hamiltonian vector def} to expand $\boldsymbol{L}$ in the local adapted coordinate basis. The restriction of $\boldsymbol{L}$ to the constraint hypersurface $\massshell$ is thus given by
\begin{equation} \label{eq:unhatted Liouville restriction}
    \boldsymbol{L}(\lambda)\approx\mathcal{N}(\lambda)\,\boldsymbol{X}_\mathcal{C}.
\end{equation}
Using Eq.~\eqref{eq:explicit hamiltonian vector def}, $\boldsymbol{X}_\mathcal{C}$ can be expanded as
\begin{equation} \label{eq:unhatted X_C m(x)}
    \boldsymbol{X}_\mathcal{C}=2g^{\mu\nu}p_\nu\dfrac{\Dif}{\partial x^\mu}-2m(x)\dfrac{\partial m(x)}{\partial x^\mu}\dfrac{\partial}{\partial p_\mu},
\end{equation}
where we have introduced
\begin{equation} \label{eq:unhatted horizontal basis}
    \dfrac{\Dif}{\partial x^\mu}:=\dfrac{\partial}{\partial x^\mu}+\Gamma^{\rho}{}_{\sigma\mu}p_\rho\dfrac{\partial}{\partial p_\sigma}.
\end{equation}

We can anticipate that $\boldsymbol{L}$ is conformally invariant, i.e.
\begin{equation} \label{eq:Liouville invariance}
    \hat{\boldsymbol{L}}(\lambda)=\boldsymbol{L}(\lambda).
\end{equation}
Let us prove this explicitly. First, we introduce
\begin{equation}
    \hat{\boldsymbol{L}}(\lambda):=\boldsymbol{X}_{\hat{\mathcal{H}}_1(\lambda)}=\hat{\mathcal{N}}(\lambda)\,\boldsymbol{X}_{\hat{\mathcal{C}}}+\hat{\mathcal{C}}\,\boldsymbol{X}_{\hat{\mathcal{N}}(\lambda)},
\end{equation}
where, here and in what follows, $\hat{\mathcal{H}}_1(\lambda)$ denotes the (conformally invariant) one-particle Hamiltonian written in terms of hatted quantities, i.e.~as given in Eq.~\eqref{eq:hatted H_1}. Invoking the transformation rules \eqref{eq:constraint transformation rule} and \eqref{eq:N transformation} for $\hat{\mathcal{C}}$ and $\hat{\mathcal{N}}$ (respectively), and using \eqref{eq:explicit hamiltonian vector def} one more time, it is straightforward to show that
\begin{equation}
    \hat{\mathcal{N}}(\lambda)\,\boldsymbol{X}_{\hat{\mathcal{C}}}+\hat{\mathcal{C}}\,\boldsymbol{X}_{\hat{\mathcal{N}}(\lambda)}=\mathcal{N}(\lambda)\,\boldsymbol{X}_\mathcal{C}+\mathcal{C}\,\boldsymbol{X}_{\mathcal{N}(\lambda)},
\end{equation}
meaning that Eq.~\eqref{eq:Liouville invariance} holds in the strong sense. Notice that, in spite of this, the restriction of $\hat{\boldsymbol{L}}$ to $\massshell$,
\begin{equation} \label{eq:hatted Liouville restriction}
    \hat{\boldsymbol{L}}(\lambda)\approx\hat{\mathcal{N}}(\lambda)\,\boldsymbol{X}_{\hat{\mathcal{C}}}.
\end{equation}
is only weakly conformally invariant, as the expansion of Eq.~\eqref{eq:hatted Liouville restriction} using Eq.~\eqref{eq:explicit hamiltonian vector def} reveals that
\begin{equation}
    \hat{\mathcal{N}}(\lambda)\,\boldsymbol{X}_{\hat{\mathcal{C}}}\approx\mathcal{N}(\lambda)\,\boldsymbol{X}_\mathcal{C};
\end{equation}
compare the right-hand side with \eqref{eq:unhatted Liouville restriction}. Notice that
\begin{equation} \label{eq:X_C transformation}
    \boldsymbol{X}_{\hat{\mathcal{C}}}=\Omega^{-2}\,\boldsymbol{X}_\mathcal{C}+\mathcal{C}\,\boldsymbol{X}_{\Omega^{-2}}\approx\Omega^{-2}\,\boldsymbol{X}_\mathcal{C}.
\end{equation}

Having established the conformal invariance of $\boldsymbol{L}$, we reserve the symbol $\hat{\boldsymbol{L}}$ for the Liouville vector expressed in terms of hatted-frame quantities, as done with $\hat{\mathcal{H}}_1(\lambda)$. Moreover, for future reference we include here the following expression for $\boldsymbol{X}_{\hat{\mathcal{C}}}$:
\begin{equation} \label{eq:hatted X_C m(x)}
    \boldsymbol{X}_{\hat{\mathcal{C}}}=2\hat{g}^{\mu\nu}p_\nu\dfrac{\hat{\Dif}}{\partial x^\mu}-2\hat{m}(x)\dfrac{\partial \hat{m}(x)}{\partial x^\mu}\dfrac{\partial}{\partial p_\mu},
\end{equation}
where
\begin{equation} \label{eq:hatted horizontal basis}
    \dfrac{\hat{\Dif}}{\partial x^\mu}:=\dfrac{\partial}{\partial x^\mu}+\hat{\Gamma}^{\rho}{}_{\sigma\mu}p_\rho\dfrac{\partial}{\partial p_\sigma}=\dfrac{\Dif}{\partial x^\mu}+C^{\rho}{}_{\sigma\mu}p_\rho\dfrac{\partial}{\partial p_\sigma},
\end{equation}
with $\boldsymbol{C}$ the disformation tensor defined in Eq.~\eqref{eq:disformation}. The significance of the new (\emph{non}-coordinate) basis elements $\{\Dif/\partial x^\mu\}$ and $\{\hat{\Dif}/\partial x^\mu\}$ will become apparent in the next section.

\section{Horizontal and Vertical Subspaces of $\mathscr{T}_{\xp}(\mathscr{T}^*\mathscr{M})$} \label{sec:horizontal and vertical}

Volume elements in phase space, necessary to define the integrals allowing for a statistical interpretation of a collection of particles, can be constructed using the symplectic structure already present on $\mathscr{T}^*\mathscr{M}$ only \cite{Sarbach:2013fya}. Despite this, in the case of particles having a constant mass in a fixed conformal frame, the aforementioned volume elements can also be derived from a metric tensor on phase space \cite{Sarbach:2013uba,ThesisRioseco}, the so-called \emph{Sasaki metric}, first introduced in Ref.~\cite{Sasaki1958}. This metric arises naturally from the decomposition of the set of phase-space vectors as a direct sum of two subspaces, known as \emph{horizontal} and \emph{vertical}. However, as previously mentioned, this decomposition is adapted only to particles which are free in a certain frame. As such, in this section, we generalise the construction of vertical and horizontal subspaces, first to multiple conformal frames, and then to particles of non-constant mass. Eventually, these results will allow us to define the corresponding generalisation of the Sasaki metric, dubbed `tilted' Sasaki metric, on each frame.

\subsection{Unhatted Horizontal and Vertical Subspaces}

The set of phase-space vectors at each point $\xp$ in $\mathscr{T}^*\mathscr{M}$ (i.e.~$\mathscr{T}_{\xp}(\mathscr{T}^*\mathscr{M})$, the tangent space at $\xp$ to the cotangent bundle of the base manifold $\mathscr{M}$) is spanned by the non-coordinate basis
\begin{equation} \label{eq:unhatted D basis}
    \Left\{\Left.\dfrac{\Dif}{\partial x^\mu}\Right|_{\xp},\Left.\dfrac{\partial}{\partial p_\mu}\Right|_{\xp}\Right\}.
\end{equation}
As a result, $\mathscr{T}_{\xp}(\mathscr{T}^*\mathscr{M})$ admits the following direct-sum decomposition:
\begin{equation} \label{eq:vertical + unhatted horizontal decomposition}
    \mathscr{T}_{\xp}(\mathscr{T}^*\mathscr{M})=\mathscr{V}_{\xp}\oplus\mathscr{H}_{\xp},
\end{equation}
where
\begin{equation}
    \mathscr{V}_{\xp}:=\ker\dif\pi|_{\xp}=\lin\Left\{\Left.\dfrac{\partial}{\partial p_\mu}\Right|_{\xp}\Right\}
\end{equation}
is the so-called \emph{vertical subspace}, whereas
\begin{equation} \label{eq:unhatted horizontal subspace}
    \mathscr{H}_{\xp}:=\ker K|_{\xp}=\lin\Left\{\Left.\dfrac{\Dif}{\partial x^\mu}\Right|_{\xp}\Right\}
\end{equation}
is known as the \emph{unhatted horizontal subspace}. In the last expression, $K$ denotes the so-called \emph{unhatted connection map}, whose action on a phase-space vector of the form
\begin{equation} \label{eq:vector in phase space Z}
    \boldsymbol{Z}=X^\mu\Left.\dfrac{\partial}{\partial x^\mu}\Right|_{\xp}+Y_\mu\Left.\dfrac{\partial}{\partial p_\mu}\Right|_{\xp}\in\mathscr{T}_{\xp}(\mathscr{T}^*\mathscr{M}),
\end{equation}
is given by
\begin{equation} \label{eq:unhatted connection map}
    \Left.K(\boldsymbol{Z})\Right|_{\xp}:=\left(Y_\mu-\Gamma^\rho{}_{\sigma\mu}p_\rho X^\sigma\right)\dif x^\mu|_{\xp}.
\end{equation}
The fact that the sum in Eq.~\eqref{eq:vertical + unhatted horizontal decomposition} is direct entails that every phase-space vector $\boldsymbol{Z}\in\mathscr{T}_{\xp}(\mathscr{T}^*\mathscr{M})$ can be split as
\begin{equation}
    \boldsymbol{Z}=\boldsymbol{Z}^\ver+\boldsymbol{Z}^\hor,
\end{equation}
with $\boldsymbol{Z}^\ver\in\mathscr{V}_{\xp}$ and $\boldsymbol{Z}^\hor\in\mathscr{H}_{\xp}$ both being unique. More precisely,
\begin{eqnarray}
    \boldsymbol{Z}^\hor &\equiv& Z^{\hor\mu}\Left.\dfrac{\Dif}{\partial x^\mu}\Right|_{\xp}, \\
    \boldsymbol{Z}^\ver &\equiv& Z^\ver_\mu\Left.\dfrac{\partial}{\partial p_\mu}\Right|_{\xp},
\end{eqnarray}
where the unhatted horizontal and unhatted vertical components of $\boldsymbol{Z}$ are respectively given by
\begin{eqnarray}
    Z^{\hor\mu} &:=& X^\mu, \label{eq:ZHmu unhatted} \\
    Z^\ver_\mu &:=& Y_\mu-\Gamma^\rho{}_{\sigma\mu}p_\rho X^\sigma \label{eq:ZVmu unhatted}
\end{eqnarray}
in terms of the components $X^\mu$ and $Y_\mu$ of $\boldsymbol{Z}$ in the adapted coordinate basis \eqref{eq:adapted vector basis}, as given in Eq.~\eqref{eq:vector in phase space Z}. This decomposition is adapted to particles which are free in the unhatted frame, for the following reasons. On the one hand, from \eqref{eq:unhatted X_C m(x)}, one has that $\boldsymbol{X}_{\mathcal{C}}$ (and hence the restriction of $\boldsymbol{L}$ to $\massshell$) is unhatted horizontal if and only if $m(x)=\const$ On the other, all horizontal phase-space vectors are tangent to $\massshell|_{m(x)=\const}$, whereas all vectors tangent to $\massshell|_{m(x)=\const}$ are horizontal and all vectors normal to $\massshell|_{m(x)=\const}$ are purely vertical \cite{Acuna-Cardenas:2021nkj}.

Finally, since it will appear later, we include here the basis of $\mathscr{T}^*_{\xp}(\mathscr{T}^*\mathscr{M})$ dual to \eqref{eq:unhatted D basis}, which is
\begin{equation}
    \Left\{\Left.\Dif p_\mu\Right|_{\xp},\Left.\dif x^\mu\Right|_{\xp}\Right\},
\end{equation}
with
\begin{equation}
    \Dif p_\mu|_{\xp}:=\dif p_\mu|_{\xp}-\Gamma^{\rho}{}_{\sigma\mu}\,p_\rho\,\dif x^\sigma|_{\xp}.
\end{equation}

\subsection{Hatted Horizontal Subspace}

There exists a completely analogous decomposition adapted to particles which are free in the hatted frame, i.e.~those satisfying $\hat{m}(x)=\const$ More precisely, defining the \emph{hatted connection map},
\begin{eqnarray}
    \Left.\hat{K}(\boldsymbol{Z})\Right|_{\xp} &:=& \left(Y_\mu-\hat{\Gamma}^\rho{}_{\sigma\mu}p_\rho X^\sigma\right)\dif x^\mu|_{\xp} \label{eq:hatted connection map} \\
    &=& \Left.\hat{K}(\boldsymbol{Z})\Right|_{\xp}-C^\rho{}_{\sigma\mu}p_\rho X^\sigma\dif x^\mu|_{\xp}, \nonumber
\end{eqnarray}
with $\boldsymbol{Z}$ given again by \eqref{eq:vector in phase space Z}, and the corresponding \emph{hatted horizontal subspace},
\begin{equation}
    \hat{\mathscr{H}}_{\xp}:=\ker\hat{K}|_{\xp}=\lin\Left\{\Left.\dfrac{\hat{\Dif}}{\partial x^\mu}\Right|_{\xp}\Right\},
\end{equation}
with $\{\hat{\Dif}/\partial x^\mu\}$ given by \eqref{eq:hatted horizontal basis}, one then has the alternative direct-sum decomposition
\begin{equation} \label{eq:vertical + hatted horizontal decomposition}
    \mathscr{T}_{\xp}(\mathscr{T}^*\mathscr{M})=\mathscr{V}_{\xp}\oplus\hat{\mathscr{H}}_{\xp}~.
\end{equation}
Note that the definition of the vertical subspace is frame-invariant, since the projection operator $\pi$ does not require the metric tensor in order to be defined. As such, there exist unique $\boldsymbol{Z}^{\hat{\ver}}\in\mathscr{V}_{\xp}$ and $\boldsymbol{Z}^{\hat{\hor}}\in\hat{\mathscr{H}}_{\xp}$ such that, given any phase-space vector $\boldsymbol{Z}$,
\begin{equation}
    \boldsymbol{Z}=\boldsymbol{Z}^{\hat{\ver}}+\boldsymbol{Z}^{\hat{\hor}},
\end{equation}
with
\begin{eqnarray}
    \boldsymbol{Z}^{\hat{\hor}} &\equiv& Z^{\hat{\hor}\mu}\Left.\dfrac{\hat{\Dif}}{\partial x^\mu}\Right|_{\xp}, \\
    \boldsymbol{Z}^{\hat{\ver}} &\equiv& Z^{\hat{\ver}}_\mu\Left.\dfrac{\partial}{\partial p_\mu}\Right|_{\xp},
\end{eqnarray}
and
\begin{eqnarray}
    Z^{\hat{\hor}\mu} &:=& X^\mu, \label{eq:ZHmu hatted} \\
    Z^{\hat{\ver}}_\mu &:=& Y_\mu-\hat{\Gamma}^\rho{}_{\sigma\mu}\,p_\rho X^\sigma. \label{eq:ZVmu hatted}
\end{eqnarray}
Combining Eqns.~\eqref{eq:ZHmu unhatted}, \eqref{eq:ZVmu unhatted}, \eqref{eq:ZHmu hatted} and \eqref{eq:ZVmu hatted}, one finds the relation between the hatted and unhatted vertical and horizontal components of $\boldsymbol{Z}$:
\begin{eqnarray}
    Z^{\hat{\hor}\mu} &=& Z^{\hor\mu}, \label{eq:hatted and unhatted ZHmu} \\
    Z^{\hat{\ver}}_\mu &=& Z^\ver_\mu-C^\rho{}_{\sigma\mu}p_\rho Z^{\hor\mu}. \label{eq:hatted and unhatted ZVmu}
\end{eqnarray}
We therefore conclude that (i) both the hatted and unhatted horizontal components of any phase-space vector coincide (and are furthermore equal to component $X^\mu$ in local adapted coordinates), whereas the hatted and unhatted vertical components generically differ, in spite of the conformal invariance of the vertical subspace. The previous relations further entail that (ii) vectors which are vertical in one frame remain vertical in the other, and their vertical components in both frames agree: without loss of generality, if $Z^{\hor\mu}=0$ $\Rightarrow$ $Z^{\hat{\hor}\mu}=0$ and $Z^{\hat{\ver}}_\mu=Z^\ver_\mu$; and (iii) in general, phase-space vectors which are purely horizontal in one frame develop a vertical component in another. For example, in the case of particles having a non-constant mass as seen from the hatted frame (such as particles which are free in the unhatted frame), the restriction to $\massshell$ of their Liouville vector features a non-vanishing hatted vertical component, cf.~\eqref{eq:hatted X_C m(x)}, and vice versa.

As done in the last subsection, we also provide an expression for the basis of $\mathscr{T}^*_{\xp}(\mathscr{T}^*\mathscr{M})$ dual to
\begin{equation} \label{eq:hatted D basis}
    \Left\{\Left.\dfrac{\hat{\Dif}}{\partial x^\mu}\Right|_{\xp},\Left.\dfrac{\partial}{\partial p_\mu}\Right|_{\xp}\Right\}.
\end{equation}
The corresponding dual basis is:
\begin{equation}
    \Left\{\Left.\hat{\Dif} p_\mu\Right|_{\xp},\Left.\dif x^\mu\Right|_{\xp}\Right\},
\end{equation}
where
\begin{eqnarray}
    \hat{\Dif} p_\mu|_{\xp} &:=& \dif p_\mu|_{\xp}-\hat{\Gamma}^{\rho}{}_{\sigma\mu}\,p_\rho\,\dif x^\sigma|_{\xp} \nonumber \\
    &=& \Dif p_\mu|_{\xp}-C^{\rho}{}_{\sigma\mu}\,p_\rho\,\dif x^\sigma|_{\xp}.
\end{eqnarray}

\subsection{Unhatted Tilted Horizontal Subspace}

Given a particle of variable mass $m(x)$ as measured in the unhatted frame, we define the corresponding \emph{unhatted tilted connection map} $\tilde{K}$ through the following action on a phase-space vector $\boldsymbol{Z}$ at $\xp$:
\begin{equation} \label{eq:unhatted tilted connection map}
    \Left.\tilde{K}(\boldsymbol{Z})\Right|_{\xp}:=\Left.K(\boldsymbol{Z})\Right|_{\xp}-\boldsymbol{\Delta}\Left(\dif\pi\Left.(\boldsymbol{Z})\Right|_{\xp}\Right)\,\boldsymbol{p},
\end{equation}
where
\begin{equation}
    \boldsymbol{\Delta}\equiv\dif\ln{m(x)}.
\end{equation}
The adjective `tilted' refers to the extra piece depending on the gradient of the particle's mass as measured in the unhatted frame, which vanishes for particles that are free in that frame. Equation \eqref{eq:unhatted tilted connection map} can be written as
\begin{equation}
    \Left.\tilde{K}(\boldsymbol{Z})\Right|_{\xp}=Z^\tver_\mu\,\dif x^\mu|_{\xp},
\end{equation}
where 
\begin{equation} \label{eq:ZVmu unhatted tilted}
    Z^\tver_\mu:=Z^\ver_\mu-Z^{\hor\rho}\,\dfrac{\partial\ln{m(x)}}{\partial x^\rho}\,p_\mu,
\end{equation}
with $Z^{\hor\mu}$ and $Z^\ver_\mu$ as presented in Eqns.~\eqref{eq:ZHmu unhatted} and \eqref{eq:ZVmu unhatted}, respectively. Mirroring \eqref{eq:unhatted horizontal subspace}, we then define the \emph{unhatted tilted connection map} as
\begin{equation} \label{eq:unhatted tilted horizontal subspace}
    \tilde{\mathscr{H}}_{\xp}:=\ker\tilde{K}|_{\xp},
\end{equation}
which corresponds to the set of phase-space vectors $\boldsymbol{Z}$ having vanishing $Z^\tver_\mu$, as per Eq.~\eqref{eq:ZVmu unhatted tilted}. It turns out that
\begin{equation}
    \tilde{\mathscr{H}}_{\xp}=\lin\Left\{\Left.\dfrac{\tilde{\Dif}}{\partial x^\mu}\Right|_{\xp}\Right\},
\end{equation}
where
\begin{equation} \label{eq:unhatted tilted horizontal basis}
    \dfrac{\tilde{\Dif}}{\partial x^\mu}:=\dfrac{\Dif}{\partial x^\mu}+\dfrac{\partial\ln{m(x)}}{\partial x^\mu}\,p_\rho\,\dfrac{\partial}{\partial p_\rho}.
\end{equation}
One can immediately check that, with these definitions,
\begin{eqnarray}
    \mathscr{V}_{\xp}\cap\tilde{\mathscr{H}}_{\xp}=\{0|_{\xp}\},
\end{eqnarray}
and since the non-coordinate basis
\begin{equation} \label{eq:unhatted tilted D basis}
    \Left\{\Left.\dfrac{\tilde{\Dif}}{\partial x^\mu}\Right|_{\xp},\Left.\dfrac{\partial}{\partial p_\mu}\Right|_{\xp}\Right\}
\end{equation}
spans the entire $\mathscr{T}_{\xp}(\mathscr{T}^*\mathscr{M})$, we then have the direct-sum decomposition
\begin{equation} \label{eq:vertical + unhatted tilted horizontal decomposition}
    \mathscr{T}_{\xp}(\mathscr{T}^*\mathscr{M})=\mathscr{V}_{\xp}\oplus\tilde{\mathscr{H}}_{\xp}.
\end{equation}
As such, every phase-space vector at point $\xp$ can be uniquely expressed as
\begin{equation}
    \boldsymbol{Z}=\boldsymbol{Z}^\tver+\boldsymbol{Z}^\thor,
\end{equation}
with $\boldsymbol{Z}^{\tver}\in\mathscr{V}_{\xp}$ and $\boldsymbol{Z}^{\thor}\in\tilde{\mathscr{H}}_{\xp}$ given by
\begin{eqnarray}
    \boldsymbol{Z}^\thor &\equiv& Z^{\thor\mu}\Left.\dfrac{\tilde{\Dif}}{\partial x^\mu}\Right|_{\xp}, \\
    \boldsymbol{Z}^\tver &\equiv& Z^\tver_\mu\Left.\dfrac{\partial}{\partial p_\mu}\Right|_{\xp},
\end{eqnarray}
where we recall that $Z^\tver_\mu$ was introduced back in Eq.~\eqref{eq:ZVmu unhatted tilted}, and
\begin{equation} \label{eq:ZHmu unhatted tilted}
    Z^{\thor\mu}:=Z^{\hor\mu}.
\end{equation}
We thus see that the effects of tilting the unhatted connection map are similar to those of a change of frame. To be more precise, (i) the unhatted tilted horizontal and unhatted horizontal components of phase-space vectors coincide, e.g.
\begin{equation} \label{eq:unhatted tilted horizontal component of X_C}
    (X_\mathcal{C})^\thor_\mu=(X_\mathcal{C})^\hor_\mu=2g^{\mu\nu}p_\nu;
\end{equation}
(ii) vectors which are purely unhatted vertical are also purely unhatted tilted vertical (because $Z^{\hor\mu}=0$ $\Rightarrow$ $Z^{\thor\mu}=0$ and $Z^{\tver}_\mu=Z^\ver_\mu$); (iii) vectors which are purely unhatted horizontal are generically not completely unhatted tilted horizontal.

Nonetheless, there is a fundamental difference between free and non-free particles in the unhatted frame. In the latter case, the restriction of the Liouville vector to the constraint surface $\massshell$ is \emph{not} tilted horizontal. This is because, for particles of non-constant mass, $\boldsymbol{X}_\mathcal{C}$ has the following non-trivial tilted-vertical component:
\begin{equation} \label{eq:unhatted tilted vertical component of X_C}
    (X_\mathcal{C})^\tver_\mu=-2m(x)\dfrac{\partial m(x)}{\partial x^\mu}-2g^{\rho\sigma}p_\sigma p_\mu\dfrac{\partial\ln{m(x)}}{\partial x^\rho}.
\end{equation}
On shell, this tilted vertical part can be rewritten as
\begin{equation} \label{eq:unhatted tilted vertical component of X_C on-shell}
    (X_\mathcal{C})^\tver_\mu\approx g^{\rho\sigma}\tilde{F}_{\mu\rho}p_\sigma,
\end{equation}
where
\begin{equation} \label{eq:unhatted tilde F}
    \tilde{F}_{\mu\rho}:=4p_{[\rho}\dfrac{\partial\ln{m(x)}}{\partial x^{\mu]}}.
\end{equation}
This situation mirrors the case of a charged particle of constant mass in the unhatted frame, where \cite{Sarbach:2013hna,Acuna-Cardenas:2021nkj}
\begin{equation}
    (X_\mathcal{C})^\ver_\mu=q\, g^{\rho\sigma}F_{\mu\rho}p_\sigma,
\end{equation}
with $q$ being the particle's electric charge and $\boldsymbol{F}$ being the electromagnetic field. The tilted vertical component \eqref{eq:unhatted tilted vertical component of X_C on-shell} present in the variable-mass case can then be understood as arising from the particle carrying a scalar charge coupled to the scalar field $m(x)$.

To close this subsection, we note that the basis of $\mathscr{T}^*_{\xp}(\mathscr{T}^*\mathscr{M})$ dual to \eqref{eq:unhatted tilted D basis} is
\begin{equation} \label{eq:unhatted tilted D basis dual}
    \Left\{\Left.\tilde{\Dif} p_\mu\Right|_{\xp},\Left.\dif x^\mu\Right|_{\xp}\Right\},
\end{equation}
with
\begin{equation} \label{eq:unhatted Dpmu tilde}
    \tilde{\Dif} p_\mu|_{\xp}:=\dif p_\mu|_{\xp}-\dfrac{\partial\ln{m(x)}}{\partial x^\rho}\,p_\mu\,\dif x^\rho|_{\xp}.
\end{equation}

\subsection{Hatted Tilted Horizontal Subspace}

A parallel construction may instead be carried out for particles of variable mass $\hat{m}(x)$ as seen from the hatted frame. Employing the corresponding hatted quantities produces the alternative splitting
\begin{equation} \label{eq:vertical + hatted tilted horizontal decomposition}
    \mathscr{T}_{\xp}(\mathscr{T}^*\mathscr{M})=\mathscr{V}_{\xp}\oplus\hat{\tilde{\mathscr{H}}}_{\xp},
\end{equation}
where we have introduced the \emph{hatted tilted horizontal subspace},
\begin{equation}
    \hat{\tilde{\mathscr{H}}}_{\xp}:=\ker\hat{\tilde{K}}|_{\xp}=\lin\Left\{\Left.\dfrac{\hat{\tilde{\Dif}}}{\partial x^\mu}\Right|_{\xp}\Right\}.
\end{equation}
Here, the \emph{hatted tilted connection map} $\hat{\tilde{K}}$ acts on any phase-space vector $\boldsymbol{Z}$ at point $\xp$ as
\begin{equation} \label{eq:hatted tilted connection map}
    \Left.\hat{\tilde{K}}(\boldsymbol{Z})\Right|_{\xp}:=\Left.\hat{K}(\boldsymbol{Z})\Right|_{\xp}-\hat{\boldsymbol{\Delta}}\Left(\dif\pi\Left.(\boldsymbol{Z})\Right|_{\xp}\Right)\,\boldsymbol{p},
\end{equation}
with 
\begin{equation}
    \hat{\boldsymbol{\Delta}}\equiv\dif\ln{\hat{m}(x)},
\end{equation}
and we have also introduced
\begin{eqnarray}
    \dfrac{\hat{\tilde{\Dif}}}{\partial x^\mu} &:=& \dfrac{\hat{\Dif}}{\partial x^\mu}+\dfrac{\partial\ln{\hat{m}(x)}}{\partial x^\mu}\,p_\rho\,\dfrac{\partial}{\partial p_\rho} \nonumber \\
    &=& \dfrac{\tilde{\Dif}}{\partial x^\mu}+C^{\rho}{}_{\sigma\mu}p_\rho\dfrac{\partial}{\partial p_\sigma}. \label{eq:hatted tilted horizontal basis}
\end{eqnarray}
Once more, any phase-space vector $\boldsymbol{Z}$ admits the unique decomposition
\begin{equation}
    \boldsymbol{Z}=\boldsymbol{Z}^{\htver}+\boldsymbol{Z}^{\hthor},
\end{equation}
where
\begin{eqnarray}
    \boldsymbol{Z}^{\hthor} &\equiv& Z^{\hthor\mu}\Left.\dfrac{\hat{\tilde{\Dif}}}{\partial x^\mu}\Right|_{\xp}, \\
    \boldsymbol{Z}^{\htver} &\equiv& Z^{\htver}_\mu\Left.\dfrac{\partial}{\partial p_\mu}\Right|_{\xp},
\end{eqnarray}
and
\begin{eqnarray}
    Z^{\hthor\mu} &:=& Z^{\hat{\hor}\mu}, \label{eq:ZHmu hatted tilted} \\
    Z^{\htver}_\mu &:=& Z^{\hat{\ver}}_\mu-Z^{\hat{\hor}\rho}\,\dfrac{\partial\ln{\hat{m}(x)}}{\partial x^\rho}\,p_\mu. \label{eq:ZVmu hatted tilted}
\end{eqnarray}
It is straightforward to establish the relation between the hatted and unhatted tilted vertical and tilted horizontal components of $\boldsymbol{Z}$ by combining Eqns.~\eqref{eq:ZVmu unhatted tilted}, \eqref{eq:ZHmu unhatted tilted}, \eqref{eq:ZHmu hatted tilted} and \eqref{eq:ZVmu hatted tilted}:
\begin{eqnarray}
    Z^{\hthor\mu} &=& Z^{\thor\mu}, \label{eq:hatted and unhatted tilted ZHmu} \\
    Z^{\htver}_\mu &=& Z^\tver_\mu-C^\rho{}_{\sigma\mu}p_\rho Z^{\thor\mu}. \label{eq:hatted and unhatted tilted ZVmu}
\end{eqnarray}
There relations are analogous to \eqref{eq:hatted and unhatted ZHmu} and \eqref{eq:hatted and unhatted ZVmu}. As a result, conditions similar to (i)--(iii) as outlined in the previous subsections hold in this case as well. To give an example, the hatted tilted horizontal and hatted horizontal components of a phase-space vector agree, as in
\begin{equation} \label{eq:hatted tilted horizontal component of X_C hat}
    (X_{\hat{\mathcal{C}}})^{\hthor}_\mu=(X_{\hat{\mathcal{C}}})^{\hat{\hor}}_\mu=2\hat{g}^{\mu\nu}p_\nu.
\end{equation}

For further reference, we note that the basis of $\mathscr{T}^*_{\xp}(\mathscr{T}^*\mathscr{M})$ dual to
\begin{equation} \label{eq:hatted tilted D basis}
    \Left\{\Left.\dfrac{\hat{\tilde{\Dif}}}{\partial x^\mu}\Right|_{\xp},\Left.\dfrac{\partial}{\partial p_\mu}\Right|_{\xp}\Right\}
\end{equation}
is given by
\begin{equation} \label{eq:hatted tilted D basis dual}
    \Left\{\Left.\hat{\tilde{\Dif}} p_\mu\Right|_{\xp},\Left.\dif x^\mu\Right|_{\xp}\Right\},
\end{equation}
where
\begin{eqnarray}
    \hat{\tilde{\Dif}} p_\mu|_{\xp} &:=& \dif p_\mu|_{\xp}-\dfrac{\partial\ln{\hat{m}(x)}}{\partial x^\rho}\,p_\mu\,\dif x^\rho|_{\xp}. \nonumber \\
    &=& \tilde{\Dif} p_\mu|_{\xp}-C^{\rho}{}_{\sigma\mu}\,p_\rho\,\dif x^\sigma|_{\xp}. \label{eq:hatted Dpmu tilde}
\end{eqnarray}

We close this subsection by noting that the vector $\boldsymbol{X}_\mathcal{\hat{C}}$ is \emph{not} hatted tilted horizontal, and its non-zero hatted tilted vertical part satisfies expressions mirroring Eqns.~\eqref{eq:unhatted tilted vertical component of X_C}--\eqref{eq:unhatted tilde F}.

\subsection{Relation between Tilted Horizontal Vectors and Tangents to the Mass Shell}

We previously stated that, if a particle is free in a certain conformal frame, then a phase-space vector is tangent to its mass shell if and only if it is horizontal (i.e.~annihilated by the standard connection map in that frame). It is also established in Ref.~\cite{Acuna-Cardenas:2021nkj} that, if one considers a charged particle of constant mass in a fixed conformal frame by defining a tilted connection map with a term proportional to the electromagnetic field, then there are tangents to the constraint surface which are not purely tilted horizontal. Exactly the same situation occurs for the tilted connection maps $\tilde{K}$ and $\hat{\tilde{K}}$ defined herein. In fact, we have already shown that, in a given conformal frame, the restriction of the Liouville vector to the constraint surface is not tilted horizontal (cf.~Eq.~\eqref{eq:unhatted tilted vertical component of X_C}), even though it is tangent to the mass shell.

In this brief section we offer a proof of the statement above. Consider a variable-mass particle as seen from a given conformal frame; without loss of generality, we choose the unhatted frame. One-forms normal to the particle's mass shell $\massshell$ are proportional to
\begin{equation} \label{eq:unhatted normal}
    \dif\mathcal{C}=2g^{\mu\nu}p_\nu\tilde{\Dif} p_\mu+2\mathcal{C}\dfrac{\partial\ln{m(x)}}{\partial x^\mu}\dif x^\mu.
\end{equation}
Although the second summand in Eq.~\eqref{eq:unhatted normal} vanishes on-shell, i.e.
\begin{equation} \label{eq:unhatted normal on shell}
    \dif\mathcal{C}\approx 2g^{\mu\nu}p_\nu\tilde{\Dif} p_\mu,
\end{equation}
it is essential to keep it for the purpose of this derivation, for reasons that will become apparent below. A phase-space vector $\boldsymbol{t}$ is tangent to $\massshell$ provided that
\begin{equation} \label{eq:tangent to m(x) unhatted}
    \dif\mathcal{C}(\boldsymbol{t})\approx 0.
\end{equation}
Expressing $\boldsymbol{t}$ as
\begin{equation}
    \boldsymbol{t}=t^{\thor\mu}\dfrac{\tilde{\Dif}}{\partial x^\mu}+t^\tver_\mu\dfrac{\partial}{\partial p_\mu},
\end{equation}
condition \eqref{eq:tangent to m(x) unhatted} reduces to
\begin{equation} \label{eq:tangent to m(x) unhatted reduced}
    g^{\mu\nu}p_\nu t^\tver_\mu+\mathcal{C}\dfrac{\partial\ln{m(x)}}{\partial x^\mu}t^{\thor\mu}=0.
\end{equation}
Because the second term vanishes when evaluated on the constraint surface, condition \eqref{eq:tangent to m(x) unhatted reduced} holds on shell for all $\boldsymbol{t}$s such that $t^\tver_\mu=0$, namely, for all tilted-horizontal phase-space vectors. Therefore, all tilted horizontal vectors are tangent to the mass shell. However, and precisely because the second summand in Eq.~\eqref{eq:tangent to m(x) unhatted reduced} does not vanish \emph{strongly}, but rather \emph{weakly}, then it is possible to find vectors having $t^\tver_\mu\neq0$ which fulfil condition \eqref{eq:tangent to m(x) unhatted reduced} strongly, i.e.~before evaluating $\mathcal{C}\approx 0$. This is precisely the case of $\boldsymbol{X}_\mathcal{C}$, as one may readily check. Hence, we deduce that not all tangents to the constraint are tilted horizontal in a frame where a particle has non-constant mass.

\section{Tilted Sasaki Metrics} \label{sec:sasaki metric}

To summarise the results obtained in the foregoing section, we found that every distinct conformal frame is endowed with two different direct-sum decompositions of $\mathscr{T}_{\xp}(\mathscr{T}^*\mathscr{M})$. First, we have the standard split into horizontal and vertical subspaces, which is adapted to free particles in that frame. In addition to this, we also have the decomposition into tilted horizontal and vertical subspaces, which is better suited to the study of particles whose mass varies arbitrarily in the given frame.

As explained in Refs.~\cite{Sarbach:2013uba,ThesisRioseco,Acuna-Cardenas:2021nkj}, in each conformal frame, there exists a metric tensor on phase space, the Sasaki metric \cite{Sasaki1958}, which is orthogonal with respect to the standard decomposition into horizontal and vertical subspaces. However, the standard Sasaki metrics obtained in two different frames are in fact not directly comparable, because they refer to different physical systems (remember that a particle that is free in one frame has a non-constant mass as seen from another, but the Sasaki metric \emph{of each frame} is adapted to particles of constant mass \emph{in that particular frame}).

Our goal is, in consequence, to introduce a family of generalised Sasaki metrics (again, one for each frame), which are comparable to one another, i.e.~they are all based on the \emph{same} arbitrary massive particle. This task can be accomplished by means of the tilted connection maps presented above, and hence we shall call the resulting phase-space metrics \emph{tilted Sasaki metrics}.

In an arbitrary conformal frame, say, the unhatted frame, we define the \emph{unhatted tilted Sasaki metric} $\tsasaki$ adapted to a particle of mass $m(x)$ in such frame through the following action on two arbitrary vectors $\boldsymbol{V}$ and $\boldsymbol{W}$ belonging to $\mathscr{T}_{\xp}(\mathscr{T}^*\mathscr{M})$:
\begin{eqnarray}
    \Left.\tsasaki(\boldsymbol{V},\boldsymbol{W})\Right|_{\xp} &:=& \Left.\tsasaki(\boldsymbol{V}^\thor,\boldsymbol{W}^\thor)\Right|_{\xp}+\Left.\tsasaki(\boldsymbol{V}^\tver,\boldsymbol{W}^\tver)\Right|_{\xp} \nonumber \\
    &:=& \Left.\boldsymbol{g}\Left(\dif\pi\Left.(\boldsymbol{V})\Right|_{\xp},\dif\pi\Left.(\boldsymbol{W})\Right|_{\xp}\Right)\Right|_{\xp} \nonumber \\
    &+& \Left.\boldsymbol{g}^{-1}\Left(\tilde{K}\Left.(\boldsymbol{V})\Right|_{\xp},\tilde{K}\Left.(\boldsymbol{W})\Right|_{\xp}\Right)\Right|_{\xp} \nonumber \\
    &=& g_{\mu\nu}V^{\thor\mu}W^{\thor\mu}+g^{\mu\nu}V^\tver_\mu W^\tver_\nu, \label{eq:unhatted tilted sasaki}
\end{eqnarray}
where $\tilde{K}$ is the unhatted tilted connection map, as defined in Eq.~\eqref{eq:unhatted tilted connection map},  corresponding precisely to the particle's mass $m(x)$. Applying definition \eqref{eq:unhatted tilted sasaki} to the basis vectors \eqref{eq:unhatted tilted D basis}, we obtain the following expression for $\tsasaki$:
\begin{equation}
    \tsasaki=g_{\mu\nu}\dif x^\mu\otimes\dif x^\nu+g^{\mu\nu}\tilde{\Dif}p_\mu\otimes\tilde{\Dif}p_\nu,
\end{equation}
where $\tilde{\Dif}p_\mu$ was introduced back in \eqref{eq:unhatted Dpmu tilde}. From here, it follows that $\det\tsasaki=1$ in this basis, and also that
\begin{equation}
    \tsasaki^{-1}=g^{\mu\nu}\dfrac{\tilde{\Dif}}{\partial x^\mu}\otimes\dfrac{\tilde{\Dif}}{\partial x^\nu}+g_{\mu\nu}\dfrac{\partial}{\partial p_\mu}\otimes\dfrac{\partial}{\partial p_\nu}.
\end{equation}
It is apparent from the construction of $\tsasaki$ that the expressions for the standard unhatted Sasaki metric $\sasaki$ found, for instance, in Refs.~\cite{ThesisRioseco,Acuna-Cardenas:2021nkj}, are recovered in the particular case $m(x)=\const$\footnote{
    Note that our notation for the standard Sasaki metric (`$\sasaki$') differs from that in Refs.~\cite{ThesisRioseco,Acuna-Cardenas:2021nkj} (`$\hat{g}$') because the latter is reserved for the rescaled spacetime metric.
}

Now, we consider exactly the same particle as before, this time as seen from another arbitrary conformal frame, namely the hatted one, where its mass $\hat{m}(x)$ is related to the mass in the unhatted frame via Eq.~\eqref{eq:m(x) transformation}. Using the hatted tilted connection map $\hat{\tilde{K}}$ introduced in \eqref{eq:hatted tilted connection map}, we define the \emph{hatted tilted Sasaki metric} $\htsasaki$ through:
\begin{eqnarray}
    \Left.\htsasaki(\boldsymbol{V},\boldsymbol{W})\Right|_{\xp} &:=& \Left.\htsasaki(\boldsymbol{V}^{\hthor},\boldsymbol{W}^{\hthor})\Right|_{\xp}+\Left.\htsasaki(\boldsymbol{V}^{\htver},\boldsymbol{W}^{\htver})\Right|_{\xp} \nonumber \\
    &:=& \Left.\hat{\boldsymbol{g}}\Left(\dif\pi\Left.(\boldsymbol{V})\Right|_{\xp},\dif\pi\Left.(\boldsymbol{W})\Right|_{\xp}\Right)\Right|_{\xp} \nonumber \\
    &+& \Left.\hat{\boldsymbol{g}}^{-1}\Left(\hat{\tilde{K}}\Left.(\boldsymbol{V})\Right|_{\xp},\hat{\tilde{K}}\Left.(\boldsymbol{W})\Right|_{\xp}\Right)\Right|_{\xp} \nonumber \\
    &=& \hat{g}_{\mu\nu}V^{\hthor\mu}W^{\hthor\mu}+\hat{g}^{\mu\nu}V^{\htver}_\mu W^{\htver}_\nu, \label{eq:hatted tilted sasaki}
\end{eqnarray}
where $\boldsymbol{V}$ and $\boldsymbol{W}$ are arbitrary phase-space vectors at $\xp$. In terms of the non-coordinate basis \eqref{eq:hatted tilted D basis} and its corresponding dual \eqref{eq:hatted tilted D basis dual}, one has
\begin{eqnarray}
    \htsasaki &=& \hat{g}_{\mu\nu}\dif x^\mu\otimes\dif x^\nu+\hat{g}^{\mu\nu}\hat{\tilde{\Dif}}p_\mu\otimes\hat{\tilde{\Dif}}p_\nu, \\
    \tsasaki^{-1} &=& g^{\mu\nu}\dfrac{\tilde{\Dif}}{\partial x^\mu}\otimes\dfrac{\tilde{\Dif}}{\partial x^\nu}+g_{\mu\nu}\dfrac{\partial}{\partial p_\mu}\otimes\dfrac{\partial}{\partial p_\nu},
\end{eqnarray}
and $\det\htsasaki=1$ in this basis.

As a relevant application of the definitions \eqref{eq:unhatted tilted sasaki} and \eqref{eq:hatted tilted sasaki} for $\tsasaki$ and $\htsasaki$ (respectively), let us compute the unit normal vectors to the particle's constraint surface $\massshell$ in each frame. These vectors are necessary to define the corresponding hatted and unhatted volume elements on the mass shell. Starting from Eq.~\eqref{eq:unhatted normal on shell}, we first have
\begin{equation}
    \normal:=\tsasaki^{-1}(\dif\mathcal{C},\,\cdot\,)\approx 2p_\mu\dfrac{\partial}{\partial p_\mu},
\end{equation}
whose norm, computed with respect to $\tsasaki$, is
\begin{equation} \label{eq:unhatted norm of normal vector}
    ||\normal||_{\tsasaki}:=\Left|\tsasaki(\normal,\normal)\Right|^{1/2}\approx 2m(x).
\end{equation}
Consequently, we can define the \emph{unhatted unit normal} to the mass shell $\massshell$ as
\begin{equation} \label{eq:unhatted unit normal vector}
    \nnormal:=\dfrac{\normal}{||\normal||_{\tsasaki}}\approx\dfrac{p_\mu}{m(x)}\dfrac{\partial}{\partial p_\mu}.
\end{equation}
Likewise, taking instead\footnote{
    Notice that $\dif\hat{\mathcal{C}}$ and $\dif\mathcal{C}$ are actually proportional to each other, considering that $\dif\hat{\mathcal{C}}=\Omega^{-2}\,\dif\mathcal{C}+\mathcal{C}\,\dif\Omega^{-2}\approx\Omega^{-2}\,\dif\mathcal{C}$.
}
\begin{equation} \label{eq:hatted normal}
    \dif\hat{\mathcal{C}}\approx 2\hat{g}^{\mu\nu}p_\nu\hat{\tilde{\Dif}} p_\mu,
\end{equation}
we first find that
\begin{equation}
    \hnormal\approx\normal,
\end{equation}
as well as
\begin{equation}
    ||\hnormal||_{\htsasaki}:=\Left|\htsasaki(\hnormal,\hnormal)\Right|^{1/2}\approx 2\hat{m}(x),
\end{equation}
allowing us to define the \emph{hatted unit normal} to $\massshell$ as
\begin{equation} \label{eq:hatted unit normal vector}
    \hnnormal:=\dfrac{\hnormal}{||\hnormal||_{\htsasaki}}\approx\dfrac{p_\mu}{\hat{m}(x)}\dfrac{\partial}{\partial p_\mu},
\end{equation}
whence
\begin{equation} \label{eq:unit normal transformation}
    \hnnormal\approx\Omega\,\nnormal.
\end{equation}
We see that $\hnnormal$ is of exactly the same functional form as $\nnormal$, the only difference being that unhatted quantities are replaced by the corresponding hatted ones, as expected for two quantities representing exactly the same physical magnitude in two different frames. In addition, note that this fact remains true even if the particle happens to be free in either the hatted or the unhatted frame: expressions \eqref{eq:unhatted unit normal vector} and \eqref{eq:hatted unit normal vector} reduce to the corresponding equations for the unit normal vector in e.g.~Ref.~\cite{Acuna-Cardenas:2021nkj} when $m(x)=\const$ or $\hat{m}(x)=\const$, respectively.

\section{Volume Elements on Phase Space} \label{sec:volume elements}

\subsection{Volume Element on $\mathscr{T}^*\mathscr{M}$}

Equipped with the tilted Sasaki metrics \eqref{eq:unhatted tilted sasaki} and \eqref{eq:hatted tilted sasaki}, which are adapted to an arbitrary massive particle as seen from two different conformal frames, we now turn to define the associated volume forms on $\mathscr{T}^*\mathscr{M}$, $\massshell^+$ and $\masshyp^+_x$. We restrict ourselves to the future mass shell and hyperboloid because we are only interested in particles propagating towards the future.

We begin with the hatted and unhatted volume elements in the entire phase space $\mathscr{T}^*\mathscr{M}$, from where the volume elements on $\massshell^+$ derive. Proceeding exactly as in Ref.~\cite{Acuna-Cardenas:2021nkj}, we first notice that the phase-space volume form induced by $\tsasaki$ is given by
\begin{eqnarray}
    \boldsymbol{\eta}_{\mathscr{T}^*\mathscr{M}} &:=& -\sqrt{\det\tsasaki}\,\tilde{\Dif}p_0\wedge\ldots\wedge\tilde{\Dif}p_3\wedge\dif x^0\wedge\ldots\wedge\dif x^3 \nonumber \\
    &=& -\tilde{\Dif}p_0\wedge\ldots\wedge\tilde{\Dif}p_3\wedge\dif x^0\wedge\ldots\wedge\dif x^3 \label{eq:unhatted eta phase space non-coordinate}
\end{eqnarray}
in terms of the non-coordinate basis \eqref{eq:unhatted tilted D basis dual}, while the volume element induced by $\htsasaki$ on $\mathscr{T}^*\mathscr{M}$ is
\begin{eqnarray}
    \hat{\boldsymbol{\eta}}_{\mathscr{T}^*\mathscr{M}} &:=& -\sqrt{\det\htsasaki}\,\hat{\tilde{\Dif}}p_0\wedge\ldots\wedge\hat{\tilde{\Dif}}p_3\wedge\dif x^0\wedge\ldots\wedge\dif x^3 \nonumber \\
    &=& -\hat{\tilde{\Dif}}p_0\wedge\ldots\wedge\hat{\tilde{\Dif}}p_3\wedge\dif x^0\wedge\ldots\wedge\dif x^3, \label{eq:hatted eta phase space non-coordinate}
\end{eqnarray}
where we are now using the analogous basis \eqref{eq:hatted tilted D basis dual}. Owing to the fact that we are contracting each $\tilde{\Dif}p_\mu$ in \eqref{eq:unhatted eta phase space non-coordinate} and each $\hat{\tilde{\Dif}}p_\mu$ in \eqref{eq:hatted eta phase space non-coordinate} with all the basis one-forms $\{\dif x^\mu\}$, the terms proportional to $\dif x^\mu$ in \eqref{eq:unhatted Dpmu tilde} and \eqref{eq:hatted Dpmu tilde} cancel out, and thus we are left with
\begin{equation} \label{eq:eta phase space}
    \boldsymbol{\eta}_{\mathscr{T}^*\mathscr{M}}=\hat{\boldsymbol{\eta}}_{\mathscr{T}^*\mathscr{M}}=-\dif p_0\wedge\ldots\wedge\dif p_3\wedge\dif x^0\wedge\ldots\wedge\dif x^3.
\end{equation}
The volume element on $\mathscr{T}^*\mathscr{M}$ is therefore frame invariant. This result is not surprising, given that, as widely known, Eq.~\eqref{eq:eta phase space} can be recast as
\begin{equation} \label{eq:eta phase space omega}
    \boldsymbol{\eta}_{\mathscr{T}^*\mathscr{M}}=-\dfrac{1}{4!}\bigwedge_{n=1}^4\sform,
\end{equation}
which is built using exclusively the (conformally invariant) symplectic structure on phase space. Together with Eq.~\eqref{eq:Lie derivative of omega}, Eq.~\eqref{eq:eta phase space omega} entails that
\begin{equation}
    \pounds_{\boldsymbol{X}_\mathcal{C}}\boldsymbol{\eta}_{\mathscr{T}^*\mathscr{M}}=\pounds_{\boldsymbol{X}_{\hat{\mathcal{C}}}}\boldsymbol{\eta}_{\mathscr{T}^*\mathscr{M}}=0,
\end{equation}
which is the expression of Liouville's theorem in $\mathscr{T}^*\mathscr{M}$ and, as can be seen, holds in both the hatted and the unhatted frames.

\subsection{Hatted and Unhatted Volume Elements on the Future Mass Shell}

For the purpose of computing the hatted and unhatted volume forms on the future mass shell $\massshell^+$, following Ref.~\cite{Acuna-Cardenas:2021nkj}, we introduce an orthonormal basis for vectors on spacetime $(\mathscr{M},\boldsymbol{g})$, namely,
\begin{equation}
    \Left\{\boldsymbol{e}_{\dot{\alpha}}=e_{\dot{\alpha}}{}^\mu\dfrac{\partial}{\partial x^\mu}\Right\}
\end{equation}
as well as its corresponding orthonormal dual basis for one forms,
\begin{equation} \label{eq:unhatted orthonormal one-form basis}
    \Left\{\boldsymbol{\theta}^{\dot{\alpha}}=e^{\dot{\alpha}}{}_\mu\dif x^\mu\Right\},
\end{equation}
both of which fulfil
\begin{eqnarray}
    \boldsymbol{g}(\boldsymbol{e}_{\dot{\alpha}},\boldsymbol{e}_{\dot{\beta}}) &=& \eta_{{\dot{\alpha}}{\dot{\beta}}}, \\
    \boldsymbol{\theta}^{\dot{\alpha}}(\boldsymbol{e}_{\dot{\beta}}) &=& \delta^{\dot{\alpha}}{}_{\dot{\beta}}.
\end{eqnarray}
In the hatted frame $(\mathscr{M},\hat{\boldsymbol{g}})$, we have analogous bases $\Left\{\hat{\boldsymbol{e}}_{\dot{\alpha}}\Right\}$ for spacetime vectors and $\Left\{\hat{\boldsymbol{\theta}}^{\dot{\alpha}}\Right\}$ for one-forms, respectively. Orthonormality in the hatted frame requires\footnote{
    Simultaneous orthogonality in both frames also implies that $\hat{\eta}_{{\dot{\alpha}}{\dot{\beta}}}=\eta_{{\dot{\alpha}}{\dot{\beta}}}$, and $\hat{\delta}^{\dot{\alpha}}{}_{\dot{\beta}}=\delta^{\dot{\alpha}}{}_{\dot{\beta}}$.
}
\begin{eqnarray}
    \hat{\boldsymbol{e}}_{\dot{\alpha}} &=& \Omega^{-1}\boldsymbol{e}_{\dot{\alpha}}, \\
    \hat{\boldsymbol{\theta}}^{\dot{\alpha}} &=& \Omega\,\boldsymbol{\theta}^{\dot{\alpha}}.
\end{eqnarray}
Hence, taking into account that one-form $\boldsymbol{p}$ is frame-invariant, its components in the aforementioned hatted and unhatted orthonormal bases are related by
\begin{equation} \label{eq:p dot alpha transformation}
    \hat{p}_{\dot{\alpha}}=\Omega^{-1}\,p_{\dot{\alpha}}.
\end{equation}
Let us concentrate now on the unhatted frame. The mass-shell constraint \eqref{eq:constraint m(x) unhatted} therein can be rewritten as
\begin{equation}
    p_{\dot{0}}\approx-\sqrt{m^2(x)+\delta^{\dot{a}\dot{b}}\,p_{\dot{a}}\,p_{\dot{b}}},
\end{equation}
where we remind the reader that dotted Latin indices run in the three spatial directions only. Differentiating this expression, we find
\begin{equation} \label{eq:dp_0 dot}
    \dif p_{\dot{0}}\approx\dfrac{p^{\dot{a}}}{p_{\dot{0}}}\Left[\dif p_{\dot{a}}-p_{\dot{a}}\dfrac{\partial\ln{m(x)}}{\partial x^\mu}\dif x^\mu\Right],
\end{equation}
with
\begin{equation}
    p^{\dot{\alpha}}\equiv\eta^{\dot{\alpha}\dot{\beta}}\,p_{\dot{\beta}}.
\end{equation}
Introducing
\begin{equation}
    \dot{\Gamma}^{\dot{\alpha}}{}_{\mu\dot{\beta}}:=\boldsymbol{\theta}^{\dot{\alpha}}\Left(\boldsymbol{\nabla}_{\frac{\partial}{\partial x^\mu}}\boldsymbol{e}_{\dot{\beta}}\Right),
\end{equation}
we have that
\begin{equation}
    \tilde{\Dif} p_{\dot{0}}=\dif p_{\dot{0}}-\Left[\dot{\Gamma}^{\dot{\alpha}}{}_{\mu\dot{0}}\,p_{\dot{\alpha}}+p_{\dot{0}}\dfrac{\partial\ln{m(x)}}{\partial x^\mu}\Right]\dif x^\mu.
\end{equation}
Upon inserting \eqref{eq:dp_0 dot}, and after some algebra, this expression simplifies to
\begin{equation} \label{eq:Dif p_0 dot}
    \tilde{\Dif} p_{\dot{0}}\approx\dfrac{p^{\dot{a}}}{p_{\dot{0}}}\tilde{\Dif} p_{\dot{a}}.
\end{equation}
We can now substitute \eqref{eq:Dif p_0 dot} in the unhatted tilted Sasaki metric \eqref{eq:unhatted tilted sasaki}, which reads
\begin{equation} \label{eq:unhatted tilted sasaki constraint}
    \tsasaki=g_{\mu\nu}\,\dif x^\mu\otimes\dif x^\nu+\eta^{\dot{\alpha}\dot{\beta}}\,\tilde{\Dif} p_{\dot{\alpha}}\otimes\tilde{\Dif} p_{\dot{\beta}}
\end{equation}
when using local adapted coordinates $x^\mu$ in the spatiotemporal directions and the unhatted orthonormal dual basis \eqref{eq:unhatted orthonormal one-form basis} in the momentum directions. In doing so, we find the \emph{induced unhatted metric on} $\massshell^+$,
\begin{equation} \label{eq:unhatted induced metric constraint}
    \tinduced:\approx g_{\mu\nu}\,\dif x^\mu\otimes\dif x^\nu+H^{\dot{a}\dot{b}}\,\tilde{\Dif} p_{\dot{a}}\otimes\tilde{\Dif} p_{\dot{b}},
\end{equation}
where
\begin{equation} \label{eq:H^ab}
    H^{\dot{a}\dot{b}}\equiv\delta^{\dot{a}\dot{b}}-\dfrac{p^{\dot{a}}p^{\dot{b}}}{(p_{\dot{0}})^2}.
\end{equation}
The \emph{unhatted volume element on the future mass shell} associated to $\tinduced$ is then constructed as
\begin{eqnarray}
    \tieta_{\massshell^+} &:\approx& \sqrt{-\det(g_{\mu\nu})}\,\dif x^0\wedge\dif x^1\wedge\dif x^2\wedge\dif x^3 \nonumber \\
    && \wedge\,\det(H^{\dot{a}\dot{b}})\,\tilde{\Dif} p_{\dot{1}}\wedge\tilde{\Dif} p_{\dot{2}}\wedge\tilde{\Dif} p_{\dot{3}} \nonumber \\
    &\approx& \sqrt{-\det(g_{\mu\nu})}\,\dif x^0\wedge\dif x^1\wedge\dif x^2\wedge\dif x^3 \nonumber \\
    && \wedge\,\dfrac{m(x)}{|p_{\dot{0}}|}\,\dif p_{\dot{1}}\wedge\dif p_{\dot{2}}\wedge\dif p_{\dot{3}}. \label{eq:unhatted eta constraint}
\end{eqnarray}
The preceding expression is formally identical to the one provided in Ref.~\cite{Acuna-Cardenas:2021nkj} for a free particle, with the constant mass replaced by $m(x)$. Moreover, one may readily check that the exact same expression for $\tieta_{\massshell^+}$ can be obtained by contracting the unhatted unit normal $\nnormal$ to $\massshell^+$, as given in Eq.~\eqref{eq:unhatted unit normal vector}, with the volume form in the cotangent bundle \eqref{eq:eta phase space}:
\begin{equation} \label{eq:unhatted eta constraint using normal}
    \tieta_{\massshell^+}=\iota_{\nnormal}\boldsymbol{\eta}_{\mathscr{T}^*\mathscr{M}}.
\end{equation}

Repeating the same steps in the hatted frame (which we shall omit for brevity), one finds the analogous \emph{hatted volume element on the future mass shell}, which turns out to be given by
\begin{eqnarray}
    \htieta_{\massshell^+} &:\approx& \sqrt{-\det(\hat{g}_{\mu\nu})}\,\dif x^0\wedge\dif x^1\wedge\dif x^2\wedge\dif x^3 \nonumber \\
    && \wedge\,\det(\hat{H}^{\dot{a}\dot{b}})\,\hat{\tilde{\Dif}}\hat{p}_{\dot{1}}\wedge\hat{\tilde{\Dif}}\hat{p}_{\dot{2}}\wedge\hat{\tilde{\Dif}}\hat{p}_{\dot{3}} \nonumber \\
    &\approx& \sqrt{-\det(\hat{g}_{\mu\nu})}\,\dif x^0\wedge\dif x^1\wedge\dif x^2\wedge\dif x^3 \nonumber \\
    && \wedge\,\dfrac{\hat{m}(x)}{|\hat{p}_{\dot{0}}|}\,\dif\hat{p}_{\dot{1}}\wedge\dif \hat{p}_{\dot{2}}\wedge\dif\hat{p}_{\dot{3}} \nonumber \\
    &\approx& \sqrt{-\det(\hat{g}_{\mu\nu})}\,\dif x^0\wedge\dif x^1\wedge\dif x^2\wedge\dif x^3 \nonumber \\
    && \wedge\,\dfrac{\hat{m}(x)}{|\hat{p}_{\dot{0}}|}\,\dif p_{\dot{1}}\wedge\dif p_{\dot{2}}\wedge\dif p_{\dot{3}}, \label{eq:hatted eta constraint}
\end{eqnarray}
where, in the last step, we have taken into account that the terms proportional to $\dif x^\mu$ in
\begin{equation} \label{eq:mass shell dp alpha hat dp alpha}
    \dif\hat{p}_{\dot{\alpha}}=\Omega^{-1}\,\dif p_{\dot{\alpha}}+p_{\dot{\alpha}}\,\Omega^{-2}\,\dfrac{\partial\Omega}{\partial x^\mu}\,\dif x^\mu
\end{equation}
drop by antisymmetry when computing the wedge product with the whole set of $\{\dif x^\mu\}$. As in the unhatted case, it turns out that the volume element in Eq.~\eqref{eq:hatted eta constraint} can be alternatively derived from
\begin{equation} \label{eq:hatted eta constraint using normal}
    \htieta_{\massshell^+}=\iota_{\hnnormal}\boldsymbol{\eta}_{\mathscr{T}^*\mathscr{M}},
\end{equation}
with $\hnnormal$ given by \eqref{eq:hatted unit normal vector}. Equations \eqref{eq:unhatted eta constraint}--\eqref{eq:hatted eta constraint using normal} reveal that
\begin{equation}
    \htieta_{\massshell^+}=\Omega\,\tieta_{\massshell^+},
\end{equation}
upon use of transformation rules \eqref{eq:def conformal}, \eqref{eq:m(x) transformation}, \eqref{eq:unit normal transformation}, \eqref{eq:p dot alpha transformation} and \eqref{eq:eta phase space}.

\subsection{Hatted and Unhatted Volume Elements on the Future Mass Hyperboloid}

The derivation of the hatted and unhatted volume forms on $\masshyp^+_x$ parallels the one presented in the foregoing section in the case of $\massshell^+$, the difference being that $\masshyp^+_x$ is a subset of $\mathscr{T}^*_x\mathscr{M}$ rather than the full $\mathscr{T}^*\mathscr{M}$. This means that, across $\masshyp^+_x$, the spacetime point $x$ is held fixed,\footnote{
    Note that we are expressing equalities holding only on $\masshyp^+_x$ with `$\approx$,' in a slight abuse of notation.
} and thus all relevant expressions are effectively obtained by setting $\dif x^\mu\approx 0$ in the equations previously derived for $\massshell^+$. For instance, on $\masshyp^+_x$ one has the much simplified relations:
\begin{equation} \label{eq:hyperboloid simplified relations}
    \tilde{\Dif} p_{\dot{0}}\approx\dif p_{\dot{0}}\approx\dfrac{p^{\dot{a}}}{p_{\dot{0}}}\dif p_{\dot{a}},~~~~\tilde{\Dif} p_{\dot{a}}\approx\dif p_{\dot{a}}.
\end{equation}
The \emph{unhatted metric on $\masshyp^+_x$ induced by $\tsasaki$} is obtained from Eqns.~\eqref{eq:unhatted tilted sasaki constraint}, \eqref{eq:unhatted induced metric constraint} and \eqref{eq:hyperboloid simplified relations}, and reads
\begin{equation} \label{eq:unhatted induced metric hyperboloid}
    \tilde{\boldsymbol{\mathfrak{H}}}:\approx H^{\dot{a}\dot{b}}\,\dif p_{\dot{a}}\otimes\dif p_{\dot{b}},
\end{equation}
with $H^{\dot{a}\dot{b}}$ given once again by \eqref{eq:H^ab}. The corresponding \emph{unhatted volume form on the mass hyperboloid} is thus
\begin{equation} \label{eq:unhatted eta hyperboloid}
    \tieta_{\masshyp^+_x} :\approx \dfrac{m(x)}{|p_{\dot{0}}|}\,\dif p_{\dot{1}}\wedge\dif p_{\dot{2}}\wedge\dif p_{\dot{3}}, 
\end{equation}
which is similar (and reduces) to the known volume element in the constant-mass case \cite{Acuna-Cardenas:2021nkj}, and may as well be obtained from
\begin{equation} \label{eq:unhatted eta hyperboloid using normal}
    \tieta_{\masshyp^+_x}=\iota_{\nnormal}\tieta_{\mathscr{T}^*_x\mathscr{M}},
\end{equation}
where
\begin{equation}
    \tieta_{\mathscr{T}^*_x\mathscr{M}}=-\dif p_{\dot{0}}\wedge\dif p_{\dot{1}}\wedge\dif p_{\dot{2}}\wedge\dif p_{\dot{3}}
\end{equation}
is the natural volume element on the cotangent space at $x$, and the unit normal to $\masshyp^+_x$ is still $\nnormal$ as provided in \eqref{eq:unhatted unit normal vector}. Combining \eqref{eq:unhatted eta constraint} and \eqref{eq:unhatted eta hyperboloid}, one finds the useful formula
\begin{equation} \label{eq:unhatted wedge decomposition of constraint volume}
    \tieta_{\massshell^+}\approx\boldsymbol{\eta}_\mathscr{M}\wedge\tieta_{\masshyp^+_x},
\end{equation}
where $\boldsymbol{\eta}_\mathscr{M}$ is the standard volume form on $\mathscr{M}$ associated to the unhatted spacetime metric $\boldsymbol{g}$. Notice that the `untilted' equivalent of \eqref{eq:unhatted wedge decomposition of constraint volume} is known to hold in the case of particles which are free in the unhatted frame \cite{Acuna-Cardenas:2021nkj}. It is also customary to introduce the rescaled volume element
\begin{equation} \label{eq:unhatted dvolp}
    \dvolp:=\dfrac{\tieta_{\masshyp^+_x}}{m(x)}\approx\dfrac{\dif p_{\dot{1}}\wedge\dif p_{\dot{2}}\wedge\dif p_{\dot{3}}}{\sqrt{m^2(x)+\delta^{\dot{a}\dot{b}}\,p_{\dot{a}}\,p_{\dot{b}}}},
\end{equation}
which reduces to the familiar Lorentz-invariant phase-space measure for particles having $m(x)=\const$, cf.~Ref.~\cite{Acuna-Cardenas:2021nkj}.

An analogous reasoning in the hatted frame yields 
\begin{equation} \label{eq:hatted eta hyperboloid}
    \htieta_{\masshyp^+_x} \approx \dfrac{\hat{m}(x)}{|\hat{p}_{\dot{0}}|}\,\dif\hat{p}_{\dot{1}}\wedge\dif \hat{p}_{\dot{2}}\wedge\dif\hat{p}_{\dot{3}} \approx \Omega^{-3}\,\tieta_{\masshyp^+_x}
\end{equation}
for the \emph{hatted volume element on the mass hyperboloid}. Note that, in the second equality in \eqref{eq:hatted eta hyperboloid}, we have taken into account that relation \eqref{eq:mass shell dp alpha hat dp alpha} simplifies to
\begin{equation} \label{eq:dif p alpha transformation hyperboloid}
    \dif \hat{p}_{\dot{\alpha}}\approx\Omega^{-1}\,\dif p_{\dot{\alpha}}
\end{equation}
on the mass hyperboloid. The last expression further implies that
\begin{equation} \label{eq:eta cotangent space transformation}
    \htieta_{\mathscr{T}^*_x\mathscr{M}}=\Omega^{-4}\,\tieta_{\mathscr{T}^*_x\mathscr{M}}
\end{equation}
and thus \eqref{eq:hatted eta hyperboloid} is compatible with
\begin{equation} \label{eq:hatted eta hyperboloid using normal}
    \htieta_{\masshyp^+_x}=\iota_{\hnnormal}\tieta_{\mathscr{T}^*_x\mathscr{M}},
\end{equation}
where we have once more taken into consideration the transformation rule \eqref{eq:unit normal transformation} for the unit normal vector. Moreover, from \eqref{eq:hatted eta constraint} and \eqref{eq:hatted eta hyperboloid} it follows that
\begin{equation} \label{eq:hatted wedge decomposition of constraint volume}
    \htieta_{\massshell^+}\approx\hat{\boldsymbol{\eta}}_\mathscr{M}\wedge\htieta_{\masshyp^+_x},
\end{equation}
which is the conformal transformation of \eqref{eq:unhatted wedge decomposition of constraint volume}. Last but not least, we define the hatted counterpart to \eqref{eq:unhatted dvolp},
\begin{equation} \label{eq:hatted dvolp}
    \hdvolp:=\dfrac{\htieta_{\masshyp^+_x}}{\hat{m}(x)}\approx\dfrac{\dif \hat{p}_{\dot{1}}\wedge\dif \hat{p}_{\dot{2}}\wedge\dif \hat{p}_{\dot{3}}}{\sqrt{\hat{m}^2(x)+\delta^{\dot{a}\dot{b}}\,\hat{p}_{\dot{a}}\,\hat{p}_{\dot{b}}}},
\end{equation}
which transforms as
\begin{equation} \label{eq:dvolp transformation}
    \hdvolp=\Omega^{-2}\,\dvolp
\end{equation}
as follows from Eqns.~\eqref{eq:m(x) transformation}, \eqref{eq:p dot alpha transformation},  \eqref{eq:hatted eta hyperboloid} and \eqref{eq:dif p alpha transformation hyperboloid}.

\section{Statistical Interpretation} \label{sec:statistical interpretation}

\subsection{One-Particle Distribution Function}

Having already procured all the necessary concepts and mathematical tools, we can finally proceed to find the equations governing a gas of collisionless point particles of variable mass and vanishing electric charge and spin, as seen from two distinct conformal frames.

As is customary in relativistic kinetic theory, we asume that the macroscopic evolution and properties of the gas all derive from a single function in phase-space, the so-called \emph{one-particle distribution function} \cite{Ehlers1971}. As its name suggests, the one-particle distribution function and other objects derived from it depend on quantities associated with one particle, but encapsulate the full dynamics of the entire system through a statistical interpretation up to zeroth order in interactions \cite{Vereshchagin:2017}.

To start, we consider a $6$-dimensional hypersurface $\Sigma$ in $\massshell^+$ having unit normal vector $\boldsymbol{\nu}$ and induced volume element
\begin{equation} \label{eq:unhatted eta Sigma}
    \tieta_\Sigma:=\iota_{\boldsymbol{\nu}}\tieta_{\massshell^+}
\end{equation}
as seen from the unhatted frame, and unit normal vector $\hat{\boldsymbol{\nu}}$ and induced volume element
\begin{equation} \label{eq:hatted eta Sigma}
    \htieta_\Sigma:=\iota_{\hat{\boldsymbol{\nu}}}\htieta_{\massshell^+}
\end{equation}
in the hatted frame. Now, we note that particle worldlines crossing $\Sigma$ exist independently of both the system of units chosen to measure them and the parametrisation employed in their Hamiltonian description. Hence, it seems reasonable to require the \emph{ensemble average of occupied trajectories intersecting $\Sigma$} to be both frame- and gauge-invariant:
\begin{equation}
    \#[\Sigma]:=\hat{\#}[\Sigma]:\neq\#[\Sigma](\lambda).
\end{equation}
The average number is thus to be determined using only the one-particle distribution function, which we take to be a non-negative function supported on the future mass shell $\massshell^+$. Following previous works \cite{Sarbach:2013fya,Sarbach:2013uba,Acuna-Cardenas:2021nkj}, where the statistical interpretation is constructed by analogy to a hydrodinamics example, our first proposal is that $\#[\Sigma]$ and the \emph{(putative) unhatted one-particle distribution function} $\tilde{f}$ are related by
\begin{equation} \label{eq:average number unhatted}
    \#[\Sigma]:=-\int_\Sigma\tieta_\Sigma\,\tinduced(\boldsymbol{\mathcal{J}},\boldsymbol{\nu})
\end{equation}
where the \emph{unhatted current density} $\boldsymbol{\mathcal{J}}$ is defined as
\begin{equation} \label{eq:unhatted phase space current J}
    \boldsymbol{\mathcal{J}}:=\tilde{f}\,\boldsymbol{L}.
\end{equation}
Similar expressions are proposed to hold in the hatted frame for the \emph{(putative) hatted one-particle distribution function} $\hat{\tilde{f}}$ and the \emph{hatted current density} $\hat{\boldsymbol{\mathcal{J}}}$:
\begin{equation} \label{eq:average number hatted}
    \#[\Sigma]:=-\int_\Sigma\htieta_\Sigma\,\htinduced(\hat{\boldsymbol{\mathcal{J}}},\hat{\boldsymbol{\nu}}),
\end{equation}
with
\begin{equation} \label{eq:hatted induced metric constraint}
    \htinduced:\approx\hat{g}_{\mu\nu}\,\dif x^\mu\otimes\dif x^\nu+\hat{H}^{\dot{a}\dot{b}}\,\hat{\tilde{\Dif}} p_{\dot{a}}\otimes\hat{\tilde{\Dif}} p_{\dot{b}},
\end{equation}
$\hat{H}^{\dot{a}\dot{b}}=H^{\dot{a}\dot{b}}$ as defined in Eq.~\eqref{eq:H^ab}, and
\begin{equation}
    \hat{\boldsymbol{\mathcal{J}}}:=\hat{\tilde{f}}\,\hat{\boldsymbol{L}}.
\end{equation}
Let us now assume that
\begin{equation}
    \Sigma:=\Left\{\xp\,|\,x\in\mathscr{S},\,\boldsymbol{p}\in\masshyp^+_x\Right\},
\end{equation}
where $\mathscr{S}$ is a codimension-$1$ surface in the base manifold $\mathscr{M}$, with unit normal vector $\boldsymbol{s}$ as seen from the unhatted frame and
\begin{equation} \label{eq:normal to S transformation}
    \hat{\boldsymbol{s}}=\Omega^{-1}\,\boldsymbol{s}
\end{equation}
in the hatted frame. Proceeding exactly as in the constant-mass case \cite{Sarbach:2013uba,Acuna-Cardenas:2021nkj}, one can readily show that $\boldsymbol{\nu}$ is necessarily tilted horizontal, whereas $\hat{\boldsymbol{\nu}}$ is necessarily hatted tilted horizontal. In particular,
\begin{eqnarray}
    \boldsymbol{\nu} &=& s^\mu\,\dfrac{\tilde{\Dif}}{\partial x^\mu}, \label{eq:unhatted nu} \\
    \hat{\boldsymbol{\nu}} &=& \hat{s}^\mu\,\dfrac{\hat{\tilde{\Dif}}}{\partial x^\mu}=\Omega^{-1}\,s^\mu\,\dfrac{\hat{\tilde{\Dif}}}{\partial x^\mu}, \label{eq:nu transformation}
\end{eqnarray}
where in the last step we have used Eq.~\eqref{eq:normal to S transformation}. Since $\boldsymbol{\nu}$ is tilted horizontal, and the unhatted tilted Sasaki metric is, by definition, orthogonal with respect to the decomposition \eqref{eq:vertical + unhatted tilted horizontal decomposition} (cf.~\eqref{eq:unhatted tilted sasaki}), we have that
\begin{equation}
    \tinduced(\boldsymbol{Z},\boldsymbol{\nu})=\tinduced(\boldsymbol{Z}^\thor,\boldsymbol{\nu})=\boldsymbol{g}(\boldsymbol{Z}^\thor,\boldsymbol{\nu})
\end{equation}
for any phase-space vector $\boldsymbol{Z}$ at $\xp$. Similarly, given that $\hat{\boldsymbol{\nu}}$ is hatted tilted horizontal, and that the hatted tilted Sasaki metric is orthogonal with respect to the splitting \eqref{eq:vertical + hatted tilted horizontal decomposition} (cf.~\eqref{eq:hatted tilted sasaki}), we also find
\begin{equation}
    \htinduced(\boldsymbol{Z},\hat{\boldsymbol{\nu}})=\htinduced(\boldsymbol{Z}^{\hthor},\hat{\boldsymbol{\nu}})=\hat{\boldsymbol{g}}(\boldsymbol{Z}^{\hthor},\hat{\boldsymbol{\nu}}).
\end{equation}
Choosing $\boldsymbol{Z}=\boldsymbol{L}=\hat{\boldsymbol{L}}$, the last two expressions entail
\begin{eqnarray}
    \tinduced(\boldsymbol{L},\boldsymbol{\nu}) &\approx& 2\mathcal{N}(\lambda)p_\mu s^\mu, \label{eq:unhatted h(L,nu)} \\
    \htinduced(\hat{\boldsymbol{L}},\hat{\boldsymbol{\nu}}) &\approx&  2\hat{\mathcal{N}}(\lambda)p_\mu\hat{s}^\mu~=~2\Omega\mathcal{N}(\lambda)p_\mu s^\mu, \label{eq:hatted h(L,nu)}
\end{eqnarray}
where we have made use of Eqns.~\eqref{eq:N transformation}, \eqref{eq:unhatted Liouville restriction}, \eqref{eq:hatted Liouville restriction}, \eqref{eq:unhatted tilted horizontal component of X_C}, \eqref{eq:hatted tilted horizontal component of X_C hat}, \eqref{eq:unhatted nu} and \eqref{eq:nu transformation}. Inserting \eqref{eq:unhatted h(L,nu)} and \eqref{eq:hatted h(L,nu)} back into the integrals in \eqref{eq:average number unhatted} and \eqref{eq:average number hatted}, respectively, we have that
\begin{eqnarray}
    \#[\Sigma] &=& -\int_\Sigma\tieta_\Sigma\,2\,\mathcal{N}(\lambda)\,\tilde{f}\,\boldsymbol{p}(\boldsymbol{s}) \nonumber \\
    &=& -\int_\Sigma\htieta_\Sigma\,2\,\Omega\,\mathcal{N}(\lambda)\,\hat{\tilde{f}}\,\boldsymbol{p}(\boldsymbol{s}). \label{eq:average number partially transformed}
\end{eqnarray}
We now turn to find the transformation rule for $\htieta_\Sigma$. First, we use \eqref{eq:hatted wedge decomposition of constraint volume} to rewrite \eqref{eq:hatted eta Sigma} as
\begin{equation} \label{eq:partial transformation of eta_sigma}
    \htieta_\Sigma=\iota_{\hat{\boldsymbol{s}}}\hat{\boldsymbol{\eta}}_\mathscr{M}\wedge\hat{\boldsymbol{\eta}}_{\masshyp^+_x},
\end{equation}
where we have taken into account that
\begin{eqnarray}
    \iota_{\hat{\boldsymbol{\nu}}}\dif x^\mu &=& \hat{s}^\mu, \\
    \iota_{\hat{\boldsymbol{\nu}}}\dif \hat{p}_{\dot{\alpha}} &=& 0,
\end{eqnarray}
as per the hatted tilted horizontality of $\hat{\boldsymbol{\nu}}$. Applying \eqref{eq:hatted eta hyperboloid} and \eqref{eq:normal to S transformation} and \eqref{eq:nu transformation}, and bearing in mind that
\begin{equation}
    \hat{\boldsymbol{\eta}}_\mathscr{M}=\Omega^4\,\boldsymbol{\eta}_\mathscr{M},
\end{equation}
\eqref{eq:partial transformation of eta_sigma} turns into
\begin{equation} \label{eq:eta_Sigma transformation}
    \htieta_\Sigma=\iota_{\boldsymbol{s}}\boldsymbol{\eta}_\mathscr{M}\wedge\boldsymbol{\eta}_{\masshyp^+_x}=\tieta_\Sigma,
\end{equation}
i.e.~the volume form on $\Sigma$ is frame-invariant. Notice that, in the last step, we have employed \eqref{eq:unhatted wedge decomposition of constraint volume} and \eqref{eq:unhatted eta Sigma}, and taken into account that
\begin{eqnarray}
    \iota_{\boldsymbol{\nu}}\dif x^\mu &=& s^\mu, \\
    \iota_{\boldsymbol{\nu}}\dif p_{\dot{\alpha}} &=& 0
\end{eqnarray}
since $\boldsymbol{\nu}$ is unhatted tilted horizontal, see Eq.~\eqref{eq:unhatted nu}. Owing to \eqref{eq:eta_Sigma transformation}, the second equation in \eqref{eq:average number partially transformed} becomes
\begin{equation}
    \#[\Sigma]=-\int_\Sigma\tieta_\Sigma\,2\,\Omega\,\mathcal{N}(\lambda)\,\hat{\tilde{f}}\,\boldsymbol{p}(\boldsymbol{s}), \label{eq:average number fully transformed}
\end{equation}
from where we infer the transformation rule for $\tilde{f}$, since $\#[\Sigma]$ is to be conformally invariant:
\begin{equation} \label{eq:f tilde transformation}
    \hat{\tilde{f}}:=\Omega^{-1}\tilde{f}.
\end{equation}
Nonetheless, there is a major issue with expression \eqref{eq:average number fully transformed}, namely, that the presence of the Lagrange multiplier $\mathcal{N}(\lambda)$ inside the integral renders $\#[\Sigma]$ gauge-dependent, unless we choose $\tilde{f}=\tilde{f}(x,\boldsymbol{p};\lambda)$ to be parametrisation-dependent in such a way that the product $\mathcal{N}(\lambda)\,\tilde{f}(x,\boldsymbol{p};\lambda)$ remains gauge-invariant. Because
\begin{equation}
    \mathcal{N}(\tilde{\lambda})=\dfrac{\dif\lambda}{\dif\tilde{\lambda}}\,\mathcal{N}(\lambda(\tilde{\lambda})),
\end{equation}
it would suffice to require
\begin{equation}
    \tilde{f}(x,\boldsymbol{p};\tilde{\lambda}):=\dfrac{\dif\tilde{\lambda}}{\dif\lambda}\,\tilde{f}(x,\boldsymbol{p};\lambda(\tilde{\lambda})).
\end{equation}
However, from a physical point of view, having a gauge-dependent one-particle distribution function is as inadequate as having a parametrisation-dependent $\#[\Sigma]$. As mentioned earlier, this function is supposed to encapsulate the \emph{macroscopic} properties of the gas, and thus should not depend on the fiducial world-line's parametrisation. As such, we conclude that $\tilde{f}$ \emph{cannot} represent the \emph{actual one-particle distribution function} $f$, but the latter is obtained from it. In addition, we should be able to recover the usual definition of $f$ and $\#[\Sigma]$ found in the literature for free particles in the proper-time gauge $\dif\lambda=\dif\tau$.

The solution to this conundrum is relatively straightforward, and will turn out to be endowed with many mathematical and physical benefits. First, we note that Eq.~\eqref{eq:N on-shell} can be cast as
\begin{equation} \label{eq:split N}
    \mathcal{N}(\lambda)=\dfrac{1}{2m(x)}\dfrac{\dif\tau}{\dif\lambda},
\end{equation}
where the first factor on the right-hand side is gauge-invariant, while the second is gauge-dependent. Then, we can use this expression alongside \eqref{eq:unhatted dvolp} and the second equality in \eqref{eq:eta_Sigma transformation} to rewrite \eqref{eq:average number fully transformed} in the following manner:
\begin{equation} \label{eq:average number dvolp partial}
    \#[\Sigma] = -\int_\mathscr{S}\boldsymbol{\eta}_\mathscr{S}\int_{\masshyp^+}\dvolp\,\dfrac{\dif\tau}{\dif\lambda}\,\tilde{f}(x,\boldsymbol{p};\lambda)\,\boldsymbol{p}(\boldsymbol{s}),
\end{equation}
where
\begin{equation}
    \boldsymbol{\eta}_\mathscr{S}:=\iota_{\boldsymbol{s}}\boldsymbol{\eta}_\mathscr{M}.
\end{equation}
Now, we define the \emph{gauge-invariant unhatted one-particle distribution function} $f$ as
\begin{equation} \label{eq:true distribution function}
    f\xp:=\dfrac{\dif\tau}{\dif\lambda}\,\tilde{f}(x,\boldsymbol{p};\lambda),
\end{equation}
which in addition turns out to be conformally invariant, by virtue of Eqns.~\eqref{eq:tau transformation} and \eqref{eq:f tilde transformation}. With this definition, \eqref{eq:average number dvolp partial} finally becomes
\begin{equation} \label{eq:average number final unhatted}
    \#[\Sigma] = -\int_\mathscr{S}\boldsymbol{\eta}_\mathscr{S}\int_{\masshyp^+}\dvolp\,f(x,\boldsymbol{p})\,\boldsymbol{p}(\boldsymbol{s}),
\end{equation}
which generalises the expression for the average number of trajectories found in the literature to the variable-mass case. Notice that \eqref{eq:average number final unhatted} can be formulated in terms of hatted-frame variables as
\begin{equation} \label{eq:average number final hatted}
    \#[\Sigma] = -\int_\mathscr{S}\hat{\boldsymbol{\eta}}_\mathscr{S}\int_{\masshyp^+}\hdvolp\,f(x,\boldsymbol{p})\,\boldsymbol{p}(\hat{\boldsymbol{s}}),
\end{equation}
where
\begin{equation}
    \hat{\boldsymbol{\eta}}_\mathscr{S}:=\iota_{\hat{\boldsymbol{s}}}\hat{\boldsymbol{\eta}}_\mathscr{M}=\Omega^3\,\boldsymbol{\eta}_\mathscr{S}.
\end{equation}

\subsection{Physical Observables}

The macroscopic physical properties of the gas are described by tensors in the spacetime manifold $\mathscr{M}$ defined as suitable integrals of the one-particle distribution function $f\xp$. In what follows, we will concentrate on two fundamental quantities, namely, the particle current density and the stress-energy tensor, from which a number of additional observables of interest can be derived.

\subsubsection{Particle Current Density}

In the unhatted frame, the \emph{particle current density} $\boldsymbol{J}$ is the one-form field on the spacetime manifold $\mathscr{M}$ whose action on an arbitrary spacetime vector $\boldsymbol{X}$ is defined as
\begin{equation} \label{eq:J unhatted}
    \boldsymbol{J}(\boldsymbol{X}):=\int_{\masshyp^+}\dvolp\,f(x,\boldsymbol{p})\,\boldsymbol{p}(\boldsymbol{X}).
\end{equation}
In a coordinate basis of $\mathscr{M}$, the components of $\boldsymbol{J}$ read
\begin{equation}
    J_\mu=\int_{\masshyp^+}\dvolp\,f(x,\boldsymbol{p})\,p_\mu.
\end{equation}
Separating $\boldsymbol{J}$ into its norm and normalised parts as
\begin{equation}
    \boldsymbol{J}\equiv n\,\boldsymbol{u}^\flat,
\end{equation}
one obtains two important observables: the \emph{unhatted average particle density} of the gas,
\begin{equation}
    n:=\sqrt{-\boldsymbol{g}^{-1}(\boldsymbol{J},\boldsymbol{J})},
\end{equation}
as well as the \emph{average particle four-velocity vector in the unhatted frame},
\begin{equation}\label{u-def}
    \boldsymbol{u}=\boldsymbol{g}^{-1}(\boldsymbol{u}^\flat,\,\cdot\,):=\dfrac{\boldsymbol{J}^\sharp}{n}=\dfrac{\boldsymbol{g}^{-1}(\boldsymbol{J},\,\cdot\,)}{n}.
\end{equation}
The corresponding hatted-frame counterpart of $\boldsymbol{J}$, $\hat{\boldsymbol{J}}$, is similarly given by
\begin{equation} \label{eq:J hatted}
    \hat{\boldsymbol{J}}(\boldsymbol{X}):=\int_{\masshyp^+}\hdvolp\,f(x,\boldsymbol{p})\,\boldsymbol{p}(\boldsymbol{X}),
\end{equation}
or, in components in a coordinate basis of $\mathscr{M}$,
\begin{equation}
    \hat{J}_\mu=\int_{\masshyp^+}\hdvolp\,f(x,\boldsymbol{p})\,p_\mu.
\end{equation}
Since $f\xp$ and $\boldsymbol{p}$ are frame-invariant, while $\dvolp$ transforms as stated in \eqref{eq:dvolp transformation}, the transformation rule for the particle current density is simply
\begin{equation} \label{eq:J transformation}
    \hat{\boldsymbol{J}}(\boldsymbol{X})=\Omega^{-2}\,\boldsymbol{J},
\end{equation}
from where we compute the transformation rules for the average particle density and four-velocity vector:
\begin{eqnarray}
    \hat{n} &:=& \sqrt{-\hat{\boldsymbol{g}}^{-1}(\hat{\boldsymbol{J}},\hat{\boldsymbol{J}})}~=~\Omega^{-3}\,n, \\
    \hat{\boldsymbol{u}} &:=& \dfrac{\hat{\boldsymbol{J}}^{\hat{\sharp}}}{\hat{n}}~=~\dfrac{\hat{\boldsymbol{g}}^{-1}(\hat{\boldsymbol{J}},\,\cdot\,)}{\hat{n}}~=~\Omega^{-1}\,\boldsymbol{u}, \label{eq:u transformation}
\end{eqnarray}
with
\begin{equation}
    \hat{\boldsymbol{J}}\equiv\hat{n}\,\hat{\boldsymbol{u}}^{\hat{\flat}}=\hat{n}\,\hat{\boldsymbol{g}}(\hat{\boldsymbol{u}},\,\cdot\,).
\end{equation}

Notice that, substituting definitions \eqref{eq:J unhatted} and \eqref{eq:J hatted} back in \eqref{eq:average number final unhatted} and \eqref{eq:average number final hatted}, respectively, one has
\begin{equation} \label{eq:average number final J}
    \#[\Sigma] = -\int_\mathscr{S}\boldsymbol{\eta}_\mathscr{S}\,\boldsymbol{J}(\boldsymbol{s}) = -\int_\mathscr{S}\hat{\boldsymbol{\eta}}_\mathscr{S}\,\hat{\boldsymbol{J}}(\hat{\boldsymbol{s}}).
\end{equation}
Compare this expression with the definitions of $\#[\Sigma]$ given as phase-space integrals involving the current densities $\boldsymbol{\mathcal{J}}$ and $\hat{\boldsymbol{\mathcal{J}}}$, provided back in \eqref{eq:average number unhatted} and \eqref{eq:average number hatted}, in that order. For the sake of convenience and reasons that will become apparent shortly, let us rewrite here the entire sequence of equalities:
\begin{eqnarray}
    \#[\Sigma]&=&-\int_\Sigma\tieta_\Sigma\,\tinduced(\boldsymbol{\mathcal{J}},\boldsymbol{\nu})~=~-\int_\mathscr{S}\boldsymbol{\eta}_\mathscr{S}\,\boldsymbol{J}(\boldsymbol{s})~~~~~~ \nonumber \\
    &=& -\int_\Sigma\htieta_\Sigma\,\htinduced(\hat{\boldsymbol{\mathcal{J}}},\hat{\boldsymbol{\nu}})~=~-\int_\mathscr{S}\hat{\boldsymbol{\eta}}_\mathscr{S}\,\hat{\boldsymbol{J}}(\hat{\boldsymbol{s}}).~~~~~~ \label{eq:chain of equalities J}
\end{eqnarray}
This chain of equalities allows us to find equations for $f\xp$ in each conformal frame, derived by requiring the conservation of the respective particle current density. As done in Refs.~\cite{Sarbach:2013fya,Sarbach:2013uba,Acuna-Cardenas:2021nkj}, we first assume that hypersurface $\mathscr{S}$ is the boundary of a compact spacetime region $\mathscr{U}$, i.e.~$\mathscr{S}=\partial\mathscr{U}$. Therefore, $\Sigma$ must be the boundary of a compact phase-space volume $\Xi$, i.e.~$\Sigma=\partial\Xi$, where
\begin{equation}
    \Xi=\{\xp\,|\,x\in\mathscr{U},\,\boldsymbol{p}\in\masshyp^+\}.
\end{equation}
Let us concentrate on a single frame now, for instance the unhatted one. Applying Gauss' theorem in the first line of \eqref{eq:chain of equalities J}, we have
\begin{equation} \label{eq:divs J equality unhatted}
    \int_\Xi\tieta_{\massshell^+}\,\diver_\mathcal{\massshell^+}\boldsymbol{\mathcal{J}}=\int_\mathscr{U}\boldsymbol{\eta}_{\mathscr{M}}\,\diver_\mathcal{\mathscr{M}}\boldsymbol{J},
\end{equation}
where $\diver_{\massshell^+}$ and $\diver_\mathscr{M}$ represent the divergence operators on the future mass shell and the base manifold, respectively. Because our goal is to find a conservation equation for $\boldsymbol{J}$, we seek a closed-form expression for its divergence, given in terms of the one-particle distribution function $f\xp$. To do so, the integral on the left-hand side of \eqref{eq:divs J equality unhatted} has to be transformed into an integral over $\mathscr{U}$ of a certain quantity. Then, this quantity will necessarily be equal to $\diver_\mathcal{\mathscr{M}}\boldsymbol{J}=g^{\mu\nu}\nabla_\mu J_\mu$. For this reason, it is necessary that we expand the integrand on the left-hand side of \eqref{eq:divs J equality unhatted}. By the implicit definition of the divergence operator,
\begin{equation}
    \tieta_{\massshell^+}\,\diver_\mathcal{\massshell^+}\boldsymbol{\mathcal{J}}=\pounds_{\boldsymbol{\mathcal{J}}}\tieta_{\massshell^+}.
\end{equation}
Recalling definitions \eqref{eq:unhatted phase space current J} and \eqref{eq:true distribution function}, as well as equation \eqref{eq:split N}, we can express $\boldsymbol{\mathcal{J}}$ as
\begin{equation}
    \boldsymbol{\mathcal{J}}\approx\dfrac{f\xp}{2m(x)}\boldsymbol{X}_\mathcal{C}.
\end{equation}
The properties of the Lie derivative then entail that
\begin{equation}
    \pounds_{\boldsymbol{\mathcal{J}}}\tieta_{\massshell^+}\approx\dfrac{f\xp}{2m(x)}\pounds_{\boldsymbol{\boldsymbol{X}_\mathcal{C}}}\tieta_{\massshell^+}+\dif\Left(\dfrac{f\xp}{2m(x)}\Right)\wedge\iota_{\boldsymbol{\boldsymbol{X}_\mathcal{C}}}\tieta_{\massshell^+}.
\end{equation}
We now simplify this expression by means of the formula for the contraction of a vector with a wedge product, and by noting that we are restricted to $\massshell^+$, so that
\begin{equation}
    \left[\dif\Left(\dfrac{f\xp}{2m(x)}\Right)\wedge\tieta_{\massshell^+}\right](\boldsymbol{X}_\mathcal{C},\boldsymbol{X}_1,\ldots,\boldsymbol{X}_7)=0,
\end{equation}
because the set of vectors $\{\boldsymbol{X}_\mathcal{C},\boldsymbol{X}_1,\ldots,\boldsymbol{X}_7\}$ (all eight of which are tangent to seven-dimensional $\mathscr{C}^+$) is linearly dependent. The result is:
\begin{eqnarray}
    \pounds_{\boldsymbol{\mathcal{J}}}\tieta_{\massshell^+} &\approx&  \dfrac{f\xp}{2}\Left[\dfrac{\pounds_{\boldsymbol{X}_\mathcal{C}}\tieta_{\massshell^+}}{m(x)}+\boldsymbol{X}_\mathcal{C}\Left(\dfrac{1}{m(x)}\Right)\,\tieta_{\massshell^+}\Right] \nonumber \\
    &+& \dfrac{\boldsymbol{X}_\mathcal{C}(f\xp)}{2m(x)}\,\tieta_{\massshell^+}. \label{eq:Lie derivative expansion}
\end{eqnarray}
Now, the Lie derivative remaining on the right-hand side of \eqref{eq:Lie derivative expansion} can be computed as explained in Appendix \ref{appendix:Lie derivative}, the results therein holding for any arbitrary Hamiltonian system. To be more precise, applying generic equation \eqref{eq:Lie derivative volume of Hamiltonian surface} to our specific case, we have 
\begin{equation}
    \pounds_{\boldsymbol{X}_\mathcal{C}}\tieta_{\massshell^+}\approx\boldsymbol{X}_\mathcal{C}(\ln{m(x)})\,\tieta_{\massshell^+},
\end{equation}
where we have taken into account equation \eqref{eq:unhatted norm of normal vector}. Thus, since
\begin{equation}
    \boldsymbol{X}_\mathcal{C}\Left(\dfrac{1}{m(x)}\Right)=-\dfrac{\boldsymbol{X}_\mathcal{C}(\ln{m(x)})}{m(x)},
\end{equation}
we see that the first line in \eqref{eq:Lie derivative expansion} is actually vanishing, so we are left with
\begin{equation}
    \pounds_{\boldsymbol{\mathcal{J}}}\tieta_{\massshell^+}\approx\dfrac{\boldsymbol{X}_\mathcal{C}(f\xp)}{2}\,\boldsymbol{\eta}_\mathscr{M}\wedge\dvolp,
\end{equation}
where we have made use of decomposition \eqref{eq:unhatted wedge decomposition of constraint volume} as well as definition \eqref{eq:unhatted dvolp}. As such,
\begin{equation}
    \int_\Xi\tieta_{\massshell^+}\,\diver_\mathcal{\massshell^+}\boldsymbol{\mathcal{J}}\approx\int_\mathscr{U}\boldsymbol{\eta}_{\mathscr{M}}\int_{\masshyp^+}\dvolp\,\dfrac{\boldsymbol{X}_\mathcal{C}(f\xp)}{2}.
\end{equation}
Substituting this result back in \eqref{eq:divs J equality unhatted}, we find, at last, the desired \emph{(unhatted) conservation equation for $\boldsymbol{J}$}, namely
\begin{equation} \label{eq:unhatted div_M J}
    \diver_\mathscr{M}\boldsymbol{J}\approx\int_{\masshyp^+}\dvolp\,\dfrac{\boldsymbol{X}_\mathcal{C}(f\xp)}{2}.
\end{equation}
Therefore, the particle current density $\boldsymbol{J}$ is conserved provided that the \emph{unhatted Liouville} (or \emph{collisionless Boltzmann}) \emph{equation},
\begin{equation} \label{eq:unhatted collisionless boltzmann}
    \dfrac{\boldsymbol{X}_\mathcal{C}(f\xp)}{2}\approx 0,
\end{equation}
is satisfied.\footnote{
    We stress that we are deliberately keeping the prefactor of $1/2$ in \eqref{eq:unhatted collisionless boltzmann} in light of expression \eqref{eq:unhatted X_C m(x)} for $\boldsymbol{X}_\mathcal{C}$. We shall do the same in its hatted counterpart, for similar reasons.
} Setting $m(x)=\const$, one recovers the standard collisionless Boltzmann equation found in the literature \cite{Sarbach:2013fya,Sarbach:2013uba,Acuna-Cardenas:2021nkj}, where a parameter affinely-related to (unhatted) proper time $\tau$ was employed.\footnote{
    Note that, if one sets $\dif\lambda=\dif\tau$, then $\boldsymbol{L}\approx\boldsymbol{X}_\mathcal{C}/(2m(x))$, cf.~\eqref{eq:unhatted Liouville restriction} and \eqref{eq:split N}, and that the Liouville vector as defined in Refs.~\cite{Sarbach:2013fya,Sarbach:2013uba,Acuna-Cardenas:2021nkj} differs from ours by a factor of the mass.
} However, by construction, our result is gauge-invariant regardless of whether $m(x)$ is constant or not.

Performing an entirely analogous computation in the hatted frame, one obtains the corresponding conservation equation for $\hat{\boldsymbol{J}}$,
\begin{equation} \label{eq:hatted div_M J}
    \hdiver_\mathscr{M}\hat{\boldsymbol{J}}\approx\int_{\masshyp^+}\hdvolp\,\dfrac{\boldsymbol{X}_{\hat{\mathcal{C}}}(f\xp)}{2},
\end{equation}
leading to the \emph{hatted Liouville} (or \emph{hatted collisionless Boltzmann}) \emph{equation},
\begin{equation}
    \dfrac{\boldsymbol{X}_{\hat{\mathcal{C}}}(f\xp)}{2}\approx 0,
\end{equation}
in cases where the hatted particle density current is to be conserved. The transformation rule for the current's divergence can be immediately obtained from \eqref{eq:X_C transformation}, \eqref{eq:dvolp transformation} and the right-hand side of \eqref{eq:hatted div_M J}, and reads
\begin{equation}
    \hdiver_\mathscr{M}\hat{\boldsymbol{J}}\approx\Omega^{-4}\,\diver_\mathscr{M}\boldsymbol{J}.
\end{equation}
As a crucial consistency check, we verify that the same transformation rule can be derived from the left-hand side of \eqref{eq:hatted div_M J} instead
\begin{equation}
    \hdiver_\mathscr{M}\hat{\boldsymbol{J}}=\hat{g}^{\mu\nu}\hat{\nabla}_\mu\hat{J}_\nu=\Omega^{-4}\,g^{\mu\nu}\nabla_\mu J_\nu=\Omega^{-4}\,\diver_\mathscr{M}\boldsymbol{J},
\end{equation}
where we have also used \eqref{eq:connection transformation}, \eqref{eq:disformation} and \eqref{eq:J transformation}. Notice that this is a strong equality.

\subsubsection{Stress-Energy Tensor}

In addition to the particle density current, another fundamental physical quantity characterising the gas of variable-mass particles is its \emph{stress-energy tensor}, which plays an instrumental role as a source of the gravitational field in GR and modified gravity theories. Starting in the unhatted frame, we define the macroscopic \emph{unhatted stress-energy tensor} $\boldsymbol{\stress}$ from microphysics through:
\begin{equation} \label{eq:stress-energy unhatted}
    \boldsymbol{\stress}(\boldsymbol{X}_1,\boldsymbol{X}_2):=\int_{\masshyp^+}\dvolp\,f(x,\boldsymbol{p})\,\boldsymbol{p}(\boldsymbol{X}_1)\,\boldsymbol{p}(\boldsymbol{X}_2),
\end{equation}
where $\boldsymbol{X}_1$ and $\boldsymbol{X}_2$ are arbitrary vector fields on spacetime. Consequently, the components of $\boldsymbol{\stress}$ in a coordinate basis of $\mathscr{M}$ are
\begin{equation}
    \stress_{\mu\nu}=\int_{\masshyp^+}\dvolp\,f(x,\boldsymbol{p})\,p_\mu\,p_\nu.
\end{equation}
Similarly, from the point of view of the hatted frame, the \emph{hatted stress-energy tensor} $\hat{\boldsymbol{\stress}}$ is given by
\begin{equation} \label{eq:stress-energy hatted}
    \hat{\boldsymbol{\stress}}(\boldsymbol{X}_1,\boldsymbol{X}_2):=\int_{\masshyp^+}\hdvolp\,f(x,\boldsymbol{p})\,\boldsymbol{p}(\boldsymbol{X}_1)\,\boldsymbol{p}(\boldsymbol{X}_2),
\end{equation}
or, equivalently,
\begin{equation}
    \hat{\stress}_{\mu\nu}=\int_{\masshyp^+}\hdvolp\,f(x,\boldsymbol{p})\,p_\mu\,p_\nu.
\end{equation}
As happened in the case of the particle density current, the conformal invariance of $f\xp$ and $\boldsymbol{p}$, alongside equation \eqref{eq:dvolp transformation}, indicate that
\begin{equation} \label{eq:stress-energy transformation}
    \hat{\boldsymbol{\stress}}=\Omega^{-2}\,\boldsymbol{\stress},
\end{equation}
which coincides with the transformation rule as obtained using the variational definition of the stress-energy tensor starting from some matter action.\footnote{
    Namely, if $S_\mathrm{m}$ denotes the matter action, it can be readily shown that
    \begin{equation*}
        \hat{\stress}_{\mu\nu}:=-\dfrac{2}{\sqrt{-\hat{g}}}\dfrac{\delta S_{\mathrm{m}}}{\delta\hat{g}^{\mu\nu}}=-\dfrac{2\Omega^{-2}}{\sqrt{-g}}\dfrac{\delta S_{\mathrm{m}}}{\delta g^{\mu\nu}}=:\Omega^{-2}\, \stress_{\mu\nu}.
    \end{equation*}
}

As done, for example, in Ref.~\cite{Acuna-Cardenas:2021nkj}, one may readily establish a conservation equation for the stress-energy tensor starting with the conservation equation for the particle current density. Considering the unhhated frame first, it is not difficult to realise that such an equation arises from replacing $\boldsymbol{J}$ by $\boldsymbol{\stress}(\,\cdot\,,\boldsymbol{X})$ on the left-hand side of \eqref{eq:unhatted div_M J} and $f\xp$ by $f\xp\,\boldsymbol{p}(\boldsymbol{X})$ on the right-hand side thereof, where $\boldsymbol{X}$ is an arbitrary vector on spacetime:\footnote{
    This result stems from the fact that it is possible to recover expression \eqref{eq:stress-energy unhatted} for the unhatted stress-energy tensor by substituting $\boldsymbol{J}(\boldsymbol{X})\mapsto\boldsymbol{\stress}(\boldsymbol{X}_1,\boldsymbol{X}_2)$, $f\xp\mapsto f\xp\,\boldsymbol{p}(\boldsymbol{X}_1)$ and $\boldsymbol{X}\mapsto\boldsymbol{X}_2$ in \eqref{eq:J unhatted} for the current density.
}
\begin{equation} \label{eq:unhatted div_M stress-energy partial}
    \diver_\mathscr{M}\boldsymbol{\stress}(\,\cdot\,,\boldsymbol{X})=\int_{\masshyp^+}\dvolp\,\dfrac{\boldsymbol{X}_\mathcal{C}(f\xp\,\boldsymbol{p}(\boldsymbol{X}))}{2}.
\end{equation}
In a coordinate basis of $\mathscr{M}$, the left-hand side of \eqref{eq:unhatted div_M stress-energy partial} becomes
\begin{equation} \label{eq:unhatted div_M stress-energy partial 2 lhs}
    \diver_\mathscr{M}\boldsymbol{\stress}(\,\cdot\,,\boldsymbol{X})=g^{\mu\nu}\nabla_\mu\stress_{\nu\rho}X^\rho+g^{\mu\nu}\stress_{\nu\rho}\nabla_\mu X^\rho.
\end{equation}
Meanwhile, making use of Eqns.~\eqref{eq:unhatted X_C m(x)}--\eqref{eq:unhatted horizontal basis} for $\boldsymbol{X}_\mathcal{C}$, the right-hand side of \eqref{eq:unhatted div_M stress-energy partial} evaluates to
\begin{eqnarray}
    \diver_\mathscr{M}\boldsymbol{\stress}(\,\cdot\,,\boldsymbol{X}) &=& \int_{\masshyp^+}\dvolp\,\dfrac{\boldsymbol{X}_\mathcal{C}(f\xp)}{2}\,p_\rho X^\rho \nonumber \\
    &-& \int_{\masshyp^+}\dvolp\,f\xp\,m(x)\dfrac{\partial m(x)}{\partial x^\rho} X^\rho \nonumber \\
    &+& g^{\mu\nu}\stress_{\nu\rho}\nabla_\mu X^\rho, \label{eq:unhatted div_M stress-energy partial 2 rhs}
\end{eqnarray}
after some algebra, where in the last line we have used definition \eqref{eq:stress-energy unhatted}. Combining \eqref{eq:unhatted div_M stress-energy partial 2 lhs} and \eqref{eq:unhatted div_M stress-energy partial 2 rhs}, and taking into account that $\boldsymbol{X}$ is arbitrary, we find
\begin{eqnarray}
    g^{\mu\nu}\nabla_\mu\stress_{\nu\rho} &=& \int_{\masshyp^+}\dvolp\,\dfrac{\boldsymbol{X}_\mathcal{C}(f\xp)}{2}\,p_\rho \nonumber \\
    &-& \int_{\masshyp^+}\dvolp\,f\xp\,m(x)\dfrac{\partial m(x)}{\partial x^\rho}.~~~~~~~~
\end{eqnarray}
Imposing the mass-shell constraint and using definition \eqref{eq:stress-energy unhatted}, we rewrite the last term in the following way:
\begin{eqnarray}
    g^{\mu\nu}\nabla_\mu\stress_{\nu\rho} &\approx& \int_{\masshyp^+}\dvolp\,\dfrac{\boldsymbol{X}_\mathcal{C}(f\xp)}{2}\,p_\rho \nonumber \\
    &+& g^{\mu\nu}\stress_{\mu\nu}\dfrac{\partial\ln{m(x)}}{\partial x^\rho}.
\end{eqnarray}
This expression can be cast into index-free notation as:
\begin{eqnarray}
    \boldsymbol{\nabla}^\sharp\cdot\boldsymbol{\stress} &\approx& \int_{\masshyp^+}\dvolp\,\dfrac{\boldsymbol{X}_\mathcal{C}(f\xp)}{2}\,\boldsymbol{p} \nonumber \\
    &+& \tr_{\boldsymbol{g}}\boldsymbol{\stress}\,\dif\ln{m(x)},
\end{eqnarray}
with
\begin{equation}
    \boldsymbol{\nabla}^\sharp\cdot\boldsymbol{\stress}\equiv g^{\mu\nu}\nabla_\mu\stress_{\nu\rho}
\end{equation}
and
\begin{equation}
    \tr_{\boldsymbol{g}}\boldsymbol{\stress}\equiv g^{\mu\nu}\stress_{\mu\nu}.
\end{equation}

By mirroring the foregoing calculation in the hatted frame, one finds instead
\begin{eqnarray}
    \hat{\boldsymbol{\nabla}}^{\hat{\sharp}}\cdot\hat{\boldsymbol{\stress}} &\approx& \int_{\masshyp^+}\hdvolp\,\dfrac{\boldsymbol{X}_{\hat{\mathcal{C}}}(f\xp)}{2}\,\boldsymbol{p} \nonumber \\
    &+& \tr_{\hat{\boldsymbol{g}}}\hat{\boldsymbol{\stress}}\,\dif\ln{\hat{m}(x)}, \label{eq:hatted stress-energy conservation partial}
\end{eqnarray}
where 
\begin{equation}
    \hat{\boldsymbol{\nabla}}^{\hat{\sharp}}\cdot\hat{\boldsymbol{\stress}}\equiv\hat{g}^{\mu\nu}\hat{\nabla}_\mu\hat{\stress}_{\nu\rho}
\end{equation}
and
\begin{equation}
    \tr_{\hat{\boldsymbol{g}}}\hat{\boldsymbol{\stress}}\equiv \hat{g}^{\mu\nu}\hat{\stress}_{\mu\nu}.
\end{equation}
Using \eqref{eq:connection transformation}, \eqref{eq:disformation}, \eqref{eq:m(x) transformation}, \eqref{eq:X_C transformation}, \eqref{eq:dvolp transformation} and \eqref{eq:stress-energy transformation}, the right-hand side of \eqref{eq:hatted stress-energy conservation partial} transforms into
\begin{equation}
    \hat{\boldsymbol{\nabla}}^{\hat{\sharp}}\cdot\hat{\boldsymbol{\stress}}\approx\Omega^{-4}\,\Left(\boldsymbol{\nabla}^\sharp\cdot\boldsymbol{\stress}-\tr_{\boldsymbol{g}}\boldsymbol{\stress}\,\dif\ln{\Omega}\Right).
\end{equation}
This is consistent with the well-known transformation rule for the left-hand side of \eqref{eq:hatted stress-energy conservation partial} (see, for example, Ref.~\cite{Deruelle:2010ht}), which can be obtained (without resorting to microphysics) directly from \eqref{eq:connection transformation}, \eqref{eq:disformation} and \eqref{eq:stress-energy transformation}:
\begin{equation}
    \hat{\boldsymbol{\nabla}}^{\hat{\sharp}}\cdot\hat{\boldsymbol{\stress}}=\Omega^{-4}\,\Left(\boldsymbol{\nabla}^\sharp\cdot\boldsymbol{\stress}-\tr_{\boldsymbol{g}}\boldsymbol{\stress}\,\dif\ln{\Omega}\Right).
\end{equation}
As happened in the case of the particle current density, let us note that this is a strong equality.

\section{Conclusions and Outlook}

In the present work, we have derived the kinetic theory of a collisionless gas of variable-mass particles in arbitrary conformal frames from first principles. Because these particles can be thought of as carrying a scalar charge, the results we have obtained resemble those previously found for electrically charged particles. In addition, as a crucial consistency check, we have explicitly verified that all our findings reduce to the well-known case of a free particle in a fixed conformal frame when the particle's mass is taken to be constant.

In order to achieve the aforementioned results, we first showed that the one-particle action, Hamiltonian, mass shell, mass hyperboloid and equation of motion are all conformally invariant, as is the symplectic structure on the cotangent bundle. Then, we introduced the tilted connection maps and tilted horizontal subspaces as seen from both hatted and unhatted conformal frames, and discussed the relationship between hatted and unhatted horizontal phase-space vectors and tangents to the mass shell. On the basis of the splitting into tilted horizontal and vertical subspaces, we introduced the tilted Sasaki metric, which generalises the standard Sasaki metric defined for particles with constant mass in a fixed frame to the non-free scenario under consideration. Afterwards, the tilted Sasaki metric is employed to compute the vector normal to the mass shell and hyperboloid, as well as the induced metrics and volume forms therein, again in two arbitrary conformal frames. These volume forms then allowed us to define the phase-space integrals necessary for a statistical interpretation of the gas in terms of a one-particle distribution function, as is customary in relativistic kinetic theory. By requiring the number of world-lines crossing an arbitrary hypersurface to remain gauge- and frame-invariant, we show that the one-particle distribution function must also be reparametrisation invariant, and can always be defined to be conformally invariant. Finally, two macroscopic observables are obtained from the distribution function: the particle current density and the stress-energy tensor. We prove that their transformation rules under a change of frame, as obtained through the statistical interpretation, coincide with those derived from purely macroscopic considerations, without reference to the distribution function or phase-space integrals. Conservation of the particle density current allowed us to establish the corresponding equations for the one-particle distribution function in two conformal frames, which generalise the Liouville (or collisionless Boltzmann) equation.

The relevance of our findings resides in the fact that the interpretational subtlety arising from the transformation of masses under conformal rescalings had never been taken into consideration at the level of the mathematical foundations of relativistic kinetic theory. In particular, to the best of our knowledge, it had not previously been recognised that the statistical interpretations based on particles which are free in different frames are indeed inequivalent. In contrast, the framework presented in this article is suited to the study of any specific species of variable-mass particles, as seen from an arbitrary frame. Notably, the expressions we have obtained are frame-covariant and ready for use in a wide range of practical applications, particularly in modified gravity theories where a transformation to a specific conformal frame (such as the Einstein frame) offers significant computational advantages. This is because our results allow for a simple and consistent translation of the physical observables and of the equation for the one-particle distribution function between different frames.

A natural follow-up to this work is to allow for collisions between the particles. This would enable the study of local and global equilibrium states from the perspective of two distinct conformal frames in a mathematically well-founded and rigorous manner. As such, it would become possible to determine under which circumstances equilibrium occurs in both frames simultaneously; it has been conjectured that local equilibrium states in one frame correspond to local equilibrium states in another frame \cite{Faraoni:2023gqg,Casado-Turrion:2026qvs}, but a definitive proof is yet to be provided. We intend to carry out this investigation in the future. Moreover, a number of physical applications in accretion-disc modelling, stellar physics and cosmology can be expected, given the wide applicability of kinetic theory in realistic scenarios.

\section*{Acknowledgements}

A.~C.-T.~would like to thank Apostolos Tsampodimos for some very insightful comments and discussions. A.~C.-T.~and M.~K.~acknowledge that this project has received funding from the Research Council of Lithuania (LMTLT), agreement No.~S-PD-24-135. A.~C.-T.~is also partially supported by grant PID2022-137003NB-I00 funded by MCIN/AEI/10.13039/501100011033/ and by ERDF `A way of making Europe.'. P.~B.~F. acknowledges support from the FCT-Fundação para a Ciência e Tecnologia (FCT), I.P. fellowship SFRH/BD/151475/2021 with DOI identifier 10.54499/SFRH/BD/151475/2021 during the early stages of this work. J.~J.~T.~D. acknowledges the financial support provided by FCT-Fundação para a Ciência e Tecnologia (FCT), I.P., through the Strategic Funding UID/04650/2025 and UID/04564/2025 and national funds with DOI identifiers 10.54499/2023.11681.PEX, 10.54499/2024.00249.CERN funded by measure RE-C06-i06.m02-``Reinforcement of funding for International Partnerships in Science, Technology and Innovation'' of the Recovery and Resilience Plan–RRP, within the framework of the financing contract signed between the Recover Portugal Mission Structure (EMRP) and the Foundation for Science and Technology I.P. (FCT), as an intermediate beneficiary, as well as the advanced computing projects 2024.00249.CERN.F1.

\section*{Appendix}
\appendix

\section{Non-Conservation of the Volume Element on Hamiltonian Hypersurfaces} \label{appendix:Lie derivative}

Let $\mathcal{H}$ be a Hamiltonian function on $\mathscr{T}^*\mathscr{M}$, $\boldsymbol{X}_\mathcal{H}$ its associated Hamiltonian vector field and $\dif\mathcal{H}$ the one-form normal to
\begin{equation}
    \mathscr{H}:=\{\xp\in\mathscr{T}^*\mathscr{M}\,|\,\mathcal{H}=0\},
\end{equation}
which we shall give the name of \emph{Hamiltonian hypersurface associated to $\mathcal{H}$}. Suppose $\mathscr{T}^*\mathscr{M}$ is endowed with some metric tensor field $\boldsymbol{\mathfrak{g}}$, allowing us to compute the unit normal vector to $\mathscr{H}$ as
\begin{equation}
    \boldsymbol{N}_\mathscr{H}:=\dfrac{\boldsymbol{n}_\mathscr{H}}{||\boldsymbol{n}_\mathscr{H}||_{\boldsymbol{\mathfrak{g}}}},
\end{equation}
where
\begin{equation}
    \boldsymbol{n}_\mathscr{H}:=\boldsymbol{\mathbfrak{g}}^{-1}(\dif\mathcal{H},\,\cdot\,)
\end{equation}
and we are using the standard notation
\begin{equation} \label{eq:norm g n_H}
    ||\boldsymbol{n}_\mathscr{H}||_{\boldsymbol{\mathfrak{g}}}:=\Left|\boldsymbol{\mathfrak{g}}(\boldsymbol{n}_\mathscr{H},\boldsymbol{n}_\mathscr{H})\Right|^{1/2}=\dif\mathcal{H}(\boldsymbol{N});
\end{equation}
note that we are also assuming that $||\boldsymbol{n}_\mathscr{H}||_{\boldsymbol{\mathfrak{g}}}\neq 0$. The volume form on $\mathscr{H}$ is then given by
\begin{equation} \label{eq:volume on H}
    \boldsymbol{\eta}_\mathscr{H}:=\iota_{\boldsymbol{N}}\boldsymbol{\eta}_{\mathscr{T}^*\mathscr{M}},
\end{equation}
where $\boldsymbol{\eta}_{\mathscr{T}^*\mathscr{M}}$ is the natural volume form on phase space as previously defined in \eqref{eq:eta phase space omega}.

We set out to find a generic expression for $\pounds_{\boldsymbol{X}_\mathcal{H}}\boldsymbol{\eta}_\mathscr{H}$. To do so, we first note that
\begin{equation}
    \dif\mathcal{H}\wedge\boldsymbol{\eta}_{\mathscr{T}^*\mathscr{M}}=0,
\end{equation}
since $\dif\mathcal{H}$ is a one-form and $\boldsymbol{\eta}_{\mathscr{T}^*\mathscr{M}}$ is a top form. Then, we contract both sides of this equation with $\boldsymbol{N}$, and use \eqref{eq:norm g n_H} and \eqref{eq:volume on H}, in order to find
\begin{equation} \label{eq:phase space volume norm of n}
    ||\boldsymbol{n}_\mathscr{H}||_{\boldsymbol{\mathfrak{g}}}\,\boldsymbol{\eta}_{\mathscr{T}^*\mathscr{M}}-\dif\mathcal{H}\wedge\boldsymbol{\eta}_\mathscr{H}=0.
\end{equation}
Taking the Lie derivative along $\boldsymbol{X}_\mathcal{H}$ of the prececing expression, we have
\begin{eqnarray}
    0 &=& \pounds_{\boldsymbol{X}_\mathcal{H}}||\boldsymbol{n}_\mathscr{H}||_{\boldsymbol{\mathfrak{g}}}\,\boldsymbol{\eta}_{\mathscr{T}^*\mathscr{M}} + ||\boldsymbol{n}_\mathscr{H}||_{\boldsymbol{\mathfrak{g}}}\,\pounds_{\boldsymbol{X}_\mathcal{H}}\boldsymbol{\eta}_{\mathscr{T}^*\mathscr{M}} \nonumber \\
    &-& \pounds_{\boldsymbol{X}_\mathcal{H}}\dif\mathcal{H}\wedge\boldsymbol{\eta}_\mathscr{H}-\dif\mathcal{H}\wedge\pounds_{\boldsymbol{X}_\mathcal{H}}\boldsymbol{\eta}_\mathscr{H}.
\end{eqnarray}
The second summand on the first line vanishes by virtue of Liouville's theorem, which holds for every Hamiltonian vector field due to \eqref{eq:Lie derivative of omega} and \eqref{eq:eta phase space omega}. Meanwhile, since the Lie and exterior derivatives commute,
\begin{equation}
    \pounds_{\boldsymbol{X}_\mathcal{H}}\dif\mathcal{H}=\dif(\pounds_{\boldsymbol{X}_\mathcal{H}}\mathcal{H})=\dif(\boldsymbol{X}_\mathcal{H}(\mathcal{H}))=0,
\end{equation}
cf.~\eqref{eq:generic Hamiltonian vector tangency}. In consequence, we are left with
\begin{eqnarray}
    0 &=& \pounds_{\boldsymbol{X}_\mathcal{H}}||\boldsymbol{n}_\mathscr{H}||_{\boldsymbol{\mathfrak{g}}}\,\boldsymbol{\eta}_{\mathscr{T}^*\mathscr{M}}-\dif\mathcal{H}\wedge\pounds_{\boldsymbol{X}_\mathcal{H}}\boldsymbol{\eta}_\mathscr{H} \nonumber \\
    &=& \boldsymbol{X}_\mathcal{H}(||\boldsymbol{n}_\mathscr{H}||_{\boldsymbol{\mathfrak{g}}})\,\boldsymbol{\eta}_{\mathscr{T}^*\mathscr{M}}-\dif\mathcal{H}\wedge\pounds_{\boldsymbol{X}_\mathcal{H}}\boldsymbol{\eta}_\mathscr{H}.~~~~
\end{eqnarray}
If we substitute \eqref{eq:phase space volume norm of n}, we find
\begin{eqnarray}
    0 &=& \dif\mathcal{H}\wedge\Left[\boldsymbol{X}_\mathcal{H}(\ln{||\boldsymbol{n}_\mathscr{H}||_{\boldsymbol{\mathfrak{g}}}})\,\boldsymbol{\eta}_\mathscr{H}-\pounds_{\boldsymbol{X}_\mathcal{H}}\boldsymbol{\eta}_\mathscr{H}\Right] \nonumber \\
    &\equiv& \dif\mathcal{H}\wedge\boldsymbol{\zeta}.
\end{eqnarray}
Contracting the preceding expression with $\boldsymbol{N}$ and seven vectors $\{\boldsymbol{X}_\mathcal{C},\boldsymbol{X}_1,\ldots,\boldsymbol{X}_7\}$ tangent to $\mathscr{H}$ (but otherwise arbitrary), we have:
\begin{eqnarray}
    0 &=& \iota_{\boldsymbol{N}}(\iota_{\boldsymbol{X}_1}(\ldots(\iota_{\boldsymbol{X}_7}(\dif\mathcal{H}\wedge\boldsymbol{\zeta})\ldots)) \nonumber \\
    &=& -\iota_{\boldsymbol{N}}\dif\mathcal{H}\,\boldsymbol{\zeta}(\boldsymbol{X}_1,\ldots,\boldsymbol{X}_7) \nonumber \\
    &=& -||\boldsymbol{n}_\mathscr{H}||_{\boldsymbol{\mathfrak{g}}}\,\boldsymbol{\zeta}(\boldsymbol{X}_1,\ldots,\boldsymbol{X}_7), \label{eq:contraction wedge N X_i}
\end{eqnarray}
where we have used \eqref{eq:norm g n_H} and taken into account that $\dif\mathcal{H}(\boldsymbol{X}_i)=0$. Due to the fact that, by assumption, $||\boldsymbol{n}_\mathscr{H}||_{\boldsymbol{\mathfrak{g}}}\neq 0$, we learn from \eqref{eq:contraction wedge N X_i} that
\begin{equation}
    \boldsymbol{\zeta}(\boldsymbol{X}_1,\ldots,\boldsymbol{X}_7)=0.
\end{equation}
As such, we finally conclude that, when pulling back to the Hamiltonian surface $\mathscr{H}$,
\begin{equation} \label{eq:Lie derivative volume of Hamiltonian surface}
    \pounds_{\boldsymbol{X}_\mathcal{H}}\boldsymbol{\eta}_\mathscr{H}=\boldsymbol{X}_\mathcal{H}(\ln{||\boldsymbol{n}_\mathscr{H}||_{\boldsymbol{\mathfrak{g}}}})\,\boldsymbol{\eta}_\mathscr{H}.
\end{equation}

\newpage

\bibliography{bibliography}

@PREAMBLE{
 "\providecommand{\noopsort}[1]{}" 
 # "\providecommand{\singleletter}[1]{#1}%" 
}

@article{Acuna-Cardenas:2021nkj,
    author = "Acu{\~n}a-C{\'a}rdenas, Rub{\'e}n O. and Gabarrete, Carlos and Sarbach, Olivier",
    title = "{An introduction to the relativistic kinetic theory on curved spacetimes}",
    eprint = "2106.09235",
    archivePrefix = "arXiv",
    primaryClass = "gr-qc",
    doi = "10.1007/s10714-022-02908-5",
    journal = "Gen. Rel. Grav.",
    volume = "54",
    number = "3",
    pages = "23",
    year = "2022"
}

@book{Carroll:2004st,
    author = "Carroll, Sean M.",
    title = "{Spacetime and Geometry}: {An Introduction to General Relativity}",
    doi = "10.1017/9781108770385",
    isbn = "978-0-8053-8732-2, 978-1-108-48839-6, 978-1-108-77555-7",
    publisher = "Cambridge University Press",
    month = "7",
    year = "2019"
}

@book{Misner:1973prb,
    author = "Misner, Charles W. and Thorne, K. S. and Wheeler, J. A.",
    title = "{Gravitation}",
    isbn = "978-0-7167-0344-0, 978-0-691-17779-3",
    publisher = "W. H. Freeman",
    address = "San Francisco",
    year = "1973"
}

@article{Chavanis:2019ouz,
    author = "Chavanis, Pierre-Henri",
    title = "{Statistical mechanics of self-gravitating systems in general relativity: I. The quantum Fermi gas}",
    eprint = "1908.10806",
    archivePrefix = "arXiv",
    primaryClass = "gr-qc",
    doi = "10.1140/epjp/s13360-020-00268-0",
    journal = "Eur. Phys. J. Plus",
    volume = "135",
    number = "3",
    pages = "290",
    year = "2020"
}

@article{Chavanis:2019mkk,
    author = "Chavanis, Pierre-Henri",
    title = "{Statistical mechanics of self-gravitating systems in general relativity: II. The classical Boltzmann gas}",
    eprint = "1908.10817",
    archivePrefix = "arXiv",
    primaryClass = "gr-qc",
    doi = "10.1140/epjp/s13360-020-00291-1",
    journal = "Eur. Phys. J. Plus",
    volume = "135",
    number = "3",
    pages = "310",
    year = "2020"
}

@article{Sarbach:2013uba,
    author = "Sarbach, Olivier and Zannias, Thomas",
    title = "{The geometry of the tangent bundle and the relativistic kinetic theory of gases}",
    eprint = "1309.2036",
    archivePrefix = "arXiv",
    primaryClass = "gr-qc",
    doi = "10.1088/0264-9381/31/8/085013",
    journal = "Class. Quant. Grav.",
    volume = "31",
    pages = "085013",
    year = "2014"
}

@article{Anguige:1999gh,
    author = "Anguige, K. and Tod, K. P.",
    title = "{Isotropic Cosmological Singularities: II. The Einstein{\textendash}Vlasov System}",
    eprint = "gr-qc/9903009",
    archivePrefix = "arXiv",
    doi = "10.1006/aphy.1999.5947",
    journal = "Annals Phys.",
    volume = "276",
    pages = "294--320",
    year = "1999"
}

@article{Joudioux:2019ecs,
    author = "Joudioux, J{\'e}r{\'e}mie and Thaller, Maximilian and Kroon, Juan A. Valiente",
    title = "{The conformal Einstein field equations with massless Vlasov matter}",
    eprint = "1903.12251",
    archivePrefix = "arXiv",
    primaryClass = "gr-qc",
    doi = "10.5802/aif.3414",
    journal = "Annales Inst. Fourier",
    volume = "71",
    number = "2",
    pages = "799--842",
    year = "2021"
}

@article{Lee:2023cbn,
    author = "Lee, Ho and Nungesser, Ernesto and Stalker, John and Tod, Paul",
    title = "{Well-posedness of anisotropic and homogeneous solutions to the Einstein-Boltzmann system with a conformal gauge singularity}",
    eprint = "2305.15902",
    archivePrefix = "arXiv",
    primaryClass = "gr-qc",
    doi = "10.1016/j.jde.2024.08.011",
    journal = "J. Diff. Eq.",
    volume = "411",
    pages = "640--738",
    year = "2024"
}

@article{Casado-Turrion:2026qvs,
    author = "Casado-Turri{\'o}n, Adri{\'a}n and Ferraz, Paulo B. and Kar{\v{c}}iauskas, Mindaugas and Terente D{\'\i}az, Jos{\'e} Jaime",
    title = "{Non-Minimally Coupled Warm Inflation in the Defining Frame}",
    eprint = "2608.20293",
    archivePrefix = "arXiv",
    primaryClass = "astro-ph.CO",
    month = "8",
    year = "2026"
}

@article{Dabrowski:2008kx,
    author = "Dabrowski, Mariusz P. and Garecki, Janusz and Blaschke, David B.",
    title = "{Conformal transformations and conformal invariance in gravitation}",
    eprint = "0806.2683",
    archivePrefix = "arXiv",
    primaryClass = "gr-qc",
    doi = "10.1002/andp.200810331",
    journal = "Annalen Phys.",
    volume = "18",
    pages = "13--32",
    year = "2009"
}

@article{BeltranJimenez:2017tkd,
    author = "Beltr{\'a}n Jim{\'e}nez, Jose and Heisenberg, Lavinia and Koivisto, Tomi",
    title = "{Coincident General Relativity}",
    eprint = "1710.03116",
    archivePrefix = "arXiv",
    primaryClass = "gr-qc",
    reportNumber = "NORDITA-2017-100, IFT-UAM/CSIC-17-093, ITS-ETH-2017-10",
    doi = "10.1103/PhysRevD.98.044048",
    journal = "Phys. Rev. D",
    volume = "98",
    number = "4",
    pages = "044048",
    year = "2018"
}

@article{BeltranJimenez:2015pnp,
    author = "Beltran Jimenez, Jose and Koivisto, Tomi S.",
    title = "{Spacetimes with vector distortion: Inflation from generalised Weyl geometry}",
    eprint = "1509.02476",
    archivePrefix = "arXiv",
    primaryClass = "gr-qc",
    reportNumber = "NORDITA-2015-100",
    doi = "10.1016/j.physletb.2016.03.047",
    journal = "Phys. Lett. B",
    volume = "756",
    pages = "400--404",
    year = "2016"
}

@article{Dicke:1961gz,
    author = "Dicke, R. H.",
    title = "{Mach's principle and invariance under transformation of units}",
    doi = "10.1103/PhysRev.125.2163",
    journal = "Phys. Rev.",
    volume = "125",
    pages = "2163--2167",
    year = "1962"
}

@article{Faraoni:2006fx,
    author = "Faraoni, Valerio and Nadeau, Shahn",
    title = "{The (pseudo)issue of the conformal frame revisited}",
    eprint = "gr-qc/0612075",
    archivePrefix = "arXiv",
    doi = "10.1103/PhysRevD.75.023501",
    journal = "Phys. Rev. D",
    volume = "75",
    pages = "023501",
    year = "2007"
}

@inproceedings{Brans:1996tp,
    author = "Brans, Carl H.",
    title = "{Gravity and the tenacious scalar field}",
    booktitle = "{Symposium to honor Engelbert Schucking}",
    eprint = "gr-qc/9705069",
    archivePrefix = "arXiv",
    month = "12",
    year = "1996"
}

@article{Brans:1962wsm,
    author = "Brans, Carl H.",
    title = "{Mach's Principle and the Locally Measured Gravitational Constant in General Relativity}",
    doi = "10.1103/PhysRev.125.388",
    journal = "Phys. Rev.",
    volume = "125",
    number = "1",
    pages = "388",
    year = "1962"
}

@article{Sarbach:2013fya,
    author = "Sarbach, Olivier and Zannias, Thomas",
    editor = "Ure{\~n}a-L{\'o}pez, Luis A. and Becerril-B{\'a}rcenas, Ricardo and Linares-Romero, Rom{\'a}n",
    title = "{Relativistic Kinetic Theory: An Introduction}",
    eprint = "1303.2899",
    archivePrefix = "arXiv",
    primaryClass = "gr-qc",
    doi = "10.1063/1.4817035",
    journal = "AIP Conf. Proc.",
    volume = "1548",
    number = "1",
    pages = "134--155",
    year = "2013"
}

@article{Sarbach:2013hna,
    author = "Sarbach, Olivier and Zannias, Thomas",
    editor = "Mac{\'\i}as, Alfredo and Maceda, Marco",
    title = "{Tangent bundle formulation of a charged gas}",
    eprint = "1311.3532",
    archivePrefix = "arXiv",
    primaryClass = "gr-qc",
    doi = "10.1063/1.4861955",
    journal = "AIP Conf. Proc.",
    volume = "1577",
    number = "1",
    pages = "192--207",
    year = "2015"
}

@article{Baier:2007ix,
    author = "Baier, Rudolf and Romatschke, Paul and Son, Dam Thanh and Starinets, Andrei O. and Stephanov, Mikhail A.",
    title = "{Relativistic viscous hydrodynamics, conformal invariance, and holography}",
    eprint = "0712.2451",
    archivePrefix = "arXiv",
    primaryClass = "hep-th",
    reportNumber = "BI-TP-2007-29, INT-PUB-07-45, SHEP-07-47",
    doi = "10.1088/1126-6708/2008/04/100",
    journal = "JHEP",
    volume = "04",
    pages = "100",
    year = "2008"
}

@article{Denicol:2014tha,
    author = "Denicol, Gabriel S. and Heinz, Ulrich W. and Martinez, Mauricio and Noronha, Jorge and Strickland, Michael",
    title = "{Studying the validity of relativistic hydrodynamics with a new exact solution of the Boltzmann equation}",
    eprint = "1408.7048",
    archivePrefix = "arXiv",
    primaryClass = "hep-ph",
    doi = "10.1103/PhysRevD.90.125026",
    journal = "Phys. Rev. D",
    volume = "90",
    number = "12",
    pages = "125026",
    year = "2014"
}

@article{Quiros:2012rnn,
    author = "Quiros, Israel and Garcia-Salcedo, Ricardo and Madriz Aguilar, Jose Edgar and Matos, Tonatiuh",
    title = "{The conformal transformation's controversy: what are we missing?}",
    eprint = "1108.5857",
    archivePrefix = "arXiv",
    primaryClass = "gr-qc",
    doi = "10.1007/s10714-012-1484-7",
    journal = "Gen. Rel. Grav.",
    volume = "45",
    pages = "489--518",
    year = "2013"
}

@article{Catena:2006bd,
    author = "Catena, Riccardo and Pietroni, Massimo and Scarabello, Luca",
    title = "{Einstein and Jordan reconciled: a frame-invariant approach to scalar-tensor cosmology}",
    eprint = "astro-ph/0604492",
    archivePrefix = "arXiv",
    reportNumber = "DESY-06-076",
    doi = "10.1103/PhysRevD.76.084039",
    journal = "Phys. Rev. D",
    volume = "76",
    pages = "084039",
    year = "2007"
}

@article{Postma:2014vaa,
    author = "Postma, Marieke and Volponi, Marco",
    title = "{Equivalence of the Einstein and Jordan frames}",
    eprint = "1407.6874",
    archivePrefix = "arXiv",
    primaryClass = "astro-ph.CO",
    doi = "10.1103/PhysRevD.90.103516",
    journal = "Phys. Rev. D",
    volume = "90",
    number = "10",
    pages = "103516",
    year = "2014"
}

@book{Fujii:2003pa,
    author = "Fujii, Y. and Maeda, K.",
    title = "{The scalar-tensor theory of gravitation}",
    doi = "10.1017/CBO9780511535093",
    isbn = "978-0-521-03752-5, 978-0-521-81159-0, 978-0-511-02988-2",
    publisher = "Cambridge University Press",
    series = "Cambridge Monographs on Mathematical Physics",
    month = "7",
    year = "2007"
}

@article{Jarv:2015kga,
    author = {J{\"a}rv, Laur and Kuusk, Piret and Saal, Margus and Vilson, Ott},
    title = "{Transformation properties and general relativity regime in scalar{\textendash}tensor theories}",
    eprint = "1504.02686",
    archivePrefix = "arXiv",
    primaryClass = "gr-qc",
    doi = "10.1088/0264-9381/32/23/235013",
    journal = "Class. Quant. Grav.",
    volume = "32",
    pages = "235013",
    year = "2015"
}

@article{Steinwachs:2011zs,
    author = "Steinwachs, Christian F. and Kamenshchik, Alexander Yu.",
    title = "{One-loop divergences for gravity non-minimally coupled to a multiplet of scalar fields: calculation in the Jordan frame. I. The main results}",
    eprint = "1101.5047",
    archivePrefix = "arXiv",
    primaryClass = "gr-qc",
    doi = "10.1103/PhysRevD.84.024026",
    journal = "Phys. Rev. D",
    volume = "84",
    pages = "024026",
    year = "2011"
}

@article{Kamenshchik:2014waa,
    author = "Kamenshchik, Alexander Yu. and Steinwachs, Christian F.",
    title = "{Question of quantum equivalence between Jordan frame and Einstein frame}",
    eprint = "1408.5769",
    archivePrefix = "arXiv",
    primaryClass = "gr-qc",
    reportNumber = "FR-PHENO-2014-011",
    doi = "10.1103/PhysRevD.91.084033",
    journal = "Phys. Rev. D",
    volume = "91",
    number = "8",
    pages = "084033",
    year = "2015"
}

@article{Pandey:2016unk,
    author = "Pandey, Sachin and Banerjee, Narayan",
    title = "{Equivalence of Jordan and Einstein frames at the quantum level}",
    eprint = "1610.00584",
    archivePrefix = "arXiv",
    primaryClass = "gr-qc",
    doi = "10.1140/epjp/i2017-11385-0",
    journal = "Eur. Phys. J. Plus",
    volume = "132",
    number = "3",
    pages = "107",
    year = "2017"
}

@article{Shapiro:1995kt,
    author = "Shapiro, Ilya L. and Takata, Hiroyuki",
    title = "{Conformal transformation in gravity}",
    eprint = "hep-th/9504162",
    archivePrefix = "arXiv",
    reportNumber = "DFTUZ-95-14",
    doi = "10.1016/0370-2693(95)01123-8",
    journal = "Phys. Lett. B",
    volume = "361",
    pages = "31--37",
    year = "1995"
}

@article{Maeda:1988ab,
    author = "Maeda, Kei-ichi",
    title = "{Towards the Einstein-Hilbert Action via Conformal Transformation}",
    reportNumber = "UTAP-75/88",
    doi = "10.1103/PhysRevD.39.3159",
    journal = "Phys. Rev. D",
    volume = "39",
    pages = "3159",
    year = "1989"
}

@article{Deruelle:2010ht,
    author = "Deruelle, Nathalie and Sasaki, Misao",
    editor = "Odintsov, Sergey D. and S{\'a}ez-G{\'o}mez, Diego and Xamb{\'o}-Descamps, Sebastian",
    title = "{Conformal equivalence in classical gravity: the example of 'Veiled' General Relativity}",
    eprint = "1007.3563",
    archivePrefix = "arXiv",
    primaryClass = "gr-qc",
    reportNumber = "YITP-10-61",
    doi = "10.1007/978-3-642-19760-4_23",
    journal = "Springer Proc. Phys.",
    volume = "137",
    pages = "247--260",
    year = "2011"
}

@article{Catena:2006gk,
    author = "Catena, Riccardo and Pietroni, Massimo and Scarabello, Luca",
    editor = "Sola, Joan",
    title = "{Local transformations of units in scalar-tensor cosmology}",
    eprint = "hep-th/0610292",
    archivePrefix = "arXiv",
    reportNumber = "DESY-06-195",
    doi = "10.1088/1751-8113/40/25/S34",
    journal = "J. Phys. A",
    volume = "40",
    pages = "6883--6888",
    year = "2007"
}

@article{Domenech:2016yxd,
    author = "Dom{\`e}nech, Guillem and Sasaki, Misao",
    editor = "Chen, Pisin",
    title = "{Conformal frames in cosmology}",
    eprint = "1602.06332",
    archivePrefix = "arXiv",
    primaryClass = "gr-qc",
    reportNumber = "YITP-16-11",
    doi = "10.1142/S0218271816450061",
    journal = "Int. J. Mod. Phys. D",
    volume = "25",
    number = "13",
    pages = "1645006",
    year = "2016"
}

@article{Faraoni:1998qx,
    author = "Faraoni, Valerio and Gunzig, Edgard and Nardone, Pasquale",
    title = "{Conformal transformations in classical gravitational theories and in cosmology}",
    eprint = "gr-qc/9811047",
    archivePrefix = "arXiv",
    reportNumber = "IUCAA-24-98",
    journal = "Fund. Cosmic Phys.",
    volume = "20",
    pages = "121",
    year = "1999"
}

@article{Faraoni:2023gqg,
    author = "Faraoni, Valerio and Vanderwee, Robert",
    title = "{Tolman-Ehrenfest{\textquoteright}s criterion of thermal equilibrium extended to conformally static spacetimes}",
    eprint = "2301.09021",
    archivePrefix = "arXiv",
    primaryClass = "gr-qc",
    doi = "10.1103/PhysRevD.107.064072",
    journal = "Phys. Rev. D",
    volume = "107",
    number = "6",
    pages = "064072",
    year = "2023"
}

@article{Wojnar:2022dvo,
    author = "Wojnar, Aneta",
    title = "{Fermi gas and modified gravity}",
    eprint = "2208.04023",
    archivePrefix = "arXiv",
    primaryClass = "gr-qc",
    doi = "10.1103/PhysRevD.107.044025",
    journal = "Phys. Rev. D",
    volume = "107",
    number = "4",
    pages = "044025",
    year = "2023"
}

@article{Dirac:1950pj,
    author = "Dirac, Paul A. M.",
    title = "{Generalized Hamiltonian dynamics}",
    doi = "10.4153/CJM-1950-012-1",
    journal = "Can. J. Math.",
    volume = "2",
    pages = "129--148",
    year = "1950"
}

@article{Anderson:1951ta,
    author = "Anderson, James L. and Bergmann, Peter G.",
    title = "{Constraints in covariant field theories}",
    doi = "10.1103/PhysRev.83.1018",
    journal = "Phys. Rev.",
    volume = "83",
    pages = "1018--1025",
    year = "1951"
}

@article{Sasaki1958,
    author    = "Sasaki, Shigeo",
    title     = "{On the differential geometry of tangent bundles of Riemannian manifolds}",
    doi       = "10.2748/tmj/1178244668",
    journal   = "Tohoku Math. J.",
    volume    = "10",
    number    = "3",
    pages     = "338--354",
    year      = "1958",
}

@phdthesis{ThesisRioseco,
    author = "Rioseco, Paola",
    title = "{Relativistic Kinetic Theory with Applications in Astrophysics}",
    school = "Universidad Michoacana de San Nicol\'{a}s de Hidalgo",
    year = "2019"
}

@inbook{Ehlers1971,
  author    = {J{\"u}rgen Ehlers},
  title     = {General Relativity and Kinetic Theory},
  booktitle = {General Relativity and Cosmology},
  editor    = {R. K. Sachs},
  series    = {Proceedings of the International School of Physics "Enrico Fermi", Course XLVII},
  pages     = {1--70},
  publisher = {Academic Press},
  address   = {New York},
  year      = {1971}
}

@inproceedings{Israel1972,
       author = "Israel, W.",
        title = "The relativistic Boltzmann equation",
    booktitle = "General Relativity",
         year = 1972,
       editor = "O'Raifeartaigh, L.",
        month = jan,
        pages = "201-241",
}

@book{Cercignani2002,
  author    = {Cercignani, Carlo and Kremer, Gilberto Medeiros},
  title     = {The Relativistic Boltzmann Equation: Theory and Applications},
  series    = {Progress in Mathematical Physics},
  volume    = {22},
  publisher = {Birkh{\"a}user Basel},
  address   = {Basel},
  year      = {2002},
  isbn      = {978-3-0348-8165-4}
}

@book{Kolb:1990vq,
    author = "Kolb, Edward W. and Turner, Michael S.",
    title = "{The Early Universe}",
    reportNumber = "FERMILAB-BOOK-1990-01",
    doi = "10.1201/9780429492860",
    isbn = "978-0-429-49286-0, 978-0-201-62674-2",
    publisher = "Taylor and Francis",
    volume = "69",
    month = "5",
    year = "2019"
}

@article{Iocco:2008va,
    author = "Iocco, Fabio and Mangano, Gianpiero and Miele, Gennaro and Pisanti, Ofelia and Serpico, Pasquale D.",
    title = "{Primordial Nucleosynthesis: from precision cosmology to fundamental physics}",
    eprint = "0809.0631",
    archivePrefix = "arXiv",
    primaryClass = "astro-ph",
    reportNumber = "DSF-20-2008, FERMILAB-PUB-08-216-A, IFIC-08-37",
    doi = "10.1016/j.physrep.2009.02.002",
    journal = "Phys. Rept.",
    volume = "472",
    pages = "1--76",
    year = "2009"
}

@article{Birrell:2014uka,
    author = "Birrell, Jeremiah and Yang, Cheng-Tao and Rafelski, Johann",
    title = "{Relic Neutrino Freeze-out: Dependence on Natural Constants}",
    eprint = "1406.1759",
    archivePrefix = "arXiv",
    primaryClass = "nucl-th",
    doi = "10.1016/j.nuclphysb.2014.11.020",
    journal = "Nucl. Phys. B",
    volume = "890",
    pages = "481--517",
    year = "2014"
}

@article{Bond:2024ivb,
    author = "Bond, J. Richard and Fuller, George M. and Grohs, Evan and Meyers, Joel and Wilson, Matthew James",
    title = "{Cosmic neutrino decoupling and its observable imprints: insights from entropic-dual transport}",
    eprint = "2403.19038",
    archivePrefix = "arXiv",
    primaryClass = "astro-ph.CO",
    reportNumber = "N3AS-24-011",
    doi = "10.1088/1475-7516/2024/09/014",
    journal = "JCAP",
    volume = "09",
    pages = "014",
    year = "2024"
}

@article{Ma:1995ey,
    author = "Ma, Chung-Pei and Bertschinger, Edmund",
    title = "{Cosmological perturbation theory in the synchronous and conformal Newtonian gauges}",
    eprint = "astro-ph/9506072",
    archivePrefix = "arXiv",
    doi = "10.1086/176550",
    journal = "Astrophys. J.",
    volume = "455",
    pages = "7--25",
    year = "1995"
}

@book{Mo2010,
    place={Cambridge},
    title={Galaxy Formation and Evolution},
    publisher={Cambridge University Press},
    author={Mo, Houjun and van den Bosch, Frank and White, Simon}, year={2010}
}

@article{Nascimento:2024hle,
    author = "Nascimento, Caio and Loverde, Marilena",
    title = "{Cosmological perturbation theory for large scale structure in phase space}",
    eprint = "2410.05389",
    archivePrefix = "arXiv",
    primaryClass = "astro-ph.CO",
    doi = "10.1088/1475-7516/2025/06/002",
    journal = "JCAP",
    volume = "06",
    pages = "002",
    year = "2025"
}

@article{Synge:1934zzb,
    author = "Synge, J. L.",
    title = "{The energy tensor of a continuous medium}",
    journal = "Trans. Roy. Soc. Canada",
    volume = "III28",
    pages = "127",
    year = "1934"
}

@book{synge1957relativisticgas,
  title     = {The Relativistic Gas},
  author    = "Synge, J. L.",
  year      = "1957",
  publisher = "North-Holland Publishing Company",
  address   = "Amsterdam"
}

@article{Galishnikova:2022mjg,
    author = "Galishnikova, Alisa and Philippov, Alexander and Quataert, Eliot and Bacchini, Fabio and Parfrey, Kyle and Ripperda, Bart",
    title = "{Collisionless Accretion onto Black Holes: Dynamics and Flares}",
    eprint = "2212.02583",
    archivePrefix = "arXiv",
    primaryClass = "astro-ph.HE",
    doi = "10.1103/PhysRevLett.130.115201",
    journal = "Phys. Rev. Lett.",
    volume = "130",
    number = "11",
    pages = "115201",
    year = "2023"
}

@article{AparicioResco:2016xcm,
    author = "Aparicio Resco, Miguel and de la Cruz-Dombriz, {\'A}lvaro and Llanes Estrada, Felipe J. and Zapatero Castrillo, V{\'\i}ctor",
    title = "{On neutron stars in $f(R)$ theories: Small radii, large masses and large energy emitted in a merger}",
    eprint = "1602.03880",
    archivePrefix = "arXiv",
    primaryClass = "gr-qc",
    doi = "10.1016/j.dark.2016.07.001",
    journal = "Phys. Dark Univ.",
    volume = "13",
    pages = "147--161",
    year = "2016"
}

@article{Wojnar:2018hal,
    author = "Wojnar, Aneta",
    title = "{Polytropic stars in Palatini gravity}",
    eprint = "1808.04188",
    archivePrefix = "arXiv",
    primaryClass = "gr-qc",
    doi = "10.1140/epjc/s10052-019-6555-4",
    journal = "Eur. Phys. J. C",
    volume = "79",
    number = "1",
    pages = "51",
    year = "2019"
}

@article{Lope-Oter:2023urz,
    author = "Lope-Oter, Eva and Wojnar, Aneta",
    title = "{Constraining Palatini gravity with GR-independent equations of state for neutron stars}",
    eprint = "2306.00870",
    archivePrefix = "arXiv",
    primaryClass = "gr-qc",
    doi = "10.1088/1475-7516/2024/02/017",
    journal = "JCAP",
    volume = "02",
    pages = "017",
    year = "2024"
}

@article{EventHorizonTelescope:2019dse,
    author = "Akiyama, Kazunori and others",
    collaboration = "Event Horizon Telescope",
    title = "{First M87 Event Horizon Telescope Results. I. The Shadow of the Supermassive Black Hole}",
    eprint = "1906.11238",
    archivePrefix = "arXiv",
    primaryClass = "astro-ph.GA",
    doi = "10.3847/2041-8213/ab0ec7",
    journal = "Astrophys. J. Lett.",
    volume = "875",
    pages = "L1",
    year = "2019"
}

@article{Guerrero:2021ues,
    author = "Guerrero, Merce and Olmo, Gonzalo J. and Rubiera-Garcia, Diego and G{\'o}mez, Diego S{\'a}ez-Chill{\'o}n",
    title = "{Shadows and optical appearance of black bounces illuminated by a thin accretion disk}",
    eprint = "2105.15073",
    archivePrefix = "arXiv",
    primaryClass = "gr-qc",
    doi = "10.1088/1475-7516/2021/08/036",
    journal = "JCAP",
    volume = "08",
    pages = "036",
    year = "2021"
}

@article{Sotiriou:2008rp,
    author = "Sotiriou, Thomas P. and Faraoni, Valerio",
    title = "{f(R) Theories Of Gravity}",
    eprint = "0805.1726",
    archivePrefix = "arXiv",
    primaryClass = "gr-qc",
    doi = "10.1103/RevModPhys.82.451",
    journal = "Rev. Mod. Phys.",
    volume = "82",
    pages = "451--497",
    year = "2010"
}

@article{Magnano:1993bd,
    author = "Magnano, Guido and Sokolowski, Leszek M.",
    title = "{On physical equivalence between nonlinear gravity theories and a general relativistic selfgravitating scalar field}",
    eprint = "gr-qc/9312008",
    archivePrefix = "arXiv",
    reportNumber = "TO-JLL-P-3-93",
    doi = "10.1103/PhysRevD.50.5039",
    journal = "Phys. Rev. D",
    volume = "50",
    pages = "5039--5059",
    year = "1994"
}

@book{Green:1987sp,
    author = "Green, Michael B. and Schwarz, J. H. and Witten, Edward",
    title = "{SUPERSTRING THEORY. VOL. 1: INTRODUCTION}",
    isbn = "978-0-521-35752-4",
    series = "Cambridge Monographs on Mathematical Physics",
    month = "7",
    year = "1988"
}

@book{Ortin:2015hya,
    author = "Ortin, Tomas",
    title = "{Gravity and Strings}",
    edition = "2nd ed.",
    doi = "10.1017/CBO9781139019750",
    isbn = "978-0-521-76813-9, 978-0-521-76813-9, 978-1-316-23579-9",
    publisher = "Cambridge University Press",
    series = "Cambridge Monographs on Mathematical Physics",
    month = "7",
    year = "2015"
}

@book{Vereshchagin:2017,
  author    = {Vereshchagin, Gregory V. and Aksenov, Alexey G.},
  publisher = {Cambridge University Press},
  title     = {Relativistic Kinetic Theory: With Applications in Astrophysics and Cosmology},
  year      = {2017},
  place     = {Cambridge},
}

@article{Karciauskas:2022jzd,
    author = "Kar{\v{c}}iauskas, Mindaugas and D{\'\i}az, Jos{\'e} Jaime Terente",
    title = "{Slow-roll inflation in the Jordan frame}",
    eprint = "2206.08677",
    archivePrefix = "arXiv",
    primaryClass = "gr-qc",
    doi = "10.1103/PhysRevD.106.083526",
    journal = "Phys. Rev. D",
    volume = "106",
    number = "8",
    pages = "083526",
    year = "2022"
}

@article{Diaz:2023tma,
    author = "D{\'\i}az, Jos{\'e} Jaime Terente and Kar{\v{c}}iauskas, Mindaugas",
    title = "{Comoving curvature perturbation in Jordan and Einstein frames}",
    eprint = "2305.15326",
    archivePrefix = "arXiv",
    primaryClass = "gr-qc",
    doi = "10.1103/PhysRevD.108.083535",
    journal = "Phys. Rev. D",
    volume = "108",
    number = "8",
    pages = "083535",
    year = "2023"
}

\end{document}